%% file: Quotas_revised_v3.tex
\documentclass[useAMS,usenatbib,fleqn]{aastex63}

\usepackage{mathtools}
\usepackage{listings}
\include{packages}
\include{macros}
\include{macros_aas_journals}

\usepackage{amssymb,amsmath}
\def\Msun{M_\odot}

\begin{document}

\title{QUOTAS: A new research platform for the data-driven discovery of black holes}
\shorttitle{QUOTAS: QUasar ObservaTions And Simulations}

\shortauthors{Natarajan et al.}

\author{Priyamvada Natarajan}
\affiliation{Department of Astronomy, Yale University, 52 Hillhouse Avenue, New Haven, CT 06511, USA}
\affiliation{Department of Physics, Yale University, P.O. Box 208121, New Haven, CT 06520, USA}
\affiliation{Black Hole Initiative, Harvard University, 20 Garden Street, Cambridge MA 02138, USA}

\author{Kwok Sun Tang}
\affiliation{Department of Astronomy, University of Illinois Urbana-Champaign}

\author{Robert McGibbon}
\affiliation{Institute for Astronomy, University of Edinburgh, Royal Observatory}

\author{Sadegh Khochfar}
\affiliation{Institute for Astronomy, University of Edinburgh, Royal Observatory}

\author{Brian Nord}
\affiliation{Fermi National Accelerator Laboratory, P.O. Box 500, Batavia, IL 60510, USA}
\affiliation{Kavli Institute for Cosmological Physics, University of Chicago, Chicago, IL 60637, USA}
\affiliation{Department of Astronomy and Astrophysics, University of Chicago,IL 60637, USA}

\author{Steinn Sigurdsson}
\affiliation{Department of Astronomy \& Astrophysics and Institute for
Gravitation and the Cosmos, Pennsylvania State University, 525
Davey Lab, University Park, PA 16802, USA}

\author{Joe Tricot}
\affiliation{Sandbox@Alphabet, Mountain View, CA 94043, USA}

\author{Nico Cappelluti}
\affiliation{Department of Physics, University of Miami, Coral Gables, FL, USA} 

\author{Daniel George}
\affiliation{Sandbox@Alphabet, Mountain View, CA 94043, USA}

\author{Jack Hidary}
\affiliation{Sandbox@Alphabet, Mountain View, CA 94043, USA}

\date{\today}

\begin{abstract}
We present QUOTAS a novel research platform for the data-driven investigation of super-massive black hole (SMBH) populations. While SMBH data --- observations and simulations --- have grown in complexity and abundance, our computational environments and tools have not matured commensurately to exhaust opportunities for discovery. To explore the BH, host galaxy and parent dark matter halo connection - in this pilot version - we assemble and co-locate the high-redshift, z>3 quasar population alongside simulated data at the same cosmic epochs. As a first demonstration of the utility of QUOTAS, we investigate correlations between observed SDSS quasars and their hosts with those derived from simulations. Leveraging machine learning algorithms (ML), to expand simulation volumes, we show that halo properties extracted from smaller dark-matter only simulation boxes successfully replicate halo populations in larger boxes. Next, using the Illustris-TNG300 simulation that includes baryonic physics as the training set, we populate the larger LEGACY Expanse dark matter-only box with quasars, and show that observed SDSS quasar occupation statistics are accurately replicated. First science results from QUOTAS comparing co-located observational and ML trained simulated data at z ~ 3 are presented. QUOTAS demonstrates the power of ML, in analyzing and exploring large datasets, while also offering a unique opportunity to interrogate theoretical assumptions that underpin accretion and feedback models. QUOTAS and all related materials are publicly available at the Google Kaggle platform.\footnote{The full dataset - observational data and simulation data - are available at: {\bf https://www.kaggle.com/quasarnet/quasarnet} and the codes are available at: {\bf https://www.kaggle.com/datasets/quasarnet/quasarnet/code}}
\end{abstract}


\section{Introduction \& Motivation}
\label{sec:introduction}

Observations suggest that most, if not all, galaxies in the Universe host central super-massive black holes (SMBHs). Theoretical modeling at present successfully embeds and integrates the formation, growth, evolution and impact of black holes (BHs) on their environments - into the larger framework of structure formation in the standard cosmological model - revealing a profound underlying {\bf BH-galaxy-halo connection}. Given our empirical access and focus, much attention has been paid to the BH-galaxy coupling, its origin and evolution. More exhaustive exploration of the underlying correlation with the associated dark matter halo is warranted given that structure formation in the standard paradigm is driven by dark matter. The key motivation for QUOTAS\ rests on garnering deeper insights into this trium-variate connection by co-locating and analyzing observational data and simulated data. And derive in particular new ways of interrogating our input assumptions regarding BH growth that are currently implemented in cosmological simulations.
\\
The coupling between SMBHs and their host galaxies presents as empirical correlations between the BH mass and properties of their host galactic stellar populations (c.f.\,\citep{1998AJ....115.2285M, 2000ApJ...539L..13G, 2000ApJ...539L...9F, 2002ApJ...574..740T,2006Natur.442..888S,2009ApJ...698..198G,2013ARA&A..51..511K,2013ApJ...764..184M, 2014ARA&A..52..589H}). How and when these are established are currently open questions, and two possibilities are under active consideration \citep{2012NewAR..56...93A,2014GReGr..46.1702N}. Correlations are suspected to either arise from the initial conditions that govern the formation of BH seeds \citep{2013MNRAS.432.3438A,2019PASA...36...27W}, or from the coupled evolution of galaxies and their central SMBHs and that emerges over cosmic time \citep{1998MNRAS.300..817H,2019BAAS...51c..73N} in the standard cold dark matter driven structure formation model. Discriminating between these alternatives for the origin of observed correlations and discerning the causal mechanisms that underlie them are active research areas. To address this fundamental question, comprehensive data of SMBH populations across multiple cosmic epochs is required, including at very high redshifts, where observations currently peter out. However, with the spate of new observational facilities and instruments expected to come online shortly, vast amounts of data probing the earliest cosmic epochs will soon be available. Combining the inhomogeneous data from these various independent instruments and surveys will be an exceedingly complex task. Here, we construct the QUOTAS\ platform as an ongoing, actively updated effort and present the pilot version here as a template for continuing consolidation studies. 
\\
A broad data landscape --- including numerical simulations and multi-wavelength, multi-epoch observations --- has been critical for building models that have helped understand the symbiotic growth history of BHs and their host galaxies  \citep{1998MNRAS.300..817H,2012NewAR..56...93A,2012Sci...337..544V,2014GReGr..46.1702N,2019NatAs...3...48S}. Simulating BH growth in the full cosmological context is numerically challenging due to the active interplay of disparate physical scales -- from the Mpc scale gas-flows in the cosmic web that feed down to pc-scale gas accretion onto the accreting SMBH. Despite these hurdles, there has been enormous progress with the inclusion of physical processes that cannot be implemented in an abinitio fashion in computations via the use of so-called sub-grid models (c.f.~\citep{2005MNRAS.361..776S,2012MNRAS.420.2662D,2015MNRAS.452..575S,2020NatAs...4...10G}). Statistical studies of SMBH populations that leverage observational and simulated data with a new suite of tools is needed to enable breakthroughs in the fundamental understanding of the demographics and the cosmic accretion history of BHs. The Hubble Space Telescope (HST) and surveys like the ground-based Sloan Digital Sky Survey (SDSS) have acquired data on more than half a million quasars: their archives contain not only raw data (e.g., images and spectra), but also high-level, value-added, and science-ready products --- often in the form of relational databases. HST data are archived in the Mikulski Archive for Space Telescope (MAST) archive and SDSS data is publicly available on several online platforms --- e.g., CyVerse\footnote{\url{https://www.cyverse.org/}} and the NOAO Data Lab\footnote{\url{https://datalab.noao.edu/}} --- that include additional tools to investigate data in situ. In the near future, multiple new facilities  --- e.g., including the Vera Rubin Observatory (LSST) and the Nancy Grace Roman Telescope (NGRST) --- will all contribute significant amounts of data to the already large banks available for the investigation of SMBHs and their galaxy hosts. Meanwhile, on the numerical side, next-generation peta-scale simulations are also currently being developed to complement this observational deluge. These two kinds of data are not integrated and used effectively in concert at present, and our work presented illustrates one way forward to fill this important gap with QUOTAS. 
\\
To optimally mine these data, we propose development of customized, sophisticated methodologies and analysis tools that leverage machine learning (ML) algorithms in combination with our current domain knowledge. Flexible ML methodologies are the natural set of tools that for extracting maximal information from these varied and rich data. ML is already being vigorously applied in astronomy, for example, the Event Horizon Telescope (EHT) collaboration combined data from multiple radio observatories and relied extensively on ML methods to generate the silhouette of the SMBH at the center of M87 \citep{2019ApJ...875L...1E, 2019ApJ...875L...3E}. ML has demonstrated successes in a growing number of other astrophysics applications \citep{2019BAAS...51c..14N} --- including time domain astronomy \citep{2005ASPC..347..604M} image classification using a convolutional neural network (CNN) to morphologically classify and extract properties from large observed samples of nearby galaxies \citep{2020ApJ...895..112G}; and the use of Generative Adversarial Networks (GANs) to recover features in astrophysical images of galaxies beyond the de-convolution limit \citep{2017MNRAS.467L.110S}. In simulations, trained ML algorithms have been used to successfully mimic a full hydrodynamic cosmological simulation to study the galaxy-halo connection \citep{2016MNRAS.457.1162K}; to obtain new insights into cosmological structure formation \citep{2020arXiv201110577L}; including predictions of the evolution of angular momentum \citep{2020arXiv201202201C} and characterizing galaxy formation itself \citep{multi_epoch_ml,2022arXiv220102202V}. The power and prospects of insights gleaned in astronomy using ML for data-driven discovery have been amply demonstrated \citep{ 2020arXiv200512276M,2020MNRAS.496.5116B,2020arXiv201002926D,2020ApJ...889..151N,Djorgovski+2016}.
\\
At the highest redshifts, only the most luminous quasars are detected. Therefore, we have an incomplete and biased census of the accreting BH population. Our goal is to develop tools that will help uncover the more characteristic, average quasars at these early epochs, beyond the most extreme objects, to derive a comprehensive view the quasar population and disentangle seeding, accretion and dynamical processes that shape properties of the quasar population. We demonstrate that building the BH-galaxy-halo connection with QUOTAS\, will permit prediction and understanding of the properties of the more ubiquitous and more representative BH population at high redshifts. Two integrated quantities, the Black Hole Mass Function (BHMF) and the Quasar Luminosity Function (QLF) computed directly from observational data are typically used to constrain theoretical models of BH growth evolution. In what follows, we present the design and operation of the QUOTAS\ framework, built and tailored specifically to address the BH-galaxy-halo connection. QUOTAS is an on-going endeavor, and we are continually collating, updating and ingesting new quasar data as it becomes available. While the data from multiple observational surveys is currently published, and is hence available in QUOTAS. For all surveys listed in Table~2, we have survey selection functions; survey depths and survey areas from the painstaking compilation provided in \citet{Kulkarni+2019}. Knowledge of selection functions is critical to homogenizing the database and to derive quasar population properties like the BHMF and QLF for study and analysis. Though we have collated all current $z>3$ observed quasars in QUOTAS, not all of them have complete information, for some only photometric information is available and for others additional spectroscopic information is also available. In this paper, we describe the current full landscape of BH data, outlining observations collected from various astronomical surveys and simulations. The database design and the considerations for QUOTAS\ are discussed, as are illustrations of ease-of-use cases with concrete examples and visualizations. The concrete examples discussed here are focused on the SDSS survey data as this is the largest, full homogenized data set we have; are able to compare our results with previously published work and the sample size permits comparison with simulations. For $z>3$ SDSS quasars, we combine the observational data with corresponding simulation slices from the cosmological dark matter only LEGACY suite \citep{2022MNRAS.512...27M} to demonstrate the power that ML offers in interrogating our input model assumptions while tackling the BH growth problem. We follow conventional notation, wherein accreting (actively evolving/growing) BHs detected in optical wavelengths are referred to as ``quasars'' (with luminosities $\sim 10^{15}\,L_{\odot}$) while those detected in X-ray observations are denoted as ``AGN'' (with lower luminosities ranging from $10^{10}\,L_{\odot}$ to $10^{15}\,L_{\odot}$). AGN are rare at $z \geq 3$, our chosen epochs for study in this pilot project, and the majority have identifiable optical quasar counter-parts. When referring to multi-wavelength data for accreting SMBHs, we use the term AGN, and the term quasars when referring to optical data. 
\\
 ML is yet to be deployed for the comprehensive and systematic study of SMBH populations, though a recent attempt was made in deriving BH masses from spectral data \citep{2020arXiv201115095Y}. As show in this paper, the BH growth problem offers an important case study to develop and hone new ML algorithms, despite the known challenges for ML techniques, namely, extrapolative prediction \citep{2020arXiv201110577L}; and uncertainty quantification. Deep learning-based models for instance, lack native uncertainty quantification, which degrades their ability to make predictions beyond the data that they have seen and trained on \citep{2020arXiv201110577L,2019BAAS...51g.224N,2020PhRvD.102j3509H}. 
\\
In this paper, we present the highlights of the QUOTAS\, pilot project including early science results. In Section~2, we briefly introduce accreting black holes and describe the status of our current understanding of the co-evolution of SMBHs and their host galaxies in Section~3. In Section~4, we outline the full current data landscape for quasars. The design and development of QUOTAS\ database that collates observational and simulated data in a user-friendly queryable format is presented in Section~5. The illustration of ease of use case along with sample queries and an example of the workflow is presented in Section~6. The first results from this pilot study, where we focus on the $z > 3$ high-redshift SDSS quasar population deploying ML tools to compare simulated and observational data are presented in Section~7. In this section, upon combining with simulated data, we show how our underlying BH accretion model assumptions stand to be questioned with these new ML tools. We also demonstrate the power of ML to help formulate optimal survey strategies for uncovering the lower luminosity quasar population. Finally, we conclude in Section~8 with a brief discussion of our longer term goals: to continually update observational data, and standardizing it with appropriate selection functions and to develop a suite of tailored ML analysis tools that will help make predictions to uncover the more characteristic, representative SMBH population, beyond just the most extreme objects in order to fully explore and understand the underlying BH - host galaxy - parent dark matter halo connection.

\section{The properties of accreting black holes}

 In this section, we briefly outline the current understanding of accreting BHs and their properties to provide the broader context for our project. This discussion includes a brief description of the taxonomy, and a concise summary of current physical models for quasars. We discuss the essential theoretical and observational characteristics of accreting SMBHs, that inform the presently known correlations between galaxy properties and BH mass, and the population statistics relevant for the data-driven investigations explored in this work. 
 \\
 Less than a parsec across, the nucleus of nearly every galaxy appears to harbor a central SMBH \citep{2014arXiv1410.8717G}. A galactic nucleus is considered active when a hot and geometrically thin disk of accreting gas surrounds the central SMBH. While incomplete, there exist descriptions of unified models for active galactic nuclei (AGN) that describe their dynamics and account for their properties across a large range of physical scales around the growing BH (for example the proposal to unify all Radio-Loud AGN is presented in \citep{1995PASP..107..803U}). Per these unified schemes that attempt to couple the structure with phenomenology across scales - on small scales - ultraviolet radiation that is emitted by the accretion disk illuminates small, closely orbiting gas clouds and on - larger scales - the disk is inferred to be surrounded by a puffed-up toroidal region that consists of dusty molecular gas. Accretion activity is believed to be episodic, lasting $\lesssim 10^8$ years, during the life cycle of a galaxy; during which it can produce a tremendous amount of energy, often outshining the entire stellar content of the galaxy by several orders of magnitude. Integrating the various physical scales and the coupling of astrophysical processes therein has been the aspiration of multiple modeling efforts to date. We describe the observational signatures of black hole activity that arise on full the range of scales across wavelengths that are now integrated in a convenient form in QUOTAS. 
\\
The typical spectral energy distribution (SED) that models the energy output of an accreting BH comprises weighted emission, which is produced by astrophysical processes occurring in distinct physical regions and hence scales of the object --- the accretion disk, the hot corona, and the surrounding dusty torus. The observed SED of an accreting BH spans all the way from the radio to very high-energy gamma-rays ($10^9 - 10^{30}$ Hz) and consists of a superposition of multiple components --- a tall bump at high frequencies, a Black-Body component, and a plethora of emission and absorption lines. Details of emission mechanisms and the corresponding scales on which they operate can be found in these two comprehensive reviews, \cite{2015ARA&A..53..365N,2014SSRv..183..253P}). Observed sources are organized primarily into two classes - Type Is and Type IIs - based on the width of emission lines in their spectra, a feature which tends to correlate with the viewing angle to the central BH. Type I AGN are characterized by the presence of relatively broad ($1000 - 20000\, {\rm km\, s^{-1}}$) emission lines, which are produced by a population of the closely orbiting clouds that move in Keplerian orbits. Both these clouds and the accretion disk itself reside within the broad-line region (BLR), which can most readily be detected when the surrounding torus is viewed face-on. Type II AGN meanwhile are characterized by the presence of narrow spectral emission lines ($300 - 1000\, {\rm km\, s^{-1}}$), which originate from the more quiescent clouds that reside farther away from the accretion disk in the so-called narrow line region (NLR). When the galaxy is viewed edge-on, the obscuring dusty torus prohibits a direct view into the accretion disk, but permits a view of the toroidal region. Observed Type I AGN are more luminous than Type II AGN and therefore tend to comprise the bulk of observationally detected high-redshift sources. All optical quasars are Type I AGN and the thus far X-ray detected Type II AGN at $z \geq 3$ also have optical counter-parts. A population of heavily obscured Type II AGN that are as yet undetected are expected to exist across cosmic epochs and the high-redshift end population of these sources stands to be uncovered soon by the James Webb Space Telescope (JWST). For a more detailed description and illustration of these unified models, again we refer the reader to two comprehensive reviews and references therein \citep{2015ARA&A..53..365N,2014SSRv..183..253P}. Currently available photometric and spectral data for $z > 3$ sources are included in QUOTAS.

\section{Current understanding of BH-Galaxy-Halo co-evolution}

All black holes - ranging from stellar mass to SMBHs - can be characterized by three fundamental properties: mass, spin, and charge. This simplicity and small number of parameters needed to characterize BHs makes them attractive objects for systematic study. For astrophysical BHs, the subject of this work, charge is irrelevant, as they are expected to be charge-neutral \citep{1977MNRAS.179..457Z}. Conceptual models of BH growth are therefore focused on their mass-assembly history and their spin evolution. We briefly describe the currently observed correlations that link central BHs to their host galaxy properties. The formation of galaxies is intricately coupled to the underlying properties of their parent dark matter halos, therefore, the existence of a galaxy-halo connection is well established in cosmology \cite{Wechsler+2018}. Here, our aim is to build new synthetic models that include BHs into this picture explicitly and extend this conceptual frame to fully understand the BH-galaxy-halo connection by combining observational data with simulated data using ML tools. 
\\
Observations in the local universe reveal the existence of a correlation between the central SMBH and the stellar mass, luminosity, and velocity dispersion of stars in the inner regions (bulges) of their host galaxies \citep{2000ApJ...539L...9F, 1998AJ....115.2285M, 2002ApJ...574..740T, 2004ApJ...604L..89H, 2009ApJ...698..198G, 2013ApJ...764..184M, 2013ARA&A..51..511K, 2015ApJ...801...38W, 2018ApJ...864..146B, 2019ApJ...878..101S, 2020ApJ...888...37D}. These empirical correlations appear to hold for local, inactive, dormant SMBHs over approximately five orders of magnitude in BH mass (from $10^{5} -10^{10}\,\Msun$). At face value these correlations suggest that the assembly and growth of BHs is likely linked to the evolution of their host galaxies as first explored in \cite{1998MNRAS.300..817H}. These correlations albeit detected at these late cosmic times could also reflect initial conditions, encoding for instance, the native properties of the sites where initial BH seeds form efficiently. The origin of these correlations and whether they reflect deeper causal physical processes is a subject of active current study. For instance, it is unclear if these correlations are set up initially at the earliest epochs where galaxies are seeded with the first BHs. For example, one proposed channel for the formation of massive initial BH seeds from the direct collapse of pristine gas, naturally accounts for the origin of these correlations as an imprint of the initial physical seeding conditions \citep{2006MNRAS.371.1813L, 2007MNRAS.377L..64L, 2014MNRAS.443..648A,2019MNRAS.488.3268A,2019Natur.566...85W}. Heavy seeds have been invoked to account for the formation and assembly of the rare, most massive BHs at early cosmic epochs. By contrast, light initial BH seeds appear to be sufficient to account for the more modest mass SMBHs with masses in the range of  $10^{6} - 10^{8}\,\Msun$, detected in the nearby Universe, like the one harbored at the center of our own galaxy, the Milky Way. Alternatively, these measured local correlations may simply reflect the accumulated history of the stochastic assembly process of galaxies that occurs via accretion of mass and repeated merging \citep{2007ApJ...671.1098P, 2011ApJ...734...92J,2010MNRAS.407.1016H, 2018MNRAS.474.1995R, 2019MNRAS.489..802R} over cosmic time. 
\\
Understanding if these local correlations reflect deeper physical causes, and whether astrophysical processes that self-regulate BH growth and star formation activity in a galactic nucleus operate efficiently and in tandem, is a question of fundamental importance. Whether such a correlation holds at any earlier times for accreting BHs is not observationally known at present \cite{2011Natur.474..356T}. More observational data are needed to probe if these correlations were imprinted during early cosmic epochs \citep{2013ARA&A..51..511K, 2014GReGr..46.1702N, 2019PASA...36...27W} and if so, how and why they persist to late cosmic times. The differences between competing models of BH seeding and growth are naturally more pronounced at early cosmic times \citep{2012MNRAS.425.2854A,2018MNRAS.481.3278R} closer to the seeding era, however observational constraints at these epochs are currently sparser. Detecting growing SMBHs at these early epochs has been a technical challenge, as distant accreting sources are significantly fainter. At high redshifts, we preferentially detect only the rare, highest luminosity quasars and the models have only an unrepresentative sample of quasars to anchor and for guidance to predict properties of the more ubiquitous, and as yet undetected, lower luminosity quasar population. Therefore, to develop a more complete theoretical understanding, we need comprehensive sampling of the entire range of luminosities and hence BH masses powering quasars and their associated hosts. Besides, we are yet to have detectors with the appropriate spectral sensitivity to capture emission from the early dusty universe, a situation that has just started to alter with new data trickling in from the JWST \citep{Eilers+2022}. Our pilot project work presented here, targets early epochs and aims to provide a framework for making predictions for the properties of accreting black holes that have lower luminosities than the currently detected population of high-redshift quasars. Our goal is to showcase QUOTAS\ by coupling observations in the database with simulations and developing a set of customized ML tools that will permit probing the BHMF and QLF in deeper ways. QUOTAS is a live database that we plan to continually update as more quasar survey data and individual follow-up of sources becomes available. In this first phase, we use the collated SDSS data and coeval simulation slices to demonstrate the power of co-locating these datasets.

In future work, we plan to address and fill in a critical gap using trained ML to predict the abundance and luminosities of quasars that are one to two orders of magnitude lower in luminosity than the brightest quasars currently observed.

\section{The current data landscape for quasars}

In this section, we first discuss multi-wavelength observational data, including extant data archive facilities, their content and how they are currently configured for access and use before outlining the specific sources that we use to construct QUOTAS\,. We reiterate that QUOTAS represents an active, on-going effort. Both space-based and ground-based observatories have been essential for developing a multi-wavelength picture of central SMBHs. For example, data from HST, Chandra, XMM, Herschel, and Spitzer Space Telescopes spanning X-ray to infrared wavelengths, and their cross-correlations have revealed important ways in which BH accretion is related to key host galaxy properties like stellar mass and the star formation rate \citep{Chen_2013}). Most space missions and ground-based observational surveys have been efficiently cataloging their own data to permit public access after the lapse of the data proprietary period. One important challenge in dealing with quasar data pertains to how candidates are traditionally selected from surveys, using color-color diagnostics via the drop-out technique which was in fact originally developed to detect high-redshift quasars in \cite{Warren+1991a} that exploits the spectral shape of their SEDs per templates, successfully adapted later to detect high-redshift galaxies \citep{Steidel_1999}. Therefore, at progressively higher redshifts only the most luminous quasars are detected, making it difficult to discern properties of unobserved fainter sources. This is precisely where we see the power of trained ML techniques that will extract and utilize the characteristic properties of the detected sample to predict the properties of hitherto undetected fainter sources based on sources currently seen and a set of underlying modeling assumptions. In fact, ML studies of these combined data, as we demonstrate in Section~7, in turn also permit interrogation of our underlying modeling assumptions that are incorporated into current simulations regarding BH growth. This finding shows that ML offers a novel and potent arbiter between various model assumptions as well.
\\
HST has acquired data on a large number of quasars and their galaxy hosts over more than three decades of operation. This data is efficiently archived in and distributed from the Mikulski Archive for Space Telescopes (MAST) hosted and managed by the Space Telescope Science Institute.{\footnote{\url{https://archive.stsci.edu/hst/search_retrieve.html}}} In addition to the images - raw, cleaned, and processed - MAST provides high-level science-ready data products. In particular, MAST houses the AGN SED ATLAS \citep{2019MNRAS.489.3351B}, which contains the full SEDs of 41 AGN at present, derived from multi-wavelength photometry and archival spectroscopy, that combines information from eight MAST-supported projects (HST, SWIFT-UVOT, GALEX, PanSTARRS, IUE, FUSE, HUT, WUPPE) plus an additional nine other missions/observatories for $z > 3$ quasars. Beyond this Atlas, MAST also contains 80 SEDs of accreting sources, composites that are produced by mixing the SEDs of the central regions of active sources with their host galaxy SEDs. This high-level science products database contains a total of 121 AGN whose SEDs span a wavelength range from $\sim 0.09-30.0$ microns for these $z > 3$ sources. For some sources, additionally, there is even broader wavelength coverage extending into the X-ray, far-infrared, and radio.{\footnote{\url{https://archive.stsci.edu/hlsp/index.html}}} The archival data is stored in relational database tables, which are accessed through two user interfaces --- a Java-based application (``StarView'') and a web-based application: neither of these require the user to understand the database architecture or how to use Structured Query Language (SQL) to browse and extract data.
\\
The ground-based Sloan Digital Sky Survey (SDSS) has mapped a third of the sky and provides the most detailed three-dimensional data for more than three million unique astronomical objects.{\footnote{\url{https://www.sdss.org/dr16/data_access/}}} SDSS contains more than half a million quasars detected out to $z \sim 7$ and comprises a homogeneous sample of quasars with optical spectra. Each round of the SDSS quasar surveys was motivated by a different science goal. For example, the SDSS DR7 quasar catalog consists of 105,783 spectroscopically confirmed quasars from the SDSS-I/II survey whose aim was to study the QLF and clustering properties \citep{2009ApJS..182..543A}. With every data release, the number of detected quasars has grown dramatically. The latest SDSS DR14, not only increased the number of quasars by a factor of five compared to the SDSS DR7, it also went 1.5 magnitudes fainter, enabling the probing of quasar properties over a much larger luminosity range \citep{2018ApJS..235...42A}. Though the SDSS catalogs contain the X-ray, UV, optical, IR and radio imaging properties of the quasars wherever available, they lack spectral information for the majority of sources. Spectra have been typically obtained from detailed, independent follow-up studies, as seen from the references listed in our Tables~1 \& 2. The detection of quasars and their redshift estimates are done primarily using the color-color selection of drop-outs \citep{Richards+2002}. The SDSS archive also offers a range of sophisticated science products that are available as Value Added Catalogs (VAC). {\footnote{\url{https://www.sdss.org/dr14/data_access/value-added-catalogs/}}}  The fully standardized and homogenized data from the SDSS serve as our current benchmark for the rest of the observational data that we have collated. For all the surveys lists in Table~2 - the Quasar Photometry Table - we have in hand the selection functions, survey area and survey depth - complete survey specifications from the homogenizing work presented in \citet{Kulkarni+2019}, that are needed for the construction of the QLF and the BHMF. We note, however, that the BHMFs computed from observations and presented in this work for comparison with simulations are all derived from SDSS DR7. All the currently available spectral data in QUOTAS are listed in Table~1 and the collated attributes are listed in the schematic in Fig.~2. The publicly available resource that we utilize to collect data for QUOTAS\ is the NASA-IPAC Extra-galactic Database (NED){\footnote{\url{https://ned.ipac.caltech.edu}}}. This database contains basic information --- e.g., coordinates and magnitude --- on all observed astronomical objects and references to all published papers that describe the sources, and this includes reports of detailed follow-up work. Unlike other archives like the MAST, SDSS and SDSS VAC, the NED repository does not contain customized value added properties. However, it contains the most comprehensive compilation of publication references for surveys and follow-up studies and is consistently updated. It is this feature of NED - the comprehensive bibliographic data - that we deploy for QUOTAS. In this first phase of our work, we derive the BHMF of collated SDSS quasars from the database as a function of cosmic time in redshift slices $z>3$.
\\
Due to the plethora of new observatories and instruments on the ground and in space due to come online soon, we are entering an era of an unprecedented deluge of multi-wavelength and multi-messenger data on active and dormant SMBHs. From the ground, the Vera Rubin Observatory's projected output catalog is expected to provide 15 petabytes of data over 10 years of operation \citep{2017ASPC..512..279J}. The targeted Rubin AGN Survey, in particular, is expected to uncover $N \sim 10^7$ optical quasars \citep{2009arXiv0912.0201L} from the near to the distant Universe. JWST \footnote{\url{https://www.jwst.nasa.gov/}}, has instruments with sensitivity in the mid-, near- and far-infrared wavelength (ranging from $\sim1-30$ microns), and has started opening up an entirely new window into the early Universe, revealing populations of early BHs and the first galaxies \citep{Eilers+2022}. With accumulating JWST data in the coming years, earlier theoretical predictions stand to be tested \citep{2016MNRAS.459.1432P, Natarajan_2017, 2018MNRAS.474.2617B}. The planned Laser Interferometer Space Antenna (LISA) mission\footnote{\url{https://www.elisascience.org/}} to be launched in the early 2030's also offers exciting prospects for detecting gravitational waves from binary SMBH mergers that will most likely also be accompanied by multi-wavelength electromagnetic counter-parts \citep{2019arXiv190306867C, 2019BAAS...51c.336T, 2018MNRAS.477..964K, 2020MNRAS.491.2301K}. Other relevant facilities currently providing new data on SMBHs include the Canadian Hydrogen Intensity Mapping Experiment (CHIME);\footnote{\url{https://chime-experiment.ca/}} the soon to be launched NGRST;\footnote{\url{https://roman.gsfc.nasa.gov/}} and the Euclid Space Mission\footnote{\url{https://sci.esa.int/web/euclid}}; the Square Kilometer Array (SKA);\footnote{\url{https://www.skatelescope.org/}} the Next Generation CMB Experiment Stage 4 (CMB-S4);\footnote{\url{https://cmb-s4.org/}} and the Advanced Telescope for High-ENergy Astrophysics (ATHENA)\footnote{\url{https://sci.esa.int/web/athena}}.

\subsection{Simulations of Galaxy Assembly \& Black Hole Growth}

Used together, observations and numerical simulations have enabled detailed studies of many aspects of BH physics, including the accretion process, feedback from growing BHs, the evolution of the BH population and how BHs fit into the larger picture of structure formation in the Universe \citep{springel+2005}. The power of simulations lies in the density of information that can be mined from them for all constituent particles - dark matter, baryons and BH sink particles that enable connecting physical scales, and regimes across cosmic epochs. Simulations also place BH growth in its full astrophysical context against the backdrop of dark matter driven galaxy formation in our cold dark matter dominated Universe. Simulations have shown that the energy released from the accretion process has the capacity to impact a wide range of spatial scales --- from the smallest scales where general relativity and magneto-hydrodynamic processes dominate to the largest scales wherein outflows driven by accreting black holes and their feedback are relevant \citep{2000NCimB.115..795B, 2001MNRAS.326L..41K, 2003ApJ...589..444G, 2005paoa.confE..15K, 2005ApJ...630L...5M,2011MNRAS.410..919J, 2018NatAs...2..198H}. This coupling of scales might well hold the key to understanding the observed correlations between BHs, their host galaxies and their parent dark matter halos. Due to the complexity of the involved physics and numerical cost of ab-initio simulations of BH formation and subsequent evolution, simulators have adopted a strategy in which they focus on specific aspects of the problem. For instance, in a given simulation suite, a subset or all of the relevant physics involved such as gravity, hydrodynamics \citep[e.g.][]{2018MNRAS.478..995B}, magneto-hydrodynamics and radiation transfer \citep{2005ApJ...630L...5M, 2007Ap&SS.311..117H, 2011MNRAS.418L..79T, 2018MNRAS.474L..81L} are followed and tracked. Besides this, they need to account for the initial conditions for forming/seeding BHs and the level of detail with which the environment around them is rendered. Combining information from different simulations has allowed us at present to build a consistent picture spanning a large range of length scales. Take for example, the role of jets driven by BHs,  small scale general relativistic magneto-hydrodynamic simulations are needed to investigate the connection of the BH spin to the power of radio-emitting jets \citep{2010ApJ...711...50T}. Complementary simulations by \citet{2011MNRAS.411..155G} bridge the gap to larger spatial scales and look at the impact of jets on galactic discs in hydrodynamical simulations. This permits assessing the impact of jets on the provision of fuel for the growing BH and the global star formation in the host galaxy. The latter simulations use idealized initial conditions for their setup and neglect the impact of feedback on the gas in the immediate vicinity of the central BH. Lower resolution cosmological zoom simulations on the other hand are able to resolve individual objects such as the dark matter halos that host BHs with spatial resolutions of up to tens of pc \citep{2011MNRAS.412.1341D} and at the same time they track the cosmological environment and its coupling with the BH \citep{2019MNRAS.483.3488B}. These simulations not only track BH relevant physics but also follow the evolution of the host galaxy in terms of its star formation, metal enrichment and feedback from exploding supernovae \citep{2019MNRAS.483.3488B,2019Natur.566...85W,2020MNRAS.497.3761L}. 
\\
The ever-increasing number and diversity of observed BHs and AGN requires numerical simulations that span a large parameter space of initial conditions for the formation of BHs and their associated cosmic environment. This kind of coupling with the larger scale environment goes beyond what current zoom simulations can computationally provide. As quasars and AGN are extremely rare objects, rare even compared to galaxies in the cold dark matter structured Universe, simulations in which their formation and growth are tracked self-consistently over cosmic time are extremely computationally challenging. To alleviate this fundamental mismatch, trade-offs are often made between the simulation volume and resolution. For example, large scale cosmological simulations of cosmic volumes of $\sim (100$ Mpc$)^3$ with coarser spatial resolution of $\sim \rm{kpc}$ have been performed \citep{2015MNRAS.446..521S}. These simulations typically include sub-grid models for the baryonic physics i.e. to model star formation, cooling of gas, chemical networks, stellar population evolution, supernovae feedback, and metal enrichment tracked via radiative transfer, but these versions can only include limited BH physics \citep{2005Natur.433..604D, 2006ApJ...645..986R, 2007MNRAS.380..877S, 2011MNRAS.412.1341D, 2014MNRAS.444.1518V, 2015MNRAS.446..521S, 2015MNRAS.452..575S, 2017MNRAS.468.3935H, 2018MNRAS.479.4056W}. To match the statistical properties of AGNs detected in large observational surveys yet larger scale simulations with volume $ >(500$ Mpc$)^3$ are needed \citep{2006MNRAS.370..645B}. Given current computational limits, such simulations have necessarily been dark matter only and are able to only incorporate simplified, semi-analytic models for the physics of BHs and galaxies that are added in by-hand post-hoc. QUOTAS offers a novel way to test the association between observed BHs with their putative host galaxies and dark matter halos assumed in semi-analytic models as well. The use of coeval simulation slices corresponding to redshifts where we have observational data permits unique tests of model assumptions including feedback from accreting BHs that appears to not only shape galaxies but also regulate the BH growth histories in quasars.
\\
We demonstrate with examples the kind of new association studies that QUOTAS\ allows us to pursue. The first is via detailed tracking of the potential assembly history of the dark matter halo and the central BH that hosts an observed $z = 3$ quasar, shown in Fig.~\ref{fig:mergerhistories}. The second illustration involves using ML tools and their application to expand available simulation volumes - to use the simulated data from a smaller volume cosmological simulation - the LEGACY mean box $(\sim 110$ Mpc$^3)$- to derive the full set of characteristic properties of halos that in turn allows us to expand the simulation box size while retaining all the statistical properties of halos including their correlations and clustering. We demonstrate the power of ML by comparing the normalizing flow model derived characteristic properties of dark matter halos from the LEGACY mean box with the statistics from the larger volume LEGACY (1 Gpc)$^3$ Expanse box. Finally, showcasing the combination of simulated and observational data at just one epoch $z \sim 3$, with the constructed BHMF we show how ML tools permit examining and interrogating assumptions that are integral to modeling the growth of SMBHs. In future work, we plan to develop additional ML tools in the next stage of our project that will allow us to further mimic larger simulated volumes that correspond to observational survey volumes and predict the putative locations for fainter as yet undetected quasars. Here we present the first key step in building up and exploring the BH-galaxy-halo connection that underpins the formation and evolution of all structures including the non-baryonic (dark matter), baryonic (galaxies) and BHs.
\\
Over the last two decades with the advent of powerful computational architectures, the size of data and complexity of numerical simulations has significantly increased. Typical simulated data sets are of Terabyte size and contain $N > 1024^3 $ computational elements that represent gas, stars, black holes and dark matter. For each of these elements, the equation of motion is solved taking into account the full physical information available in the simulation. Besides containing phase-space information on momentum and spatial coordinates, computational elements also carry further information to compute the evolution of the system in time such as e.g. the spin and mass of black holes, chemical composition of gas and stars, the present mass function of stars that is associated with an individual star particle and internal energy and density of gas. In addition, it has become standard practice to add evolutionary information to elements that are needed for sub-grid computations during the simulation or post-processing of the data. These include the birth age of a star particle to track the evolutionary state of the underlying stellar population that it represents or the maximum temperature an individual gas element achieves during its evolution, to name just a few. It is evident that this list could be arbitrarily expanded putting pressure on resulting data volumes. On top of this, simulation data sets include frequent snapshots with the full information of the state of all constituents - dark matter, stars, gas and BHs - and their evolution through time. Further dramatic increase in data volumes is expected over the next decade with new peta-scale codes that will increase the number of computational elements significantly. To deal with such large data volumes the optimal strategy is to focus on only a subset of the information. In our case, we are concerned with BH formation and evolution which takes place in gravitationally bound dark matter halos. Focusing only on material bound to dark matter halos reduces the data volume to roughly $\sim 20-30 \% $. One standard way to identify dark matter halos is based on linking dark matter particles by assuming a fixed linking length that is a fraction $b=0.2$ times the mean inter-particle separation \citep{1985ApJ...292..371D}. These so-called Friends-of-Friends (FOF) halos contain further sub-structure in form of sub-halos that may contain galaxies and BHs. Several different approaches have been utilized in the literature to identify subhalos based on their being gravitationally self-bound \citep{2001MNRAS.328..726S,2013ApJ...762..109B}. Halo catalogs have become the standard way of providing the information from cosmological simulations to the community. They store all the relevant information on collapsed halos; global properties of dark matter halos (e.g. total mass and angular momentum), properties of their constituent galaxies (e.g. star formation rate and metallicity), gas (temperature, pressure and entropy) and BH (e.g. mass, accretion rate and spin). This set is complemented by information on all the gas, stars, BH and dark matter in the region defined by the halo. Halo catalogs are easily ingested in SQL databases  (e.g. Millennium simulation \citet{2006Natur.440.1137S}\footnote{\url{https://www.g-vo.org}}, Illustris simulation \citep{2015A&C....13...12N}\footnote{\url{https://www.illustris-project.org}} or the EAGLE project \citep{2015MNRAS.446..521S}\footnote{\url{http://icc.dur.ac.uk/Eagle/database.php}}) which allow easy and fast access for population studies. A complementary approach is to use a web-based Application Programming Interface (API) which allows for in depth analyses of data beyond halo catalogues without the need of storing a local copy. Within QUOTAS\ we include simulation data using a combination of these approaches to maximize the ease of application for ML techniques.

\section{QUOTAS\,: Database Design \& Development}

With  QUOTAS\, our goal has been to create an optical quasar repository of $z > 3$ quasars that includes a combination of raw and high-level products; covering a broad range of cosmic epochs; and co-locating observational and simulated data of corresponding epochs (see Tables~1 \& 2 for published references for the data). In this pilot project, focused on investigating the high redshift quasar population, we curate data for these $z > 3$ sources in three distinct tables: {\bf the Quasar Spectra Table} that contains BH mass estimates, spectroscopic data like line-widths used to derive independent BH masses and errors when available, and quantities like the accretion luminosities that also depend on the BH mass (source list with references listed in Table~1); {\bf the Quasar Photometry Table} that contains all the photo-metric data measured across bands (source list with references listed in Table~2) and {\bf the Halo Properties Table} that includes the properties of the putative parent dark matter halos for the corresponding redshift slices spanning $3 < z < 7$. The distribution of the entire collated quasar population including all sources listed in Table~1 on the sky are shown in Fig.~1. Fig.~\ref{fig:db_table} depicts a color-coded schematic of these data structures outlining the stored attributes: blue boxes denote the primary data source - the NASA Extragalactic Database (NED); the red Quasar Spectra Table (Table~1) contains the properties of individual objects collected from NED and multiple additional published sources and the orange Quasar Photometry Table includes data from NED and surveys including their selection functions, survey depth and survey areas (Table~2). We use the selection functions, survey depth, areas and other details that have been computed to homogenize data from \citet{Kulkarni+2019}. They present selection functions for the surveys listed in Table~2 and these are included in the Selection Function data structure in QUOTAS as shown in the schematic of Fig.~2. For science results showcased here in this first paper, that include the computation of the BHMF we use only the SDSS data. In QUOTAS, data attributes like the photometry, redshift, luminosity, etc downloaded from NED are first flattened and combined into the respective attribute tables whose formats follow the column naming convention as described in \citep{2011ApJS..194...45S}.

\subsection{Collating the Observational Data for QUOTAS\,}

 Our primary source data set is NED which contains sources retrieved from several independent optical surveys, principally the SDSS. The following procedure was used to construct our database: first, we used the Astropy-affiliated package \texttt{astroquery} on NED to search and collect all known quasars at $z\geq3$ \footnote{\url{https://ned.ipac.caltech.edu}, DOI:10.26132/NED1}, and in the next step, we used the \texttt{By Parameters} search query to obtain basic information ---  object name, RA, DEC, redshift, photometry, spectra, images, and associated references --- for each extracted object. From painstaking manual follow-up of published NED sources, we scoured tables in papers and the accompanying Vizier catalogues site\footnote{\adb{\url{https://vizier.u-strasbg.fr/}}} for raw and derived properties of individual quasars \citep{2019MNRAS.488.1035K}. We collected object names, positions, redshifts, luminosities, masses, line widths, Eddington ratios, and references wherever available. 
 
\begin{table*}
\centering
\begin{tabular}{ccccc}
\hline 
$z$ Range & Num. Quasars & Luminosity Estimator & Vel. Disp. Estim. & Reference  \\
\hline
3.01-4.33 & 43 & \begin{tabular}{@{}c@{}} $L_{\lambda}(1350)$, $L_{\lambda}(3000)$, \\ $L_{\lambda}(5100)$, $L(H\beta)$, $L(MgII)$ \end{tabular},& $H\beta$, Mg II, C IV  & \cite{2019ApJS..241...34S} \\
3.00-5.46 & 7955 & \begin{tabular}{@{}c@{}} $L_{\lambda}(1350)$, $L_{\lambda}(3000)$, \\ $L_{\lambda}(5100)$, $L(H\beta)$, $L(MgII)$ \end{tabular} & H$\beta$, Mg II, C IV & \cite{2011ApJS..194...45S} \\
3.00-4.89 & 26571 & $L_{\lambda}(1350), L_{\lambda}(3000), L_{\lambda}(5100)$ & Mg II, C IV &\cite{2017ApJS..228....9K} \\
3.22-3.76 & 24 & $ L_{\lambda}(3000), L_{\lambda}(5100)$ & H$\beta$, Mg II & \cite{2015ApJ...799..189Z} \\
4.52-6.41 & 21 & $L_{\lambda}(3000)$ & Mg II & \cite{2011ApJ...739...56D} \\
4.66-4.92 & 40 & $L_{\lambda}(1450), L_{\lambda}(3000) $ & Mg II    & \cite{2011ApJ...730....7T} \\
5.64-6.42 & 50 & $L_{\lambda}(1350),  L_{\lambda}(1700), L_{\lambda}(3000)$ & Mg II, CIV & \cite{2019ApJ...873...35S} \\
5.77-6.31 & 6  & $L_{\lambda}(1350),  L_{\lambda}(3000)$ & Mg II, CIV & \cite{2007AJ....134.1150J} \\
5.82-6.28 & 5  & $L_{\lambda}(1350),  L_{\lambda}(3000)$ & Mg II, CIV & \cite{2007ApJ...669...32K} \\
5.90      & 1  & $L_{\lambda}(3000)$ & Mg II & \cite{2018ApJ...867...30E} \\
5.98-6.44 & 9  & $L_{\lambda}(3000)$ & Mg II & \cite{2010AJ....140..546W} \\
5.98-6.9 & 38  & $L_{\lambda}(3000)$ & Mg II, CIV & \cite{2022ApJ...941..106F} \\
5.98-6.9 & 69  & L(Ly{$\alpha$}) & Ly$\alpha$ & \cite{2022ApJS..259...18M} \\
6.13      & 1  & $L_{\lambda}(3000)$ & Mg II & \cite{2017ApJ...845..138S} \\
6.48-7.08 & 11 & $L_{\lambda}(3000)$ & Mg II & \cite{2017ApJ...849...91M} \\
6.60-7.10 & 4  & $L_{\lambda}(3000)$ & Mg II, CIV & \cite{2014ApJ...790..145D} \\
7.07      & 1  & $L_{\lambda}(1350)$ & Mg II & \cite{2019ApJ...872L...2M} \\
\hline 
\end{tabular}
\caption{The compiled list of quasars including additional properties of the SMBHs included in our database. This table comprises all reported spectroscopic data from quasar samples (surveys listed in Table 2) as well as individual quasars.}
\label{tab:bh_prop_samples}
\end{table*}

\subsection{Measured \& Derived Properties of BHs powering quasars}

Mass and spin are fundamental properties of black holes. While mass estimates are available for several thousand sources at the present time \citep{2013ApJ...764...45K,2014SSRv..183..253P,2019NatAs...3...11V}); spin measurements are available only for a handful of sources (\cite{2020arXiv201108948R,2006AN....327.1039N}). As the observed correlations are between SMBH mass and host galaxy properties, in this work, we restrict our investigation to BH masses. By construction the database structure of QUOTAS\ is kept inherently flexible to accommodate data on BH spins as and when they become available, at which time spins can be seamlessly incorporated to the other stored source attributes. QUOTAS\ includes BH masses derived from multiple techniques where available. Many independent methods have been used to derive BH masses - including from the mapping of the orbits of individual stars, which is possible only for the Milky Way \citep{1997MNRAS.291..219G,1998ApJ...509..678G}; to modeling the orbits of bulge stars from imaging and spectroscopy as performed for nearby galaxies \citep{1994AJ....107..634T} and using measurements of the speed of rotating gas using water mega-masers as tracers of the mass \citep{1995Natur.373..127M}. It has been demonstrated with local data that the luminosity of an accreting SMBH can be used to infer its BH mass across a range of redshifts.

One widely adopted method for SMBH mass determination, available for many sources in QUOTAS\, assumes that the BLR is virialized and that the motion of the emitting clouds reflects the gravitational potential of the central BH  \citep{1982ApJ...255..419B,1993PASP..105..247P}. Under this assumption, the black hole mass $M_{\bullet}$ can be estimated via: 
\begin{eqnarray}
    M_{\bullet} = f \frac{V_{\rm{vir}}^2R_{\rm BLR}}{G}, 
    \label{eq:massvirial}
\end{eqnarray}

\noindent 
where $V_{\rm{vir}}$ is the virial velocity, $R_{\rm BLR}$ is the size of the BLR, $G$ is Newton's gravitational constant, and $f$ is the virial coefficient that accounts for the geometry and kinematics of the material around the BLR \citep{2013BASI...41...61S}. The virial velocity can be estimated using the velocity dispersion derived from the width of observed BLR emission lines. The time-lagged broad-line response to variations in the continuum flux enables the measurement of the light travel time from the central ionizing source to the broad line regions, which can be used to estimate $R_{\rm BLR}$. Acquiring these time lags from reverberation data is challenging, as it requires a long observational baseline, monitoring an accreting BH for six months to a year \citep{2004ApJ...613..682P, 2017ApJ...849..146G}. In QUOTAS, we include BLR determined masses and errors therein where available. There are many interesting questions relevant to observed variable quasars, like the sources referred to as "changing look" quasars. However for these, uniform sampled variability data are not currently available, though deep-learning methods have been recently used to fill-in data gaps \citep{2020ApJ...903...54T}. Data on time variability of quasars is not included in this current pilot version of QUOTAS.

An alternative method for SMBH mass measurements, that is not predicated on the assumption of virialization of the BLR, is one in which line and continuum luminosities from the X-ray, ultraviolet, infrared, and optical wavelengths can instead be used to estimate the BLR size. For this purpose, continuum luminosities $L_{\rm cont}$ are often preferred over line luminosities, because they tend to yield a tighter correlation with the size of BLR \citep{2000ApJ...533..631K, 2005ApJ...629...61K, 2009ApJ...697..160B}. Reverberation mapping has revealed a tight correlation between the size of BLR and the continuum luminosity. This empirical scaling relationship is then used to derive black hole mass. Therefore, the continuum luminosity $L$, combined with the widths of broad emission lines ($\Delta v$), can be used to derive estimates of the mass of the black hole:
\begin{eqnarray}
    \log \left(\frac{M_{\bullet}}{{\rm{M}_{\odot}}} \right) = a + b \log \left( \frac{L}{L^{\prime}} \right) + c \log \left(\frac{\Delta v}{\Delta v^{\prime}}\right),
\label{eq:mass}
\end{eqnarray}

\noindent where $L^{\prime} = 10^{44}\, {\rm erg ~s}^{-1}$,  $\Delta v^{\prime} = 1\, {\rm km ~s}^{-1}$. In general, the values of pre-factors $a$ and $b$ depend on the choice of the luminosity and velocity dispersion estimators, and if a virialized BLR is assumed, then the coefficient of the line width is typically taken to be $c=2$ \citep{2012ApJ...753..125S}. Various cross-calibrations for this relation have been proposed \citep{Shen_2012, 2006ApJ...641..689V, 2009ApJ...699..800V, 2012MNRAS.427.3081T}. Additional calibrations are needed to reconcile with estimates from other methods \citep{2017MNRAS.465.2120C} and for the determination of robust error-bars. For example, the masses derived using the continuum luminosity are often calibrated with BH mass measurements obtained from reverberation mapping. Additional BH mass estimates are often made using individual lines like the Carbon-IV lines \citep{2006ApJ...641..689V}. In QUOTAS\ we provide all available BH mass estimates as shown in the schematic (Fig.~\ref{fig:db_table}) into {\bf the Quasar Spectra Table}, computing them when relevant measured quantities are available. We also keep track of the error bars in the mass estimates when available. We focus on SMBH masses as the BHMF is a key component of theoretical models to study BH growth that occurs via both merging and accretion. Apart from the study of individual objects, population studies of quasars can reveal aspects of their evolution across multiple cosmic epochs and are crucial for building evolutionary models. Large quasar samples derived from ambitious astronomical surveys have enabled the study of the cosmological evolution of the QLF. A critical quantity for theoretical modeling, the QLF, encapsulates the growth history of quasars through accretion. The importance of the QLF was demonstrated in early modeling work, and it continues to be vigorously studied in the field (for example see early work by \citep{1992MNRAS.259..725S,1998MNRAS.300..817H,2008A&A...490..905H}). In studies using the QLF, the growth via mergers are additionally accounted for with the modeling assumption that mergers themselves trigger additional accretion episodes. The QUOTAS\ database by design permits easy computation of both the QLF and BHMF as demonstrated in the workflow examples provided below.

The QLF is one of the key inputs to current conceptual models of BH growth (see reviews that outline the methodology by \cite{2012Sci...337..544V,2014GReGr..46.1702N} and references therein). The specifications of observational surveys needs to be taken into account to determine the QLF. Observationally determined QLFs are then compared with those predicted by theoretical models of BH growth. Calibrated with data, the honed models are extrapolated and then used to predict QLFs down to fainter luminosities compared to current observed limits and out to larger redshifts than current detections. The most recent predicted QLFs extrapolated out to $z= 9$ derived from the combination of observational data and modeling can be found in \cite{2018MNRAS.481.3278R}). The QLF provides a census of the number of sources at a given redshift and absolute magnitude $M$ and is not easy to estimate. It is computed as follows:
\begin{eqnarray}
    \phi(M, z) \approx  \sum_{j=1}^{N} \frac{1}{V_{a}^{j}} (\Delta M )^{-1},
    \label{eq:qlfa}
\end{eqnarray}
\noindent where $\Delta M $ is magnitude range \{$M$, $M+dM$\}, and $V_a^j$ is the effective volume for a source $j$. This is usually estimated with the binned-volume method \citep{1980ApJ...235..694A}, where the effective volume is estimated using:

\begin{eqnarray}
    V_{a}^{j} =  \int_{M}^{M+\Delta M} \int_{\Omega} \int_{z_{min}}^{z_{max}} dM dz f(M,z) \frac{dV}{dzd\Omega},
    \label{eq:qlfb}
\end{eqnarray}
\noindent where $f(M,z)$ is the survey sample completeness, $z_{min}$ and $z_{max}$ define the redshift bin over which the QLF is averaged, $\frac{dV}{dzd\Omega}$ is the comoving volume element, and $d\Omega$ is the survey area. The minimum $z_{min}$ is determined by the specifications of the survey. The maximum redshift $z_{max}$ depends on individual sources and is defined by the maximum redshift out to which the source would remain detectable given the survey detection limit, as specified by the selection function of the survey. QUOTAS\ collates data for computing the QLF by merging observations taken by several independent surveys while keeping track of the individual selection functions of the surveys - these are stored in {\bf the Quasar Photometry Table} (see the schematic in Fig.~\ref{fig:db_table}).
\\
The BHMF is a critical diagnostic that permits understanding of the mass assembly history of black holes over cosmic time. We briefly outline its computation from quasar survey data, and in this work, focus exclusively on deriving the BHMF from SDSS data to illustrate. The first step in computing the BHMF involves evaluating the detectable volume of quasars in a given survey, using its design specifications and the computed QLF as noted above. Using for instance, the SDSS DR3 data, we calculate the volume $V_{\rm bin}$, the effective volume within which a quasar of a given luminosity would be detectable. This is done with the data from our photometry data table. $V_{\rm bin}$ is defined as follows:
\begin{eqnarray}
 V_{\rm bin} = \int_{M_{\rm min}}^{M_{\rm max}}\,dM\,\int_{z_{\rm min}}^{z_{\rm max}}\,dz\,f(M,z)\,\frac{dV}{dz},   
\end{eqnarray}
where $f(M,z)$ is the survey selection function, for quasars as a function of magnitude $M$ and redshift $z$; $M_{\rm min}$ and $z_{\rm min}$ are the magnitude and redshift of the observed quasar and $M_{\rm max}$ and $z_{\rm max}$ is the magnitude of the survey depth limit and the maximum detectable redshift given the luminosity of the quasar and $dV/dz$ in the redshift evolution of the volume element in the LCDM Universe which in turn is a function of the luminosity distance $d_L(z)$; $\Omega_{\rm m}$ and $\Omega_{\Lambda}$. For the next step we look up the corresponding BH mass for the quasar from the spectroscopic data table and then bin in logarithmic mass intervals. Finally, we combine the volume calculation to compute the volume density of quasars that falls within these mass bins. For each quasar in QUOTAS\, its computed value of $V_{\rm bin}$ is also stored using the $f(M,z)$ for the survey in which it is detected. Every survey has a selection function, and as noted previously, we use the compilation from \citet{Kulkarni+2019} in QUOTAS. In this work, we use the selection function for SDSS quasars and the compute the BHMF for SDSS quasars, which is shown in Fig.~\ref{fig:bhmf_z} and Fig.~12. As part of our ongoing, continual assembly of QUOTAS, we are updating additional attributes to our tables as they become available. Our database is currently fully updated for SDSS quasars.
\\
Instead of archiving only the reported SMBH mass, in QUOTAS\, all the parameters used to derive the mass, like the line-widths, and line luminosities are also tabulated when available. The redshift distribution of quasar counts in the database are plotted in Fig~\ref{fig:distributions}. We also include serendipitously discovered high-redshift quasars into QUOTAS. We have carefully assigned unique object identifiers and recorded relevant information like the survey flux limits, redshift limits, sky area, magnitude limit for sources in the database whenever available as described above. To date, QUOTAS\ contains the most comprehensive and growing compilation of optical quasars and contains in excess of 30000 unique objects. Quasars in Table~2 in our database contain sufficient parameters to compute their BHMFs and the QLFs, the two key quantities needed for comparison with models of the assembly history of black holes. In this work, we present comparison of the observationally determined BHMF with the one derived from the corresponding simulated slice in QUOTAS\ for SDSS quasars as this is the largest uniform sample that permits straightforward comparison with simulations (see Section~7.3 below and Fig.~12).

 \begin{table*}
  \centering
  \begin{tabular}{cllrr}
    \hline
    $z$ range & Survey & Reference & Number & {Area} \\
    & & & of quasars & {(deg$^2$)} \\
    \hline
    2.2--3.5 & BOSS DR9 & \cite{2013ApJ...773...14R} & 23301 & 2236.0 \\
    3.7--4.7 & SDSS DR7 & \cite{2010AJ....139.2360S} & 1785 & 6248.0 \\
    3.6--5.2 & NDWFS & \cite{2011ApJ...728L..26G} & 12 & 1.71 \\
    3.8--5.3 & DLS & \cite{2011ApJ...728L..26G} & 12 & 2.05 \\
    4.7--5.4 & SDSS+WISE & \cite{2016ApJ...829...33Y} & 99 & 14555.0 \\
    4.7--5.5 & SDSS DR7 & \cite{2013ApJ...768..105M} & 103 & 6248.0 \\
    4.7--5.5 & --- Stripe 82 & \cite{2013ApJ...768..105M} & 59 & 235.0 \\
    5.7--6.5 & SDSS Main & \cite{2016ApJ...833..222J} & 24 & 11240.0 \\
    5.7--6.5 & --- Overlap & \cite{2016ApJ...833..222J} & 10 & 4223.0 \\
    5.7--6.5 & --- Stripe 82 & \cite{2016ApJ...833..222J} & 13 & 277.0 \\
    5.8--6.6 & CFHQS Deep & \cite{2010AJ....139..906W} & 1 & 4.47 \\
    5.8--6.6 & --- Very Wide & \cite{2010AJ....139..906W} & 16 & 494.0 \\
    5.8--6.5 & Subaru High-$z$ Quasar & \cite{2015ApJ...798...28K} & 2 & 6.5 \\
    4.0--6.5 & CANDELS GOODS-S & \cite{2015AA...578A..83G} & 19 & 0.047 \\
    6.5--7.4 & UKIDSS & \cite{2011Natur.474..616M} & 1 & 3370.0 \\
    6.5--7.4 & UKIDSS & \cite{2015ApJ...801L..11V} & 1 & 3370.0 \\
    6.5--7.4 & \begin{tabular}{@{}c@{}}ALLWISE+UKIDSS \\ +DECaLS\end{tabular} & \cite{2018Natur.553..473B} & 1 & 2500.0 \\
    \hline
  \end{tabular}\\
  \caption{
  The list of quasar survey samples for which we have photometric data that are included in our database QUOTAS\, and for which we also have selection functions, survey area, survey depths and completeness estimates from \citep{Kulkarni+2019}.
    \label{tab:qlf_samples}
}
\end{table*}

\subsection{Collation of Simulation data for QUOTAS\,}

We ingest simulated data from the LEGACY suite\footnote{Details of the LEGACY suite runs are available at: https://sites.google.com/site/tmoxwiki/legacysuite/infolegacy.} runs, a set of large-scale cosmological dark matter-only simulations \citep{2022MNRAS.512...27M} 
into QUOTAS. We outline the key properties of the runs from the LEGACY suite used in this pilot project work in Table~3. The LEGACY simulations consist of a parent run with a $(2.3 ~\rm{Gpc})^3$ volume of $2048^3$ particles at a mass resolution of $M_{\rm{DM}} =5.43 \times 10^{10} \rm{M}_{\odot}$ (referred to in Table~3 as the {\bf parent} run) accompanied by a set of higher resolution zoom-in simulations with a volume of $(119 ~\rm{Mpc})^3$ and mass resolution of $M_{\rm{DM}} =1.32 \times 10^{7} \rm{M}_{\odot}$ which corresponds to an effective particle number count of $32768^3$ (labeled as the {\bf mean} run in Table~3) and another further zoom-in simulation with $ V = 1 ~\rm{Gpc}^3$ and  $M_{\rm{DM}} =6.78 \times 10^{9} \rm{M}_{\odot}$ to capture sufficient large scale modes \citep{2019MNRAS.489.1684K} (labeled as the {\bf Expanse} run in Table~3). The initial conditions of the simulation are created using the public code MUSIC \citep{2011MNRAS.415.2101H}, adopting the following values for cosmological parameters based on the results from the WMAP satellite \citep{2013ApJS..208...19H}: $\Omega_{\Lambda} =0.715$,  $\Omega_{\rm{M}} =0.285$, $\Omega_{\rm{b}} =0.045$, $h =0.695$, $\sigma_8 =0.828$. For each of the simulations (parent and zoom-ins) gravitationally bound structures and sub-structures were determined using the halo finder ROCKSTAR \citep{2013ApJ...762..109B}. Particles and halo/sub-halo properties are stored in 300 snapshots from $z=20$, covering the redshift range of interest in this study, $z \geq 3$, though slices are available down to $z=0$.

\begin{table*}
  \centering
  \begin{tabular}{cllrr}
    \hline
    {Run Name} & {Box Size (Mpc/h)} & {DM particle mass in $\Msun$} & {Effective number of particles} & {Comment} \\
    \hline
    {Parent} & {1600} & {$5.43 \times 10^{10}$} & {${2048}^3$} & {Full Box} \\
    {Mean} & {83} & {$1.32 \times 10^{7}$} & {${32768}^3$} & {Zoom-in Box} \\
    {Expanse} & {703.1} & {$1.32 \times 10^{7}$} & {${4096}^3$} & {Zoom-in Box} \\
    \hline
  \end{tabular}\\
  \caption{
  The list of LEGACY suite runs used in this work. Halo catalogs from the Parent, Mean and Expanse runs are included in the Halo Properties Table in QUOTAS\,.
    \label{tab:qlf_samples}
}
\end{table*}

The halo catalogs provide detailed information - 81 individual attributes for each identified halo - and all these characteristic parameters are stored in QUOTAS. We also store detailed evolutionary information for each halo in finely spaced time steps for each snapshot that is used to generate merging histories for individual halos. This data structure allows us to uniquely identify for each halo - its progenitors and ancestors. This information provides further leverage in establishing links to evolutionary models of BH growth providing the BH-halo connection piece. QUOTAS\ permits association of the host dark matter halo hierarchy for each observed quasar, an essential link for our goal of probing the underlying BH-galaxy-halo connection. Assuming for instance, that the most massive simulated dark matter halo at $z=3$ hosts the brightest observed quasar at $z = 3$, from the data in QUOTAS\, we can trace its detailed merger history as shown in Fig.~\ref{fig:mergerhistories}. In the case shown, the host dark matter halo has a mass of $9.3 \times 10^{12}$ M$_{\odot}$ at $z=3$. 
 
 \begin{table*}
  \centering
  \begin{tabular}{ccc}
  \hline
    Name & Units & Description \\
  \hline
    $z$                 & -             & redshift \\
    $M_{\rm{vir}}$      & M$_{\odot}$   & virial mass \\
    $r_{\rm{vir}} $     & kpc           & virial radius \\
    $x, ~y, ~z$         & Mpc           & position \\
    $v_x, ~v_y, ~v_z$   & $\rm{km ~ s^{-1}}$      & peculiar velocity \\
    $ J_x, ~J_y, ~J_z$  & M$_{\odot}$ Mpc $\rm{km ~ s^{-1}}$  & angular momentum \\
    $ \dot{M}_{\rm{acc}}$ & M$_{\odot}/$yr & accretion rate onto halo \\
    $I_{\rm{desc}} $ & - & descendant halo ID \\
    $T/|U|$ & - & virial parameter \\
    $x_i, ~y_i, ~z_i$         & Mpc           & position of each particle \\
    $v_{x,i}, ~v_{y,i}, ~v_{z,i}$   & $\rm{km ~ s^{-1}}$     & peculiar velocity of each particle \\
   \hline
  \end{tabular}\\
  \caption{
  Examples of dark matter halo properties stored in QUOTAS.
  \label{table:dm_halo_properties}
  }
\end{table*}

The QUOTAS\, platform and developed tools are publicly available for open access at the Google Kaggle platform. The complete observational data and simulation data are available at:\\
{\bf https://www.kaggle.com/quasarnet/quasarnet}\\ and the codes are available at:\\ {\bf https://www.kaggle.com/datasets/quasarnet/quasarnet/code} \\ 
and examples of use are available at:\\
{\bf https://www.kaggle.com/code/quasarnet/bhmf-quasarnet; \\ 
https://www.kaggle.com/code/quasarnet/data-exploration-with-quasarnet.} 

\section{Demonstration of ease of use \& Workflow: Querying QUOTAS}

To facilitate querying the database, searches can be performed with SQL-like \texttt{BigQuery} on \texttt{Google Public Dataset} on Kaggle.\footnote{{\url{https://www.kaggle.com/datasets/hisunnytang/halo-sims}}}. With the data hosted on the public Google Kaggle Platform, researchers can directly access and interact with the data through a \texttt{Colab} notebook available at {\url{https://www.kaggle.com/datasets/hisunnytang/halo-sims/code}} that we provide and perform analyses directly on the dataset. The subset of photometry data for instance, can be accessed for all quasars with the following query:
\\
\begin{lstlisting}[language=Sql]
%%bigquery --project black hole-database df
SELECT
qtable.*,
ptable.*
FROM `black hole-database.ml_test.qso` as qtable
JOIN `black hole-database.ml_test.photometry` as ptable
ON qtable.object_name = ptable.source_object_name
\end{lstlisting}

The Quasar Spectra Table and the Quasar Photometry Table are stored similarly, therefore, simply replacing ``qtable'' with ``qlftable'' will yield the existing photometry data for all quasars across surveys.

\begin{lstlisting}[language=Sql]
%%bigquery --project black hole-database df
SELECT
qlftable.*,
ptable.*
FROM `black hole-database.ml_test.qso` as qlftable
JOIN `black hole-database.ml_test.photometry` as ptable
ON qlftable.object_name = ptable.source_object_name
\end{lstlisting}

{\bf The Quasar Photometry Table} includes the selection function for every source from \citet{Kulkarni+2019} and with it we can calculate the effective volume for each source and therefore generate QLFs in a straightforward fashion.\footnote{For the plots shown here the following values of standard cosmological parameters was adopted:$\Omega_{\rm m} = 0.3$; $\Omega_{\Lambda} = 0.7$ and $H_0 = 70$ km/sec/Mpc.}  Next, we illustrate ease of use and utility of QUOTAS\, and describe the computation of the BHMF. To derive the BHMF, we use the measurements of BH masses and the flux limits for every observed source compiled in QUOTAS, to calculate the spatial density of objects. In quasar surveys, every detected quasar can be assigned a detection probability and a detection volume in a given redshift bin, in accordance with the survey detection limit and its completeness. These quasar properties are combined with the derived spatial density to compute the BHMF \citep{2010AJ....140..546W, 2016ApJ...833..222J}. This provides a swift estimate of the BHMF with a single line query to QUOTAS. We illustrate the ease of use by plotting the resulting estimate of the BHMF in Fig.~\ref{fig:distributions}, that is volume complete. \citep{Kelly_2010} derived the BHMF from the SDSS DR3 data release, here we demonstrate the ease-of-use to generate the updated BHMF with the full current available SDSS DR7 quasar data. The plotted Fig.~\ref{fig:distributions} clearly reveals that there appears to be an upper limit for BH masses across redshift bins. Interestingly, such a limit is theoretically predicted and the rendering of data compiled in QUOTAS\ aligns with modeling expectations, as per eqn.~7 in \citep{PN_UMBH2009}.

\section{First science results from QUOTAS}

We present early science results from (i) mining the observational data stored in QUOTAS\, reporting on findings of $z>3$ SDSS quasar properties and correlations with their host properties; (ii) illustrate the use and power of ML to expand dark matter only simulation volumes that faithfully reproduce the underlying statistical properties of halos; (iii) show how ML can be used to "populate" these larger dark matter only simulation volumes after training on full hydrodynamic simulations that include baryonic physics and (iv) compare SDSS quasar properties with the ML derived properties from the "populated" simulation. Finally, we demonstrate the power of ML to interrogate key assumptions of BH growth, in this case, the accretion paradigm, in the sub-grid models that are currently implemented in simulations. We focus primarily on $z \sim 3$ as the benchmark redshift to explore and compare properties of observed SDSS quasars ($3 < z < 3.5$) and the corresponding $z = 3.25$ LEGACY simulation output stored in QUOTAS\,.

For this work, the homogenized, largest observational data set that is available is from the SDSS. For the SDSS, we have access to all survey properties including the selection function and survey specification – we have in hand all attributes that permit easy calculation of the QLF and the BHMF in QUOTAS. For data collated from other surveys, listed in Table~2, selection functions are curated from \citet{Kulkarni+2019}. In this first paper we showcase results for the BHMF computed for SDSS quasars. With all the information on selection function and completeness from SDSS and large number of sources, we are able to mimic the relevant cuts in our simulated LEGACY slice that is populated with quasars at $z=3$.

\subsection{Exploration of SDSS quasar properties and correlations}

To start with, we demonstrate exploration of the observational data stored in {\bf the Quasar Spectra Table} and {\bf Quasar Photometry Table} convolved with the corresponding Selection Functions in QUOTAS\,. In Fig.~\ref{fig:bh_properties}, we plot derived BH properties - BH mass and the Eddington ratio for the SDSS quasar population from a sample query to QUOTAS\, in 5 redshift bins ranging from $3 < z < 7$. New insights from this exercise include a noticeable precipitous decline in the number of quasars with redshift that constrains the general shape of the QLF and its slope as a function of redshift after taking into account the effect of incompleteness and our preferential detection of the brightest quasars with increasing redshift. In contrast, to what we find at these higher redshifts, at lower redshifts $z < 2$ for instance, although a similar decline in the abundance of quasars is detected, that does not reflect our inability to detect sources that exist but rather signals the true deficit of actively accreting BHs at these late times when galactic nuclei are gas-poor and much of the gas in galaxies has already been consumed by star-formation. Bright as the currently detected quasars are, we note from the top panel of Fig.~\ref{fig:bh_properties} that the majority of the detected quasar population is still accreting at sub-Eddington rates. We caution that the dusty super-Eddington sources that are yet to be uncovered at these early epochs are missing from our current optical datasets, and therefore also in QUOTAS\,. We reiterate that QUOTAS\ includes all X-ray detected Type I AGN, since they all have optical counter-parts. Given the current limits of X-ray instruments on the Chandra telescope and XMM-Newton telescopes, accreting sources peter out at $z > 3$ where the Type II AGN inventory is expected to pick up. This is where we expect JWST and NGRST to start filling in the landscape in the very near future.
\\
As a function of redshift, as expected, our current observational surveys detect progressively brighter quasars. Reading off the peaks of the contour plots in Fig.~\ref{fig:bh_properties}, we see that the bolometric luminosity of the average detected quasar at $z=3$ is $3.6 \times 10^{46}\,{\rm erg\,s^{-1}}$ while by $z=4$ it is an order of magnitude higher at $2.8 \times 10^{47}\,{\rm erg\,s^{-1}}$. The typical measured BH mass powering the corresponding average detected quasar exceeds $10^9\,\Msun$ at all redshifts beyond $z>3$. This demonstrates clearly that we sample different sub-portions of the QLF and BHMF as a function of redshift, and are preferentially accessing the most extreme objects at earlier epochs. This biased observational view severely limits our ability to model the growth history of the overall BH population. And it is this limitation that we would like to circumvent with ML techniques that will permit us to predict and hence detect the lower luminosity, more representative quasar population at every epoch. This will finally yield a more complete picture of BH growth and galaxy assembly. 
\\
As noted earlier, correlations between the properties of central SMBHs and their host galaxy luminosity, stellar mass and bulge velocity dispersion are well documented for the dormant, local SMBH population. QUOTAS\, offers a unique opportunity to explore and discover new correlations between active SMBHs and their host galaxies. Turning now to Fig.~\ref{fig:bhprop_pairplot}, we demonstrate once again the ease of visualizing data and the ability to explore correlations between multiple measured quantities from QUOTAS. Several new science results and insights are obtained from just this single analysis. Take for instance, the anti-correlation seen between the FWHM of CIV lines versus Eddington ratio plotted in the second panel on the bottom row. In a recent paper, \cite{Marziani+2019} discuss that a correction needs to be applied to the FWHM of CIV derived BH masses that depends on the Eddington ratio for a lower redshift population of quasars $0<z<3$. Our plot from QUOTAS\ clearly demonstrates and supports that such a correction may be required and suggests that the \cite{Marziani+2019} finding also extends to higher redshift quasars at $z>3$ studied here. With these set of first results, we illustrate the power of QUOTAS\ to reveal correlations and their evolution with redshift as seen in the diagonal kernel density plots as a function of cosmic epoch. 

\subsection{Tailored ML tools to expand simulation volumes}

We show how the explicit association of observed quasars at $z > 3$ with their putative parent dark matter halo sites and their history shown in Fig.~\ref{fig:mergerhistories} can be used for improved modeling and providing insights into the BH-galaxy-halo connection. However, as noted previously there are several important limitations that prevent direct comparison of observational data with simulations. We start by deploying ML to circumvent the limitations arising from small simulation volumes and resolution due to the prohibitive computational cost of including baryonic physics in larger box simulations. It is numerically inexpensive to scale up and perform large volume N-body dark matter only simulations, and generate halo catalogs from volumes that would permit detailed comparison to the comparable volumes probed by observational surveys. This is where we use ML to train on the smaller volume full hydrodynamical simulation that includes baryonic physics, in this case the Illustris-TNG300\footnote{Details of the Illustris-TNG300 set of simulations can be found at: {https://www.tng-project.org/}.} simulation to then ``populate" a larger volume dark matter only simulation with SMBHs from the LEGACY suite stored in QUOTAS.  As a first step in this process, we perform an important task, i.e. use ML to extract the key, characteristic statistical properties of the dark matter halos from a subset, the training set, of dark matter halos, in the LEGACY mean box (see Table~4) and then see if we can successfully recover the overall statistical properties - beyond the training set - in the larger volume (lower mass resolution) LEGACY $(1 {\rm Gpc})^3$ Expanse box.

Since our goal is to explore the BH-galaxy-halo connection, we ultimately need to relate simulated dark matter halos, the putative sites for BH formation and growth, to the population of observed quasars at every epoch. To conduct the association studies between observed quasars and their potential parent dark matter halos, we require simulated volumes to match the large observational survey volumes. Directly simulating corresponding volumes with the full baryonic physics implemented is not computationally possible currently. Given that quasars are rare objects, in order to circumvent the restrictions of the simulated box size for our planned association studies, we present the application of the ML normalizing flow algorithm to expand our sample of simulated halo catalogs. 
\\
Therefore, we first utilize ML methods to generate mock simulated surveys that are better volume matched to the data. To do so, we first need to characterize ensemble properties of the simulated halo population to mimic the halo population in a significantly larger volume box that successfully reproduces its statistical properties. The underlying statistical properties of the halo population of the simulated LEGACY mean run at $z=3.25$ stored in QUOTAS\ are derived using the class of generative ML models referred to as normalizing flows \citep{rezende2016variational,2019arXiv191202762P,2019arXiv190809257K}. Normalizing flow provides a non-parametric fit to the joint data distribution and offers a powerful way to generate new halos by sampling latents. The observed variables $X$, in our case, the properties of each simulated halo such as its mass, angular momentum (modeled via its spin parameter), mass accretion redshift (redshift at which half the mass of the halo is assembled) etc, are mapped to a latent variable $z \sim p_z(z)$ of the same dimension drawn from a known distribution (usually gaussian) through an invertible transformation parameterized by a neural network, i.e. $z = f_{\theta}(X)$. The marginal likelihood of $p(X)$ can be obtained with the changes of variables given by:
\begin{eqnarray}
p_X(X;\,\,\, \theta) = p_z | \rm{det} \frac{\partial z}{\partial x}| = p_z(f^{-1}_{\theta}(x) ) | \rm{det} (\frac{\partial f^{-1}_{\theta}(x)}{\partial x})|.
\end{eqnarray}
The model is trained by maximizing the log likelihood of the estimate under the observed sample distribution $X$. Once trained, the model can be used to generate as yet unseen data by sampling the latent ($z'$) space from the base distribution and transforming it through the invertible transform $X' = f^{-1}_{\theta}(z')$.  Although, several other kinds of transformations have been proposed \citep{dinh2017density, kingma2017improving, papamakarios2018masked}, we use a version of Normalizing Flows as they are exactly invertible and are particularly well attuned to deriving the posterior probability distributions from which larger volume statistics for the halo population can be generated for our future association studies. In other astronomical applications, normalizing flows have been shown to successfully constrain the posterior estimate of the distance estimates from Gaia Data \citep{cranmer2019modeling}; performing likelihood free Bayesian inference on cosmological parameters \citep{Alsing_2019} and in constraining the re-ionization history of the universe \citep{Hort_a_2020}. We adopt the Real Non-Volume Preserving model (RealNVP) \citep{dinh2017density} to fit our simulated halo properties tables. The results of this crucial first test step in the combined exploitation of simulated and observed data are shown in Fig.~\ref{fig:halo_prop}. The properties of the generated host halo population (orange) from sampling the trained normalizing flow model on the LEGACY mean runs are distributed very similarly to those of true halo population (blue) stored in the Halo Properties Table from the larger volume zoom-in, the LEGACY $(1 {\rm Gpc})^3$ Expanse cosmological simulation  at $z=3.25$ in QUOTAS. This coherence in recovered halo properties demonstrates that the trained model can be used further for downstream ML tasks to explore the quasar - host halo connection exploiting the large observational datasets currently in hand and expected deluge in the near-future. 
\\
As seen in Fig.~\ref{fig:halo_prop}, the generated host halo population reproduces all properties of interest of the true halo population from the LEGACY $(1 {\rm Gpc})^3$ simulation slice at $z = 3.25$ in QUOTAS\ extremely well. The use of normalizing flows therefore permits us to train on a smaller volume simulation (the $z=3.25$ slice from the LEGACY mean run) and accurately extend our simulated catalog to correctly generate the rarer dark matter halos that would only appear in significantly larger volume simulations (in this case, the $z=3.25$ slice of the LEGACY $(1 {\rm Gpc})^3$ Expanse run). ML permits us to extend and expand our reach to capture the parent dark matter hosts of the rare, bright observed high-redshift quasars that would only appear in significantly larger volume simulations. In future work, we plan to use this trained model to generate halo catalogs in volumes that correspond more closely to those of observational surveys for further detailed association studies.

\subsection{Application of ML tools to probe the BH-galaxy-halo connection}

In the next step, we employ an ML algorithm to populate the dark matter only Legacy simulation volume with accreting BHs. To do this we train a model to learn the relationship between BHs and their host halos using a full physics simulation. The publicly available IllustrisTNG hydrodynamical simulations \citep{2019ComAC...6....2N} incorporate baryonic processes by way of sub-grid model recipes for galaxy formation and evolution in which BHs grow via accretion and mergers \citep{2017MNRAS.465.3291W}. Full information about black holes and their associated parent halos is therefore available to train an ML model. As this is a supervised learning problem with tabular data, we use a Random Forest algorithm \citep{random_forests}.

This approach uses an ensemble of decision trees to make a prediction. Each decision tree is constructed top-down from a root node. At each node the data is split into two bins based on the values of its input parameters. The splits are chosen such that the weighted average of the MSE of the two bins is minimized. This partitioning of the data results in each leaf node at the bottom of the tree containing a small subset of the data, where almost all members of the subset have a similar output value. Predictions from decision trees are based on the assumption that test data points will have a similar output value to the other members of the leaf node it is placed into. A random forest is made up of a number of decision trees. There is a bootstrapping procedure such that each decision tree within the forest is trained on a randomly generated subset of the training data. Further randomness is added in that for each split only a subset of input features can be used. The prediction from a random forest is the average prediction of its component decision trees. A major advantage of random forests is that they are significantly less prone to overfitting data compared with a single decision tree. This results from the randomness added when training the individual decision trees.

Specifically we choose the extremely randomised tree (ERT) ensemble algorithm, which has been shown in several previous works \citep[e.g.][]{2016MNRAS.457.1162K, 2021MNRAS.504.4024M} to be suitable for this task. The ERT algorithm adds in additional randomization by computing a random split for each feature at each node, rather than the optimal split. We follow the approach in \cite{multi_epoch_ml}, who used halo properties from a wide range of redshifts as model inputs, and showed how this can provide information about the times when galaxy properties are determined. As we are predicting properties at a high redshift, and so do not have a large number of prior snapshots, we adopt the base model from \cite{multi_epoch_ml}.  We train the model using the IllustrisTNG300 simulation volume, with the BH mass and accretion rate as our target variables. We then apply the trained model to the dark matter only Legacy $(1 {\rm Gpc})^3$ Expanse simulation. This gives us a catalog with approximately 40x the number of black holes that are present in the largest IllustrisTNG simulation.
\\
To test the robustness of this ML procedure for our specific task, in Fig.~\ref{fig:illustris-legacy_number}, in the left panel, we plot the QLF derived from the training set - the Illustris-TNG300 box - with that from the ML populated LEGACY $(1 {\rm Gpc})^3$ Expanse box at $z = 3.25$. We note that the QLF is in excellent agreement for the full sample and the sub-sample demonstrating that the deployed ML algorithm successfully enables us to expand simulation volumes while accurately capturing the occupation fraction of accreting BHs. In the right panel of Fig.~\ref{fig:illustris-legacy_number}, we compare and find excellent agreement in the number density of accreting SMBHs in the Illustris-TNG300 slice and the larger volume LEGACY $(1 {\rm Gpc})^3$ Expanse box. Further, in Fig.~\ref{fig:illustris-legacybhlum}, we dissect the accreting SMBH population in these two simulations in 3 BH mass bins and once again find excellent agreement. Notable in these histograms is the bi-modal luminosity distribution that starts to emerge at higher BH masses (as seen in the middle and right panels of the plot). 
\\
Confident that the extended Random Forest ML algorithm has statistically populated the LEGACY $(1 {\rm Gpc})^3$ Expanse box in concordance with the Illustris-TNG300 simulation, we delve deeper into studying BH-parent dark matter halo connection. Therefore, next, we proceed to compare the BHMF derived from observations - SDSS quasars in the redshift bin $3 < z < 3.5$ in this instance - from QUOTAS\, with that derived from the ML populated LEGACY $(1 {\rm Gpc})^3$ Expanse box. However, before doing so, we need to pay attention to the fact that the optically bright SDSS quasars collated in QUOTAS\ are a subset (Type I's) of the full population of accreting black holes (AGN) and therefore we need to appropriately scale the accreting black hole population to determine the fraction of Type I AGN's in the simulation by classifying them. We partition the simulated population into Type I and Type II AGN using an independent empirical finding by \citep{2019ApJ...871..240A}. First, detailed multi-wavelength studies of AGN find that by $z \sim 3$, close to 90\% of the sources are obscured Type II AGNs (not optically bright quasars), and their fraction decreases slightly and monotonically as function of X-ray luminosity in the range $L_X \sim 10^{42} - 10^{46}$ erg$\,{\rm s}^{-1}$ (for the latest census see \cite{2019ApJ...871..240A} and references therein). Adopting this empirically derived fraction from \citep{2019ApJ...871..240A}, with an X-ray to bolometric luminosity correction factor, we scale the number of accreting BHs populated in the LEGACY $(1 {\rm Gpc})^3$ Expanse box to mimic the SDSS quasars (Type I AGN) at this epoch. With this selection implemented, we plot the NFW concentration parameters for the dark matter halos that host quasars (the Type I AGN) in the LEGACY $(1 {\rm Gpc})^3$ Expanse slice at $z=3.25$ in Fig.~\ref{fig:lum_clustering_comp}. Combining AGN abundance with clustering measurements from large X-ray surveys \cite{Powell_2018,2020ApJ...891...41P} report that Type I AGN occupy dark matter halos with concentration parameters $ < c_{\rm NFW} < 10.14$ and Type II are found in $ 10.14 < c_{\rm NFW} < 100$. Plotting the concentration parameter distribution as a function of bolometric accreting luminosity for sources in the LEGACY $(1 {\rm Gpc})^3$ Expanse box, we find that our selection of Type I's in the box is in excellent agreement with empirical findings as shown in Fig.~\ref{fig:lum_clustering_comp}. We note that all our sources selected as quasars are indeed consistent with the empirically determined classification criteria derived by \cite{Powell_2018}, as all selected sources lie leftward of the white dashed line that marks $c_{\rm NFW}\,<\,10.14$) in the Expanse box.
\\
In order to compute the BHMF we need the selection function and survey specifications including depth and completeness, that are available for SDSS in QUOTAS.
In Fig.~\ref{fig:illustris-legacybhlum}, we compare the BHMF of SDSS quasars and the equivalent simulated population from QUOTAS. The mass functions match extremely well at the turnover point, which corresponds to a bolometric luminosity cut of $L \sim 10^{46.5}$ erg$\,{\rm s}^{-1}$. In Fig.~\ref{fig:accretion_test}, the dashed vertical line marks the location of BHs that would be accreting at the Eddington limit. Therefore, we see clearly that every accreting BH with mass $M_{\rm bh} > 10^{8.4} \Msun$ in the simulation is accreting at sub-Eddington rates. At BH masses greater than $10^{9}\,\Msun$, neither the amplitude nor the slope match well, as the BH population is underestimated in simulations. This indicates, as expected, that the simulation under produces the rarest, most massive black holes which likely signals incompleteness arising from the smaller volume compared to the SDSS survey volume footprint. What is however completely unexpected is that even at lower masses, $M_{\rm bh} < 10^{8.5}\Msun$, BHs that are detected by the SDSS $z \sim 3$ are missing from the simulations. While at the high mass end of the BHMF, this deficit of accreting SMBHs in the simulations could be partially attributed to a mismatch in sampling volumes between the SDSS and the simulation box, or as potentially arising as a limitation of our training set. However, the mismatch at lower masses, namely the lack of accreting BHs in mass range $M_{\rm bh} < 10^{8.5}\Msun$ in simulations is glaring and strongly suggests that the adopted sub-grid accretion and feedback prescriptions in simulations are suspect. Note that accretion in our training set Illustris-TNG300 simulations is capped at the Eddington limit, so unsurprisingly no super-Eddington sources are predicted in these simulations and therefore none are found the LEGACY $(1 {\rm Gpc})^3$ Expanse box either. Our comparison clearly reveals that lower mass black holes in the simulation box are not luminous enough, as they are not accreting at high enough rates to survive the luminosity cut. This suggests that the sub-grid model for accretion adopted in Illustris-TNG300 does not accurately  capture the accretion in observed quasars. The high accretion rate tail of the distribution of accretion rates appears to be missing in the simulations. Note that even though the LEGACY $(1 {\rm Gpc})^3$ Expanse box given its larger volume, encapsulates a larger range and diversity of formation and assembly histories for SMBHs, it replicates the issue with accretion rates found in the Illustris-TNG300. Our work suggests that the sub-grid, multi-mode BH feedback operating in a thermal 'quasar' mode at high accretion states, and a kinetic 'wind' mode at low accretion states adopted in Illustris-TNG300 is over-efficient and appears to choke accretion onto BHs prematurely. Note that current state-of-the-art simulations like the Illustris-TNG300 successfully reproduce observed properties like the BHMF at $z=0$ and the integrated mass density in BHs over cosmic time at epochs $z <5$. Therefore, these simulations with unique recipes for feedback from growing BHs, model for gas accretion onto them and prescriptions for star formation reproduce integrated quantities well. However, as we show above, QUOTAS\ permits a unique diagnosis of the mass assembly history over time, the differential mass accumulation in the BH population by focusing on a slice at $z \sim 3$. The mismatch we find between SDSS quasars and their simulated counter-parts points to the fact that current sub-grid models of accretion and feedback do not reproduce the mass build-up over time accurately. More specifically, the probability distribution of accretion rates onto black holes with masses between $10^8$ and $10^{8.75}$ M$_{\odot}$ does not match observations, and that this is not just a cosmic variance issue.
\\
In recent work \citet{2022MNRAS.511.3751H}, have shown that nearly none of the currently deployed sub-grid models for AGN feedback across simulation suites adequately captures the underlying physics and matches the observational data at $z > 5$. Here we, show that the problem starts to appear earlier, at even more modest redshifts, of $z \sim 3$. This growing divergence shows that the undetected lower luminosity quasar population at modest to high-redshift needs to be critically filled in to fully understand BH growth and evolution. Analyzing and comparing results on BH growth and assembly across redshift in several large-scale independent cosmological simulations like the Illustris-TNG100, Illustris-TNG300, Horizon-AGN, EAGLE, and SIMBA suites, \cite{2022MNRAS.511.3751H}, show that while all of them predict the same BHMF and relation between BH mass and host galaxy stellar mass $M_*$ at $z=0$ in agreement with local observations, their predictions disagree at higher redshifts. For instance, there is much disagreement on whether BHs at $z > 4$ are overmassive or undermassive at fixed host galaxy stellar mass with respect to the $z=0$ $M-{\rm bh}-M_{*}$ relation. And as can be seen in Fig.~5 of their paper, none of the simulations reproduce the observed accretion luminosities of observed SDSS quasars at $z \geq 5.8$. Only the Illustris-TNG300 and SIMBA simulations produce enough SMBHs with $M_{\rm bh} > 10^8\,\Msun$, however, they both clearly under-predict the luminosities of these sources at $z \sim 6$. Aside from the disagreements between various independent simulations arising from the range of feedback prescriptions and sub-grid models they each adopt, here we show explicitly that even at $z \sim 3$ closer to epochs when simulations reproduce local observations; and where the BHMF is observationally well-sampled, the Illustris-TNG300 simulation for instance, does not adequately  capture the growth rate of SMBHs that current SDSS observations provide. That is, the  mass accretion history of the average SMBH that would lead to the production of a quasar with a bolometric luminosity in the range of $L_{\rm bol} \sim 10^{45-46}\,{\rm erg\,s^{-1}}$, even at modest redshifts of $z \sim 3$ are not currently reproduced by simulations. The AGN feedback and BH growth prescriptions adopted in simulations do not get the growth build up over time correct and analysis with QUOTAS\, reveals that this is already evident by $z \sim 3$. This strongly suggests that re-examination of our current sub-grid theoretical models is warranted, specifically, the models adopted for gas accretion and feedback. It is evident that BH gas feeding rates are insufficient and therefore less aggressive feedback models that do not scupper the accretion flow are needed in order to facilitate uninterrupted gas supply into the centers of galactic nuclei.
\\
In addition to permitting the analysis of large, complex data-sets, ML techniques as we illustrate above also help us interrogate key theoretical model assumptions. We have presented a specific instance where observational data from QUOTAS\ simultaneously leveraged with ML populated LEGACY simulations has afforded deeper diagnosis of the sub-grid model for BH accretion and its implementation in simulations. Our key science result from this first deployment of the pilot version of QUOTAS\ has revealed that two key coupled theoretical assumptions, namely the accretion rate and efficiency of AGN feedback, need to be revisited and the current sub-grid models deployed in simulations are too efficient in driving feedback and hence end up suppressing the growth of low mass black holes. Essentially, new ML techniques applied simultaneously to analyze the co-located observational and simulation data in QUOTAS\, have clearly demonstrated that SMBHs in simulations even at modest redshifts of $z \sim 3$ fail to accrete, shine and grow at high enough rates to match observations of quasars at these cosmic epochs. 
\\
While our work has revealed that BH populations in simulations simply do not grow enough, QUOTAS\, offers a deeper look that will enable us to probe if the association between SMBHs and their corresponding dark matter halos can accurately reproduce the observed clustering properties of the quasar population. This can be tested easily by comparing clustering of observed quasars with the simulated population in the LEGACY $(1 {\rm Gpc})^3$ box. Comparison of the clustering properties via the 2-point correlation function will help investigate the robustness of the association between SMBHs and their parent halos in the SDSS data and the LEGACY $(1 {\rm Gpc})^3$ Expanse box. In Fig.~\ref{fig:clustering_comp}, we show that the clustering properties of optical SDSS quasars in this redshift range, $z \sim 3$, are well reproduced by the accreting SMBH population at the corresponding redshift slice ($z=3.25$) in the LEGACY $(1 {\rm Gpc})^3$ Expanse box. This clearly suggests that  simulations accurately capture the statistics of the association of SMBHs with their parent dark matter halos inferred from clustering studies of SDSS quasars. Therefore simulated quasars occupy the right parent dark matter halos as inferred from observations, and  where they fail is in how they grow. The mass assembly history over time, that is represented by the accretion rate and its evolution for the SMBH population is suspect in simulations. Comparison of the predictions of simulations by \citep{2022MNRAS.511.3751H} reveals that not only do they progressive diverge from each other at higher redshifts, but that they also progressively deviate from observations. We have demonstrated the power of QUOTAS\ to produce new science insights on our fundamental input theoretical model assumptions, the implementation of accretion physics in the adopted in sub-grid models. This too with the analysis of the observed and simulated quasar population at just one epoch $z \sim 3$. The prospects and science returns from the comprehensive future analysis and application of the tests explored here to even earlier epochs with the publicly available user-friendly research platform QUOTAS\ look to be extremely promising.

\subsection{Probing the environments of the brightest quasars}

Finally, one of the fundamental science questions that motivated us to create QUOTAS\ was the ability to devise future survey strategies leveraging ML, simulations and observational data to uncover the lower luminosity, more characteristic, accreting BH population in the modest and high-redshift Universe. We have established above that one key feature of SDSS quasars that our ML trained simulated quasars accurately reproduce is the occupation statistics, namely, that we get the association between quasars and their parent dark matter halos right as evidenced by (i) the agreement of the correlation function determined from the two samples (as shown in Fig.~\ref{fig:clustering_comp}) and (ii) the NFW concentration parameter distribution of the parent halos in the LEGACY $(1 {\rm Gpc})^3$ Expanse box (as shown in Fig.~\ref{fig:lum_clustering_comp}). Therefore, we are well placed to infer an observational survey strategy that will help with uncovering the fainter quasar population. To do this, leveraging the established accuracy of the occupation statistics, we construct a novel neighborhood statistic and evaluate its robustness using QUOTAS\,. The proposed statistic is derived by enumerating the BHMF  found in the neighborhood of a bright quasar powered by a $10^9\,\Msun$ SMBH in annular rings of varying radius, 8 and 16 comoving Mpc respectively. As we change the radial region under consideration, the mass function of BHs within that annular radius changes. In Fig.~\ref{fig:neighborhood_bhmf}, we plot this neighborhood occupation statistic for quasars in the ML populated LEGACY $(1 {\rm Gpc})^3$ Expanse box. Included in this plot is the mean neighborhood occupation statistic, shown with a solid line, derived from averaging over a much larger annulus with radius $r > 40$ comoving Mpc. We clearly see that there is a preferential excess of lower luminosity sources well above the mean value of the neighborhood statistic in the vicinity of a bright quasar. A strong excess of sources is detected even for sources with $L_{\rm bol} > 10^{46}\,{\rm erg\,s^{-1}}$ (just a factor of 3 lower in luminosity), compared to the cut at $L_{\rm bol} > 10^{46.5}\,{\rm erg\,s^{-1}}$.  This suggests that the optimal observational strategy to enlarge the high redshift quasar sample, and probe beyond just the brightest quasars, would be to survey regions around these most luminous currently detected sources.

\section{Conclusions \& Future Directions}

In this paper, we present the new research platform QUOTAS\ that for the first time comprises co-located observational and simulated data for the study of the growth and assembly history of SMBHs over cosmic time. QUOTAS\, is an on-going project to collect, and homogenize observational quasar data and their simulated counter-parts.  Leveraging the existing Normalizing Flow and Random Forest ML algorithms, we provide a template for how we can expand simulated samples of quasars to better match observational survey volumes that permit detailed studies of quasar properties and their environments. In this paper, outlining the pilot phase of our project, we describe the successful compilation of the new database; and provide a proof of concept demonstration of ease of use and first science results from querying QUOTAS exploring the SDSS data. With this pilot project, we demonstrate that the profound connection between black holes, their host galaxies and parent dark matter halos in the context of the standard model for structure formation in the Universe can be and needs to be more deeply investigated in order to fully understand how black hole growth unfolds in the larger cosmological context.

We present a collated database of high-redshift $z>3$ optical quasars in a format ready for application of ML techniques, and have made it available publicly on the Google Kaggle platform. Collecting data from many disparate observational surveys, we make available detailed information and properties where available for individual objects tailored to address the problem of the mass assembly history of the earliest black holes. While full exploration of the BH-galaxy-halo connection with further customized ML tools is the longer term goal of this live database that we plan to continually update, here we present the QUOTAS\ database and some early results. We anticipate that QUOTAS\, will be actively used by the astronomy community to investigate quasar populations and BH growth models. We report on the critical first steps that leverage the QUOTAS\ platform to construct key quantities of theoretical interest - the BHMF and the QLF - by querying the collated data. Picking $z \sim 3$ as the benchmark redshift, we then compare the Illustris-TNG300 trained ML algorithm used to populated dark matter-only LEGACY $(1 {\rm Gpc})^3$ Expanse simulation slice with observed SDSS quasars. We find that the occupation fraction of BHs; and the clustering properties of the associated parent dark matter halos is accurately reproduced. We demonstrate a use case with the BHMF derived from SDSS data and ML trained extended LEGACY simulations. Comparing the BHMFs we find that the simulations under-predict the accretion luminosities for BH masses below $M_{\rm bh} < 10^{8.75}\,\Msun$. This deficit is also clearly reflected in the comparison of the census of the luminosity of accreting BHs as a function of BH mass. Essentially, analysis with QUOTAS\, reveals that the sub-grid models deployed to model the physics of accretion in the Illustris-TNG300 suite are unable to match the observed mass assembly history of BH populations over time. The accretion luminosities are too low, suggesting that AGN feedback is likely too aggressive and causes BHs to stunt their own growth prematurely. Detection of the lower luminosity quasar population is needed to anchor models and sharpen our understanding of BH growth over cosmic time, therefore, we deployed ML once again to help devise optimal observational survey strategies to uncover this population. ML tools we have shown permit us to question some of the basic assumptions embedded in our theoretical modeling of accretion processes in simulations. So despite simulations matching $z = 0$ observations of the properties of BHs and their hosts, by modest redshifts of just $z \sim 3$, we find that the predictions of simulations are out of line with observed properties of SDSS quasar populations. In this work, we demonstrate how tailored ML tools can be leveraged to visualize, analyze and compare observational and simulated data of the quasar population.  
\\
This successful combination of simulation and observational data sets in a single platform by epoch, we show, allows seamless application of ML techniques. In follow-up work, we plan to expand the data sets with the inclusion of spectral data; addition of the high-z AGN sources that will be detected in upcoming infra-red surveys by JWST; and expand with populating host galaxy properties in the LEGACY $(1 {\rm Gpc})^3$ Expanse simulation to study correlations between the BH and the host galaxy, namely, correlations like $M_{\rm bh}-M_{*}$ and beyond. The ultimate goal of this project is to make full use of the powerful capabilities of ML to analyze these rich, complex multi-dimensional data sets together. The power of co-located simulated and observational data-sets in QUOTAS\ coupled with the expected improvements in computational infrastructure will allow us to before long conduct mock surveys in ML aided expanded simulation volumes that mimic the exact footprint of observational surveys. Adding in more attributes for halo properties in the near future will allow the generation of expansive mock catalogs corresponding to galaxy surveys of varying depths and volumes. This will allow us to connect across data types and test against more detailed model predictions beyond the BHMF and the QLF, for instance, the occupation fraction and generate maps of larger scale density of environments around SMBH hosts. In this work, with the introduction of a novel neighborhood statistic, we propose that one reliable detection strategy to uncover the lower luminosity quasar population is to perform targeted, deeper surveys around currently detected luminous SDSS quasars. Our ML propelled work has provided this new insight that has the potential to open up discovery space for the lower luminosity quasar population that will significantly constrain our theoretical models. Our future plan is to further leverage these current tools, build upon them using ML to predict the detailed properties of the host galaxies of the quasars and utilize them to devise additional detection strategies for the as yet undetected lower luminosity quasars at $3 < z < 7$ to fill in the current gaps in our knowledge of this population. Information on lower luminosity quasars is critically needed to hone our understanding of accreting black holes and their mass assembly history over cosmic time.  In future work, we intend to explore the interplay between observations and simulations and generate predictions for the more representative undetected quasar populations and their properties using ML algorithms that will enable deeper exploration of the BH-galaxy-halo connection.

\section{Acknowledgments}

PN gratefully acknowledges support from the Black Hole Initiative (BHI) by grants from the Gordon and Betty Moore Foundation and the John Templeton Foundation. She thanks her colleagues at the BHI and members of the next generation Event Horizon Telescope (ngEHT) Science Working Group on Black Holes and their Cosmic Context for many useful conversations on evolving black hole populations and is grateful to Alphabet-X for technical support and computational resources for this project. PN acknowledges conversations on databases with Sanjay Sarma and Brian Subirana during the early stages of this project. PN and KST thank Rick Ebert at the Infra-Red Processing and Analysis Center (IPAC) at the California Institute of Technology for his help with accessing the NED database. KST thanks Frank Wang at Google for his help with the Google Cloud Platform. SK acknowledges use of the ARCHER UK National Super-computing Service (http://www.archer.ac.uk) for running the LEGACY simulation. BN acknowledges support from the Fermi National Accelerator Laboratory, managed and operated by Fermi Research Alliance, LLC under Contract No. DE-AC02-07CH11359 with the U.S. Department of Energy. The U.S. Government retains and the publisher, by accepting the article for publication, acknowledges that the U.S. Government retains a non-exclusive, paid-up, irrevocable, world-wide license to publish or reproduce the published form of this manuscript, or allow others to do so, for U.S. Government purposes. SS acknowledges the Aspen Center for Physics where parts of this work were done, which is supported by National Science Foundation grant PHY-1607611.

\section{Data Availability}

All data and associated codes for QUOTAS\, are publicly available on the Google Kaggle platform for community use. The complete observational data and simulation data sets are available at:\\
{\bf https://www.kaggle.com/quasarnet/quasarnet}\\ and the developed related codes are all available at:\\ {\bf https://www.kaggle.com/datasets/quasarnet/quasarnet/code} \\ 
and examples of use are available at:\\
{\bf https://www.kaggle.com/code/quasarnet/bhmf-quasarnet; \\ 
https://www.kaggle.com/code/quasarnet/data-exploration-with-quasarnet.}

\bibliography{references}

\begin{figure*}[ht!]
    \centering
    \includegraphics[width=\textwidth]{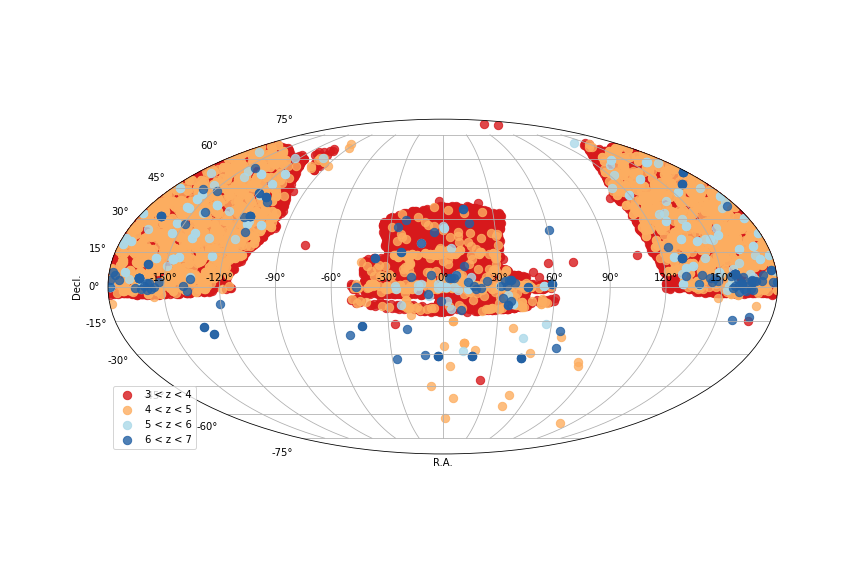}
    \caption{The distribution of all quasars collated in QUOTAS\ on the sky that are listed in Table~2. Sources are color-coded according to their redshifts: $3 < z < 4$ - red;  $4 < z < 5$ - orange; $5 < z < 6$ - cyan and $6 < z < 7$ - blue.}
    \label{fig:qlf_radec}
\end{figure*}

\begin{figure*}[!ht]
    \centering
    \includegraphics[width=0.9\textwidth]{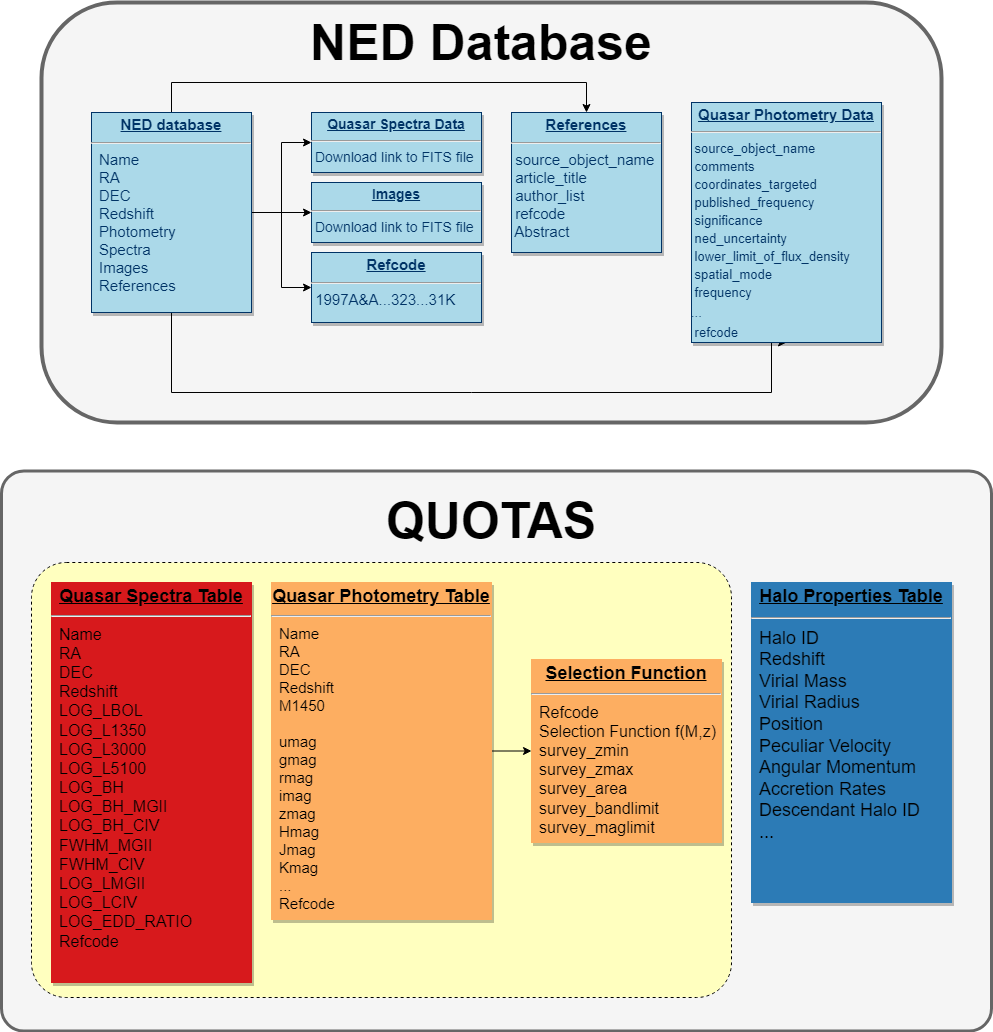}
    \caption{Schematic of the data structures and derived custom data tables that comprise QUOTAS\: The information sourced from NED is shown in blue, the collated data in Quasar Spectra Table in red; the Quasar Photometry data table in orange and the data table for Halo Properties derived from the simulated LEGACY catalogs is shown in blue. The stored attributes in each of the data tables is listed.}
\label{fig:db_table}
\end{figure*}

\begin{figure*}[ht]
\centering
\begin{center}
\includegraphics[width=0.4\textwidth]{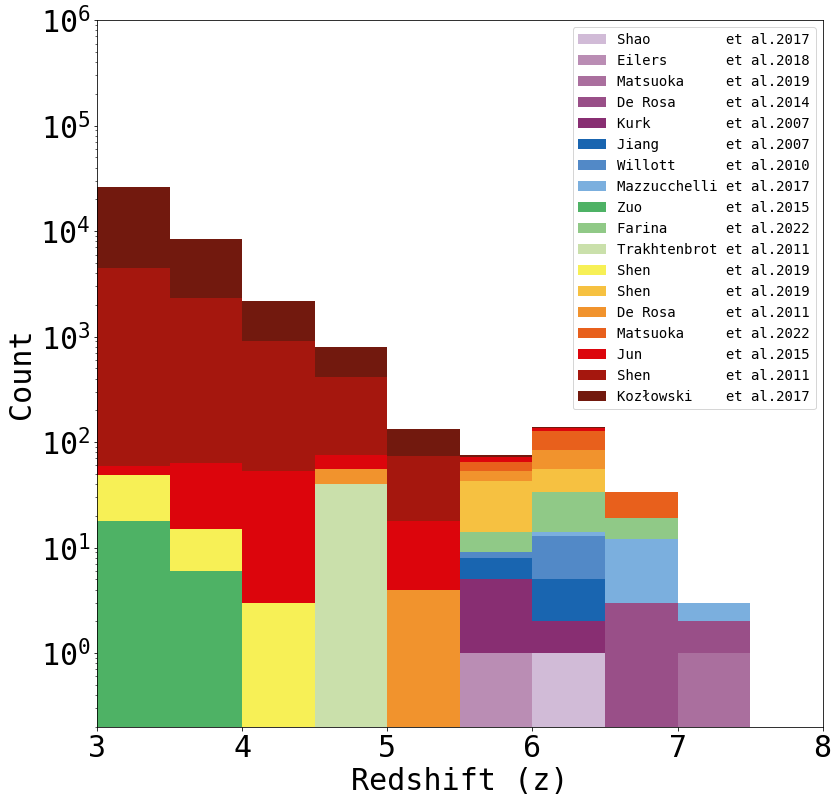}
\end{center}
\caption{BH/Quasar Population Demographics - the number of quasars plotted as a function of redshift collected from all surveys listed in Table~1 (referenced in the inset) that are in the QUOTAS database.} 
\label{fig:distributions}
\end{figure*}

\begin{figure*}[ht]
\centering
\begin{center}
\includegraphics[width=0.9\textwidth]{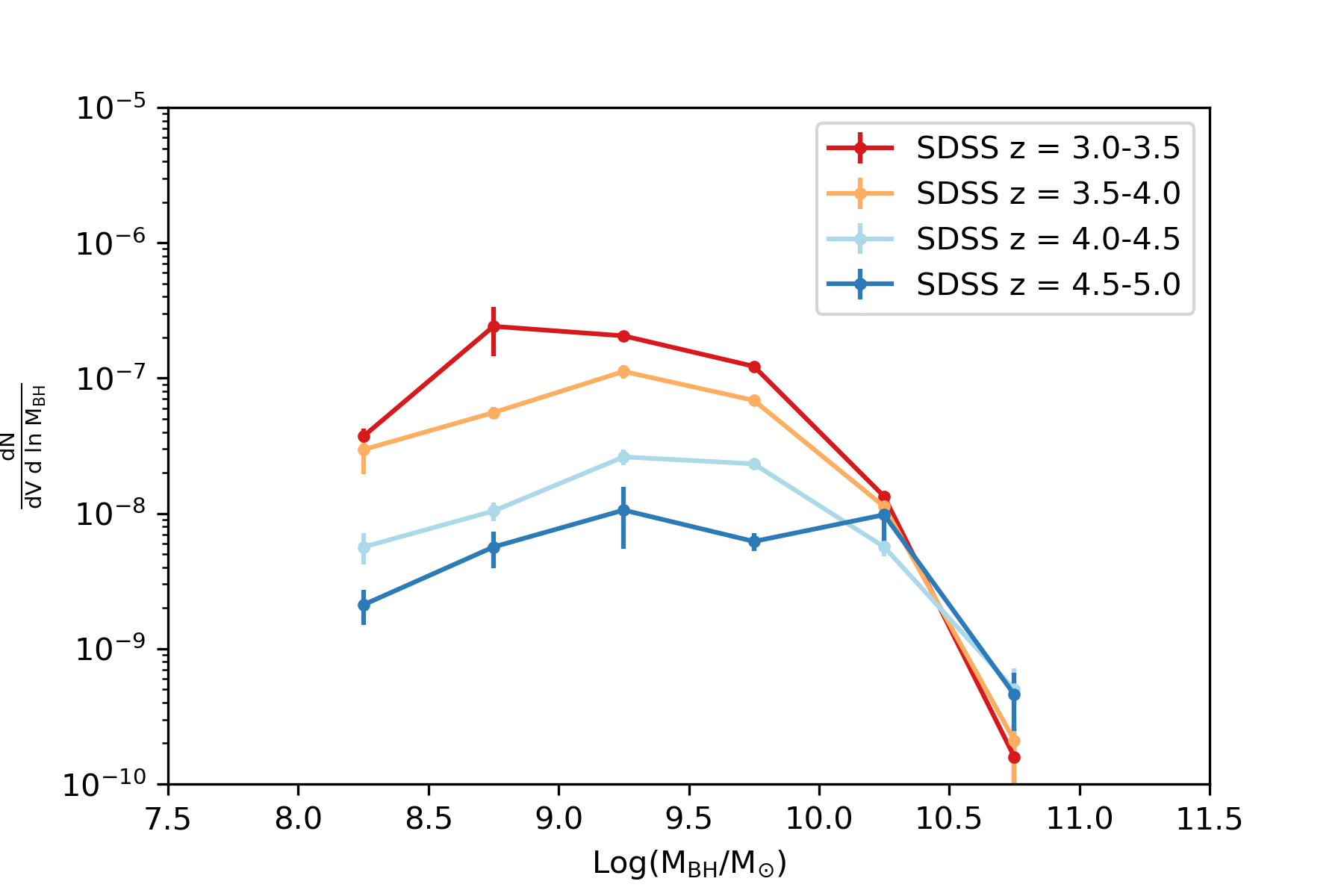}
\end{center}
\caption{BHMF for SDSS computed from QUOTAS: The BHMF is an important quantity for calibrating/constraining theoretical models of black hole mass assembly and growth and it can be derived in one step from the collated tables in QUOTAS\, when all relevant attributes are available to estimate quantities in eqns. (3-5). The BHMFs queried from QUOTAS\ sourced from the SDSS survey (SDSS DR7 for the $z < 5$ quasars) are plotted.} 
\label{fig:bhmf_z}
\end{figure*}

\begin{figure}[ht]
\centering
\includegraphics[width=0.75\textwidth,keepaspectratio]{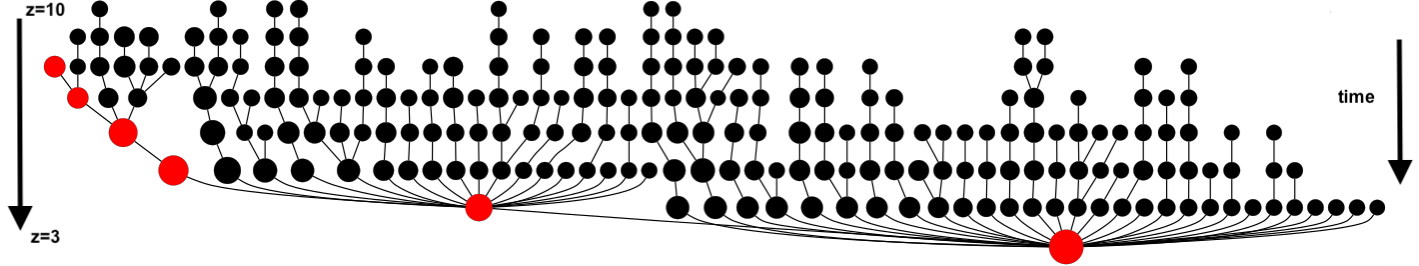}
\caption{Assembly history studies permitted by QUOTAS\: the merger and assembly history of a dark matter halo that likely hosts the brightest observed quasar at $z = 3$ tracked through time generated from QUOTAS\ is illustrated here. The size of the circle is proportional to the logarithm of the dark matter halo mass. Red circles mark the most massive progenitor of a halo at each time step. In this plot only the progenitor halos with mass $ \geq 5 \times 10^9$ M$_{\odot}$ and only time steps with $\Delta z \sim 1$ are shown.}
\label{fig:mergerhistories}
\end{figure}

\newpage

\begin{figure*}[ht]
    \centering
    \includegraphics[width=0.9\textwidth]{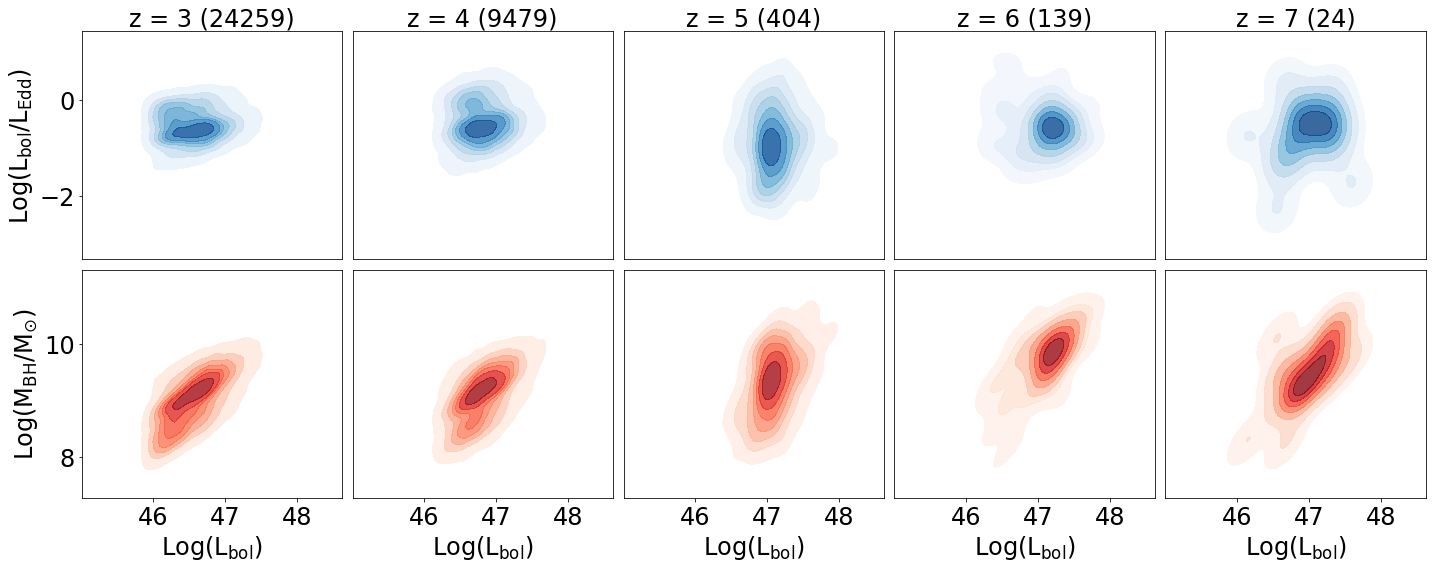}
    \caption{Key derived statistical properties of the population of observed $z > 3$ quasars relevant to modeling their growth history over cosmic time - BH mass, bolometric luminosity and Eddington ratio in 5 redshift slices for sources in Table~1. The number in parentheses denotes the number of observed quasars in the redshift bin. These data were drawn from QUOTAS\ collated in the Quasar Spectra Table and the Quasar Photometry Table per the schematic shown in Fig.~\ref{fig:db_table}. Each column corresponds to redshift bins ranging from: $3 < z < 3.5$; $3.5 < z < 4.5$; $4.5 < z < 5.5$; $5.5 < z < 6.5$; and $z > 6.5$ respectively.}
    \label{fig:bh_properties}
\end{figure*}

\begin{figure*}[!ht]
    \centering
    \includegraphics[width=0.9\textwidth]{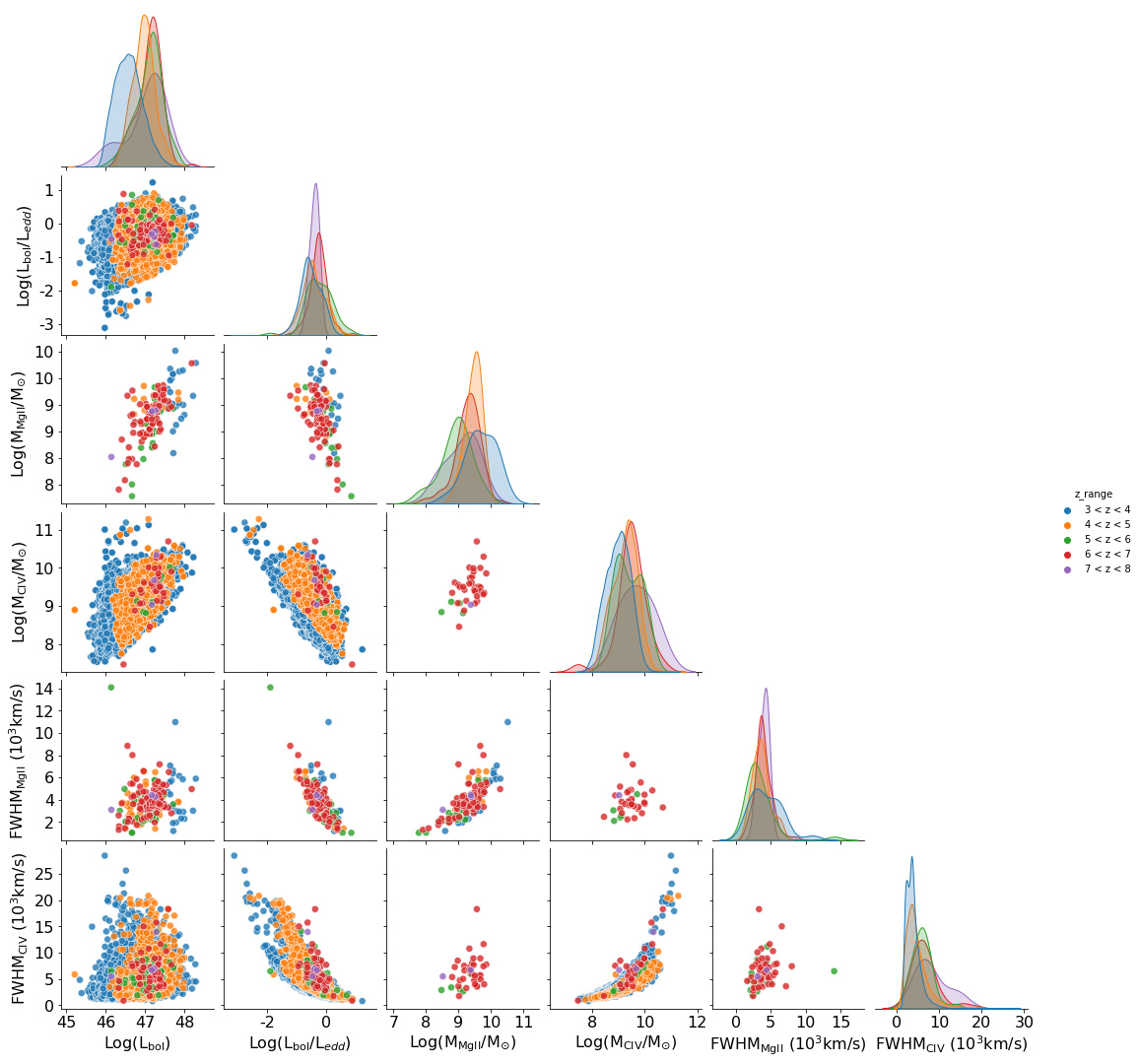}
    \caption{Preliminary exploration of correlations using QUOTAS\ from information in Table~1: correlations between the bolometric luminosity; the Eddington ratio; BH mass determined from MgII line-widths; BH mass determined from CIV line-widths; the FWHM of MgII lines and the FWHM of CIV lines queried from the Quasar Spectra Table. The data are color-coded according to redshift as follows: $3 < z < 4$ (orange);  $4 < z < 5$ (blue); $5 < z < 6$ (cyan) and $6 < z < 7$ (pink). Along the diagonal, we plot the kernel density estimations of the distributions for each of these parameters. The off-diagonal panels show the cross-correlation between different parameters.}
\label{fig:bhprop_pairplot}
\end{figure*}

\begin{figure*}[!ht]
    \centering
    \includegraphics[width=0.9\textwidth]{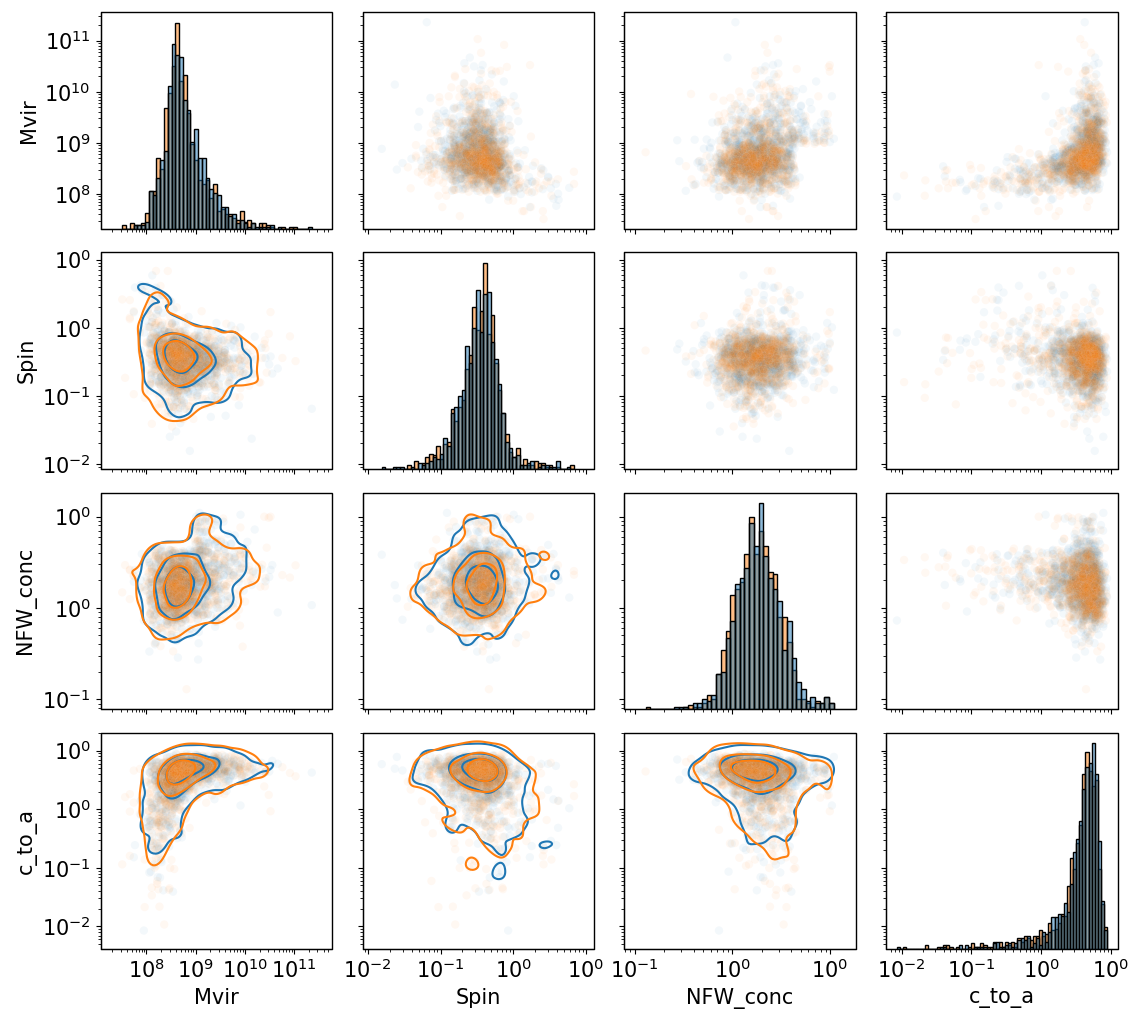}
    \caption{Derived underlying statistical properties of dark matter halos at redshift $z=3$ using the normalizing flow ML algorithm. These prior distributions for halo properties are derived from the Halo Properties table in QUOTAS. The distributions of "true" halo properties are colored in blue; the orange points are generated by sampling with the trained normalizing flow model. Here we select the halo mass, halo spin, halo NFW concentration, and halo ellipticity as key characteristic properties to train the normalizing flow model. This is the first of a set of many tailored ML tools that will be used to fill in the key missing link to probe the black hole-host galaxy- parent dark matter halo connection.}
    \label{fig:halo_prop}
\end{figure*}

\begin{figure*}[ht]
    \centering
    \includegraphics[width=0.45\textwidth]{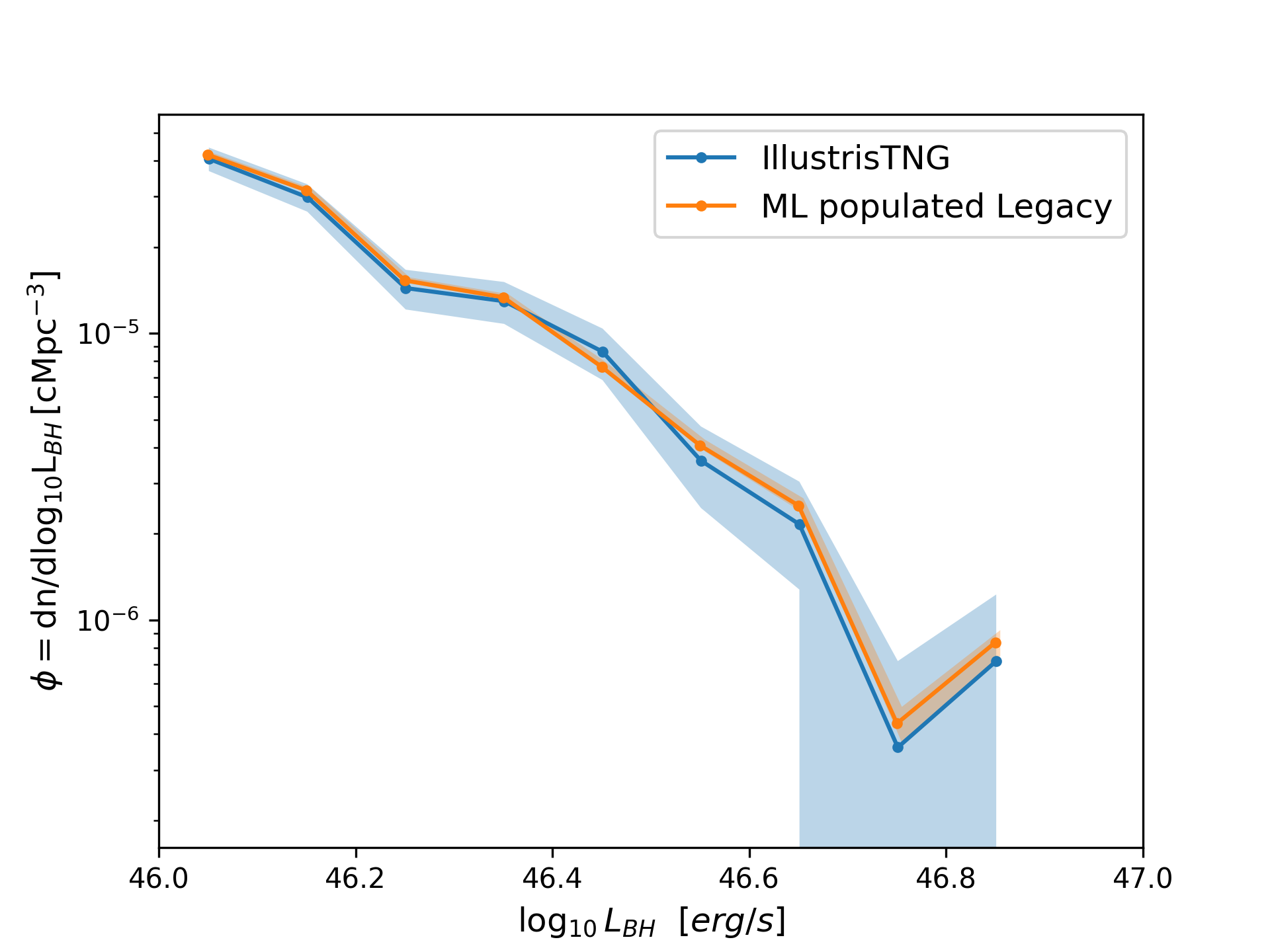}
    \includegraphics[width=0.45\textwidth]{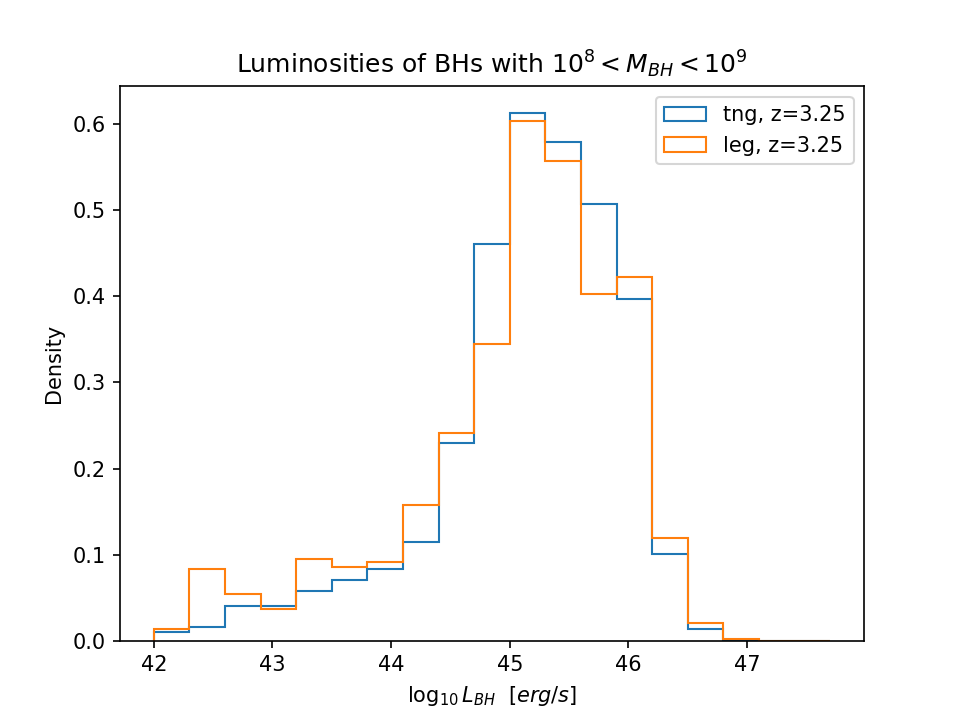}
    \caption{Left Panel: Comparison of the QLF derived for sources with bolometric luminosity $L > 10^{45}$ erg${\rm s}^{-1}$ from the Illustris-TNG300 trained ML used to populate the larger LEGACY 1 Gpc Expanse run box from the $z = 3.25$ slice. Right Panel: Comparison of the census of accreting SMBHs - the number density of quasars as a function of bolometric Luminosity in the training set (the Illustris-TNG300 box) and the larger volume LEGACY 1 Gpc Expanse box at $z=3.25$. There is excellent agreement showing that the ML algorithm (ERT Random Forest) deployed here accurately reproduces the number statistics of SMBHs in the larger simulation box slice.}
    \label{fig:illustris-legacy_number}
\end{figure*}

\begin{figure*}[ht]
    \centering
    \includegraphics[width=0.9\textwidth]{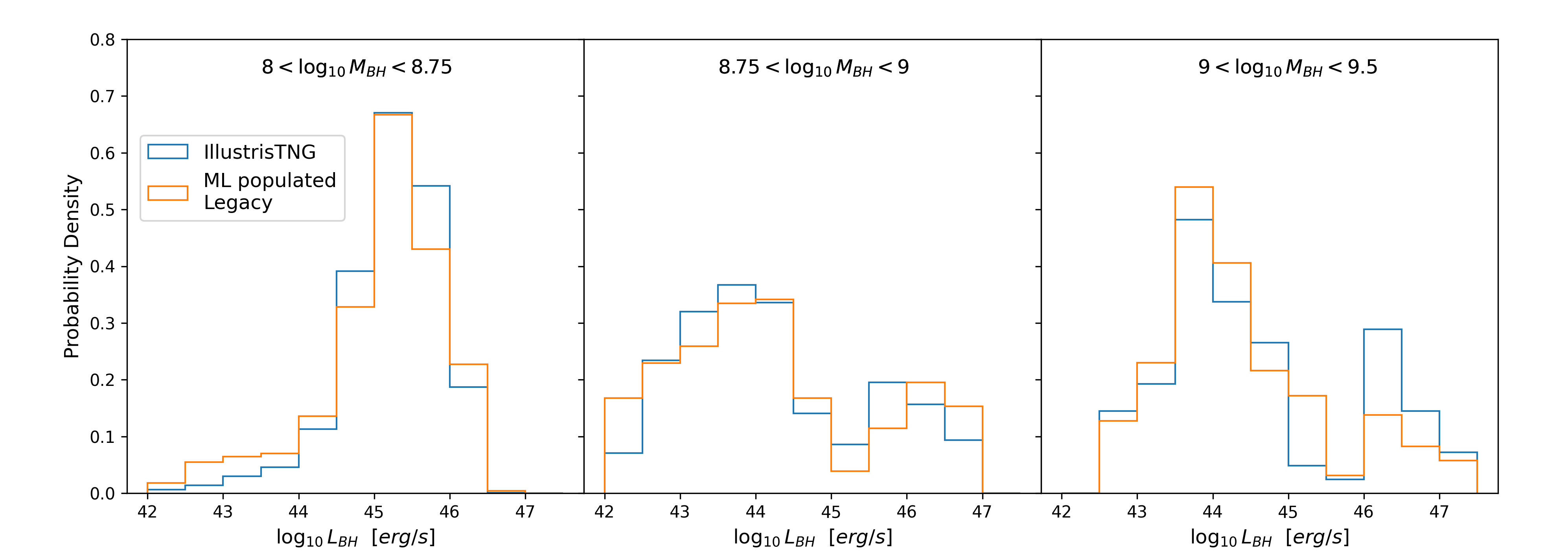}
    \caption{Inventory of quasar population at $z \sim 3$: sources with bolometric luminosity $L > 10^{42}$ erg${\rm s}^{-1}$ from the Illustris-TNG300 trained ML used to populate the larger LEGACY 1 Gpc Expanse run box from the $z = 3.25$ slice in 3 mass bins.}
    \label{fig:illustris-legacybhlum}
\end{figure*}

\begin{figure*}[ht]
    \centering
    \includegraphics[width=0.9\textwidth]{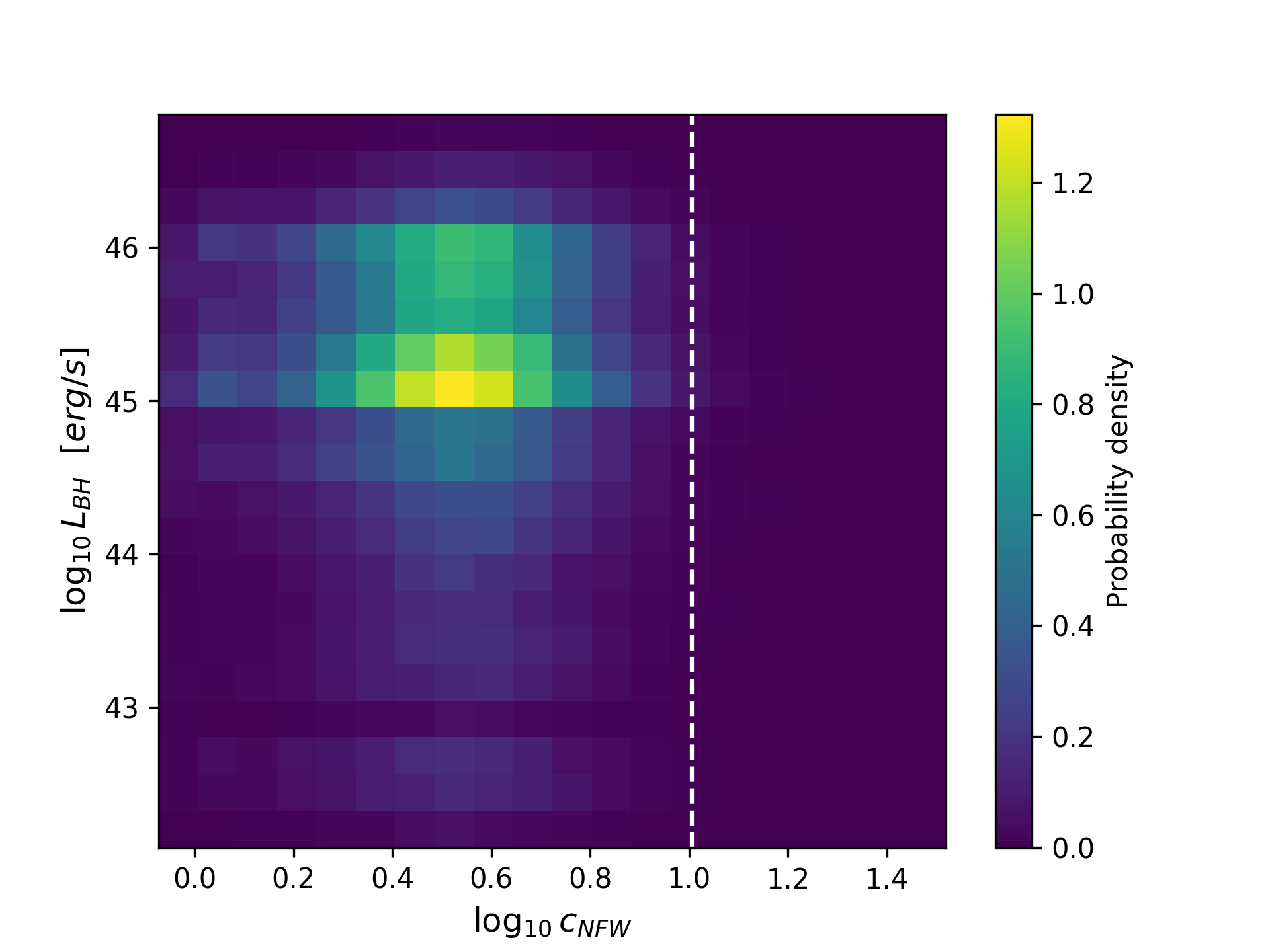}
    \caption{Robustness of the selection of accreting quasars in the ML populated LEGACY 1 Gpc Expanse box from the $z = 3.25$ slice as Type I's. To test if the selected quasars inhabit the right parent dark matter halos, here we plot the distribution of the NFW fit concentration parameters for their parent halos and show that our selected sample is in excellent agreement with the observational determination combining clustering measurements for AGN reported in \cite{2020ApJ...891...41P,Powell_2018}.}
    \label{fig:lum_clustering_comp}
\end{figure*}

\begin{figure*}[ht]
    \centering
    \includegraphics[width=0.9\textwidth]{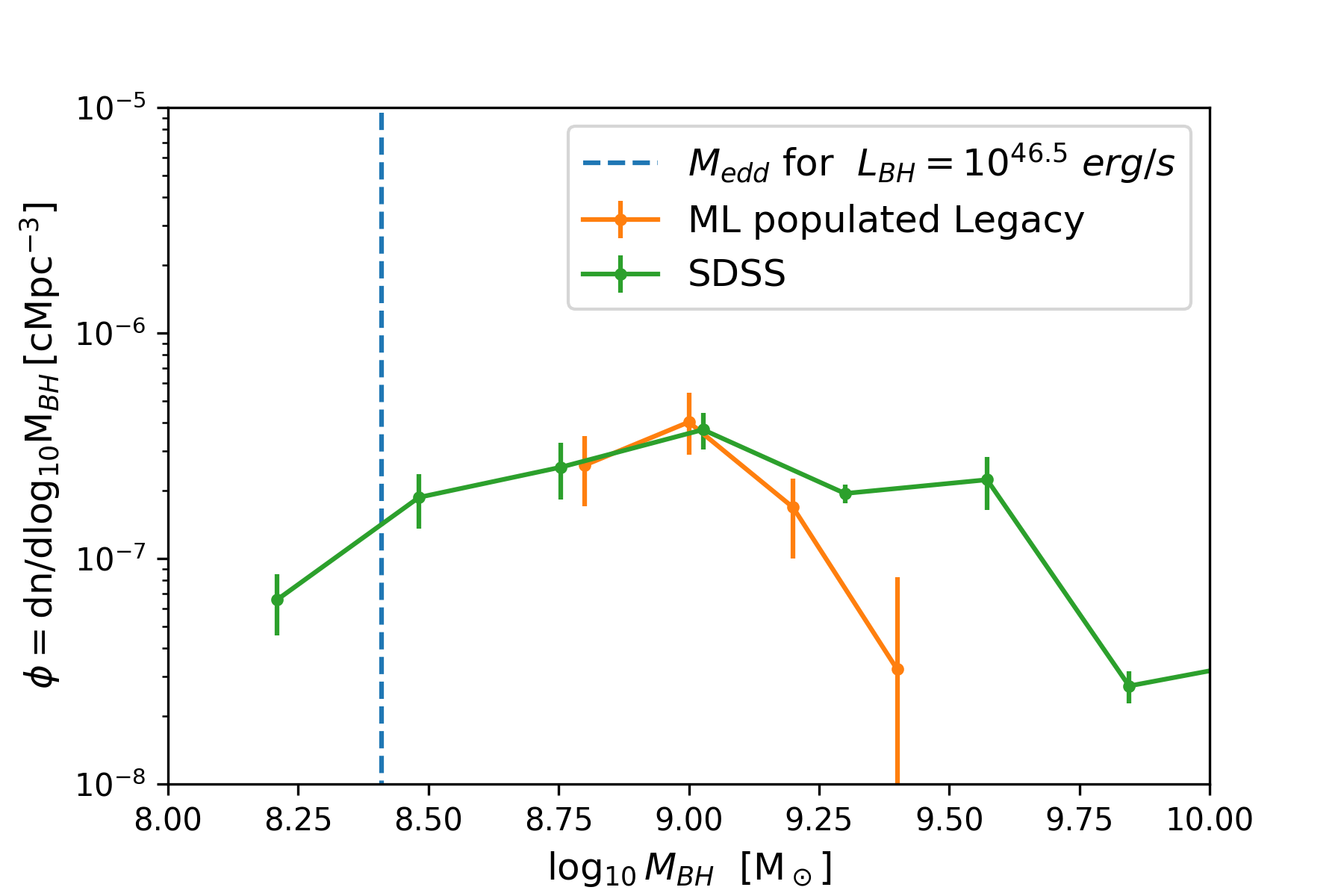}
    \caption{Comparison of the BHMF from $z = 3 - 3.5$ SDSS quasars and the ML populated LEGACY 1 Gpc box ($z = 3.25$ slice). The blue dashed line marks the BH mass that corresponds to the imposed luminosity cut of $L_{\rm bol} > 10^{46.5}\,{\rm erg\,s^{-1}}$. This cut implies that all BHs included in the census of the BHMF are accreting at sub-Eddington luminosities.}
    \label{fig:accretion_test}
\end{figure*}

\begin{figure*}[ht]
    \centering
    \includegraphics[width=0.9\textwidth]{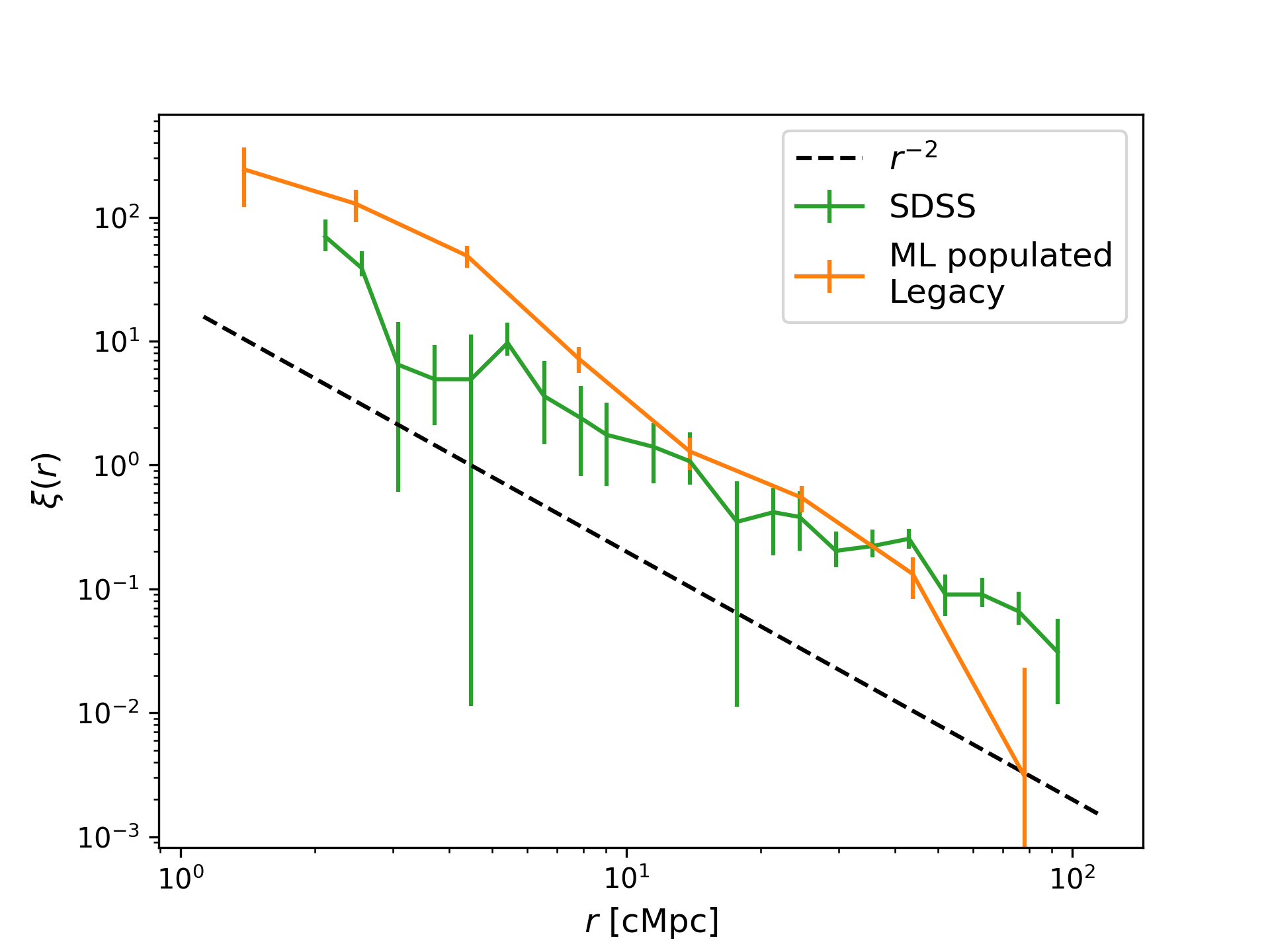}
    \caption{Comparison of the clustering of quasars from SDSS and the ML populated LEGACY 1 Gpc Expanse box at $z \sim 3$. The SDSS quasars plotted here are also part of the BOSS survey, with clustering measurements reported in \cite{2015MNRAS.453.2779E}}.
    \label{fig:clustering_comp}
\end{figure*}

\begin{figure*}[ht]
    \centering
    \includegraphics[width=0.9\textwidth]{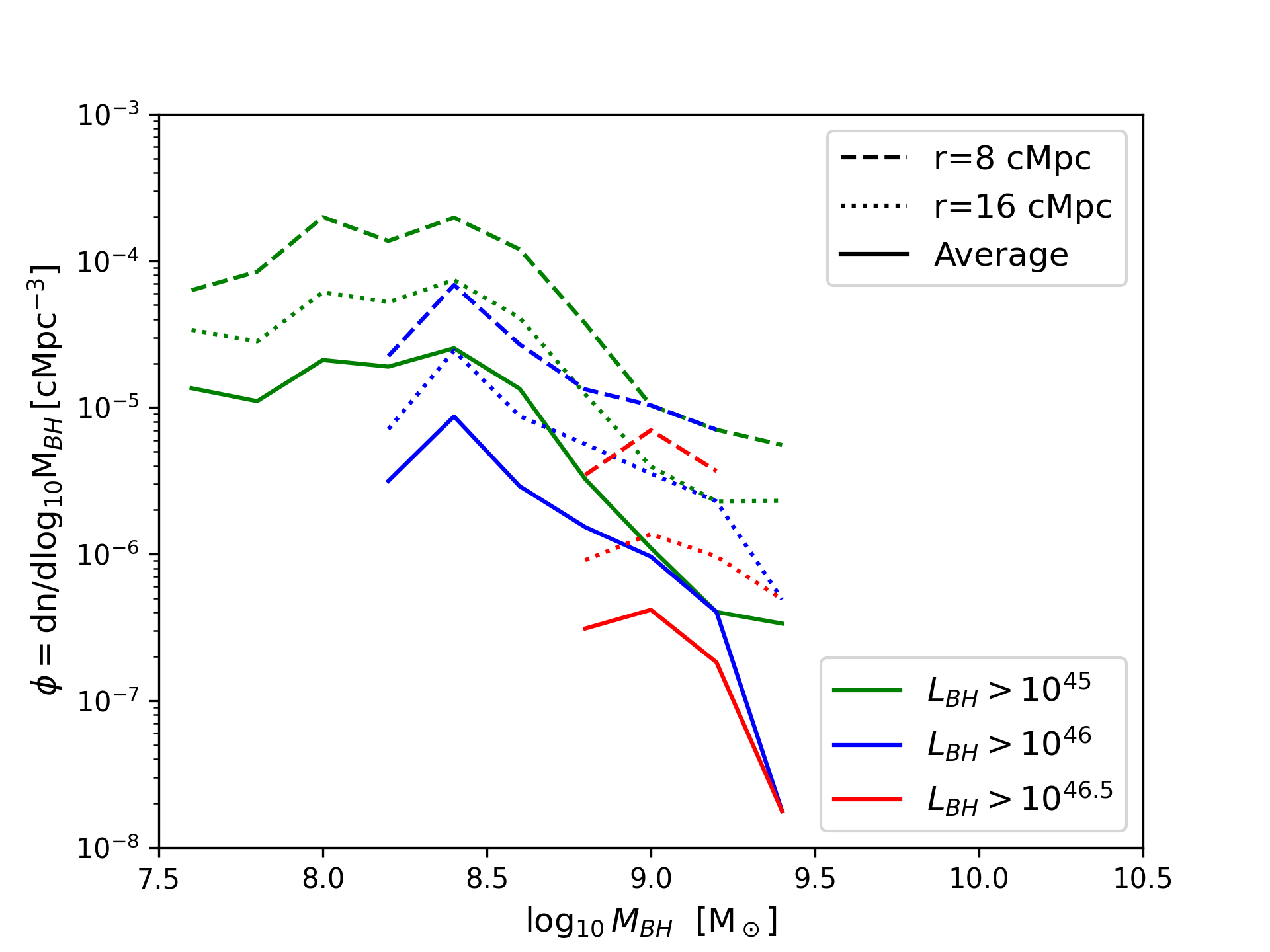}
    \caption{New neighborhood statistic at $z \sim 3$ - comparison of the BHMF of the accreting SMBH population that lies within 8, 16 and 40 comoving Mpc radius annuli around $10^9\,\Msun$ quasars in the LEGACY 1 Gpc Expanse box and SDSS quasars computed with QUOTAS.}
    \label{fig:neighborhood_bhmf}
\end{figure*}

\end{document}

%% file: macros.tex
\definecolor{codegreen}{rgb}{0,0.6,0}
\definecolor{codegray}{rgb}{0.5,0.5,0.5}
\definecolor{codepurple}{rgb}{0.58,0,0.82}
\definecolor{backcolour}{rgb}{0.95,0.95,0.92}
 
\newcommand{\adb}[1]{\textcolor{blue}{ #1}} 

\def\Msun{M_\odot}

%% file: Quotas_revised_v3.bbl
\begin{thebibliography}{}
\expandafter\ifx\csname natexlab\endcsname\relax\def\natexlab#1{#1}\fi
\providecommand{\url}[1]{\href{#1}{#1}}
\providecommand{\dodoi}[1]{doi:~\href{http://doi.org/#1}{\nolinkurl{#1}}}
\providecommand{\doeprint}[1]{\href{http://ascl.net/#1}{\nolinkurl{http://ascl.net/#1}}}
\providecommand{\doarXiv}[1]{\href{https://arxiv.org/abs/#1}{\nolinkurl{https://arxiv.org/abs/#1}}}

\bibitem[{{Abazajian} {et~al.}(2009){Abazajian}, {Adelman-McCarthy},
  {Ag{\"u}eros}, {Allam}, {Allende Prieto}, {An}, {Anderson}, {Anderson},
  {Annis}, {Bahcall}, {Bailer-Jones}, {Barentine}, {Bassett}, {Becker},
  {Beers}, {Bell}, {Belokurov}, {Berlind}, {Berman}, {Bernardi}, {Bickerton},
  {Bizyaev}, {Blakeslee}, {Blanton}, {Bochanski}, {Boroski}, {Brewington},
  {Brinchmann}, {Brinkmann}, {Brunner}, {Budav{\'a}ri}, {Carey}, {Carliles},
  {Carr}, {Castander}, {Cinabro}, {Connolly}, {Csabai}, {Cunha}, {Czarapata},
  {Davenport}, {de Haas}, {Dilday}, {Doi}, {Eisenstein}, {Evans}, {Evans},
  {Fan}, {Friedman}, {Frieman}, {Fukugita}, {G{\"a}nsicke}, {Gates},
  {Gillespie}, {Gilmore}, {Gonzalez}, {Gonzalez}, {Grebel}, {Gunn},
  {Gy{\"o}ry}, {Hall}, {Harding}, {Harris}, {Harvanek}, {Hawley}, {Hayes},
  {Heckman}, {Hendry}, {Hennessy}, {Hindsley}, {Hoblitt}, {Hogan}, {Hogg},
  {Holtzman}, {Hyde}, {Ichikawa}, {Ichikawa}, {Im}, {Ivezi{\'c}}, {Jester},
  {Jiang}, {Johnson}, {Jorgensen}, {Juri{\'c}}, {Kent}, {Kessler}, {Kleinman},
  {Knapp}, {Konishi}, {Kron}, {Krzesinski}, {Kuropatkin}, {Lampeitl},
  {Lebedeva}, {Lee}, {Lee}, {French Leger}, {L{\'e}pine}, {Li}, {Lima}, {Lin},
  {Long}, {Loomis}, {Loveday}, {Lupton}, {Magnier}, {Malanushenko},
  {Malanushenko}, {Mandelbaum}, {Margon}, {Marriner}, {Mart{\'\i}nez-Delgado},
  {Matsubara}, {McGehee}, {McKay}, {Meiksin}, {Morrison}, {Mullally}, {Munn},
  {Murphy}, {Nash}, {Nebot}, {Neilsen}, {Newberg}, {Newman}, {Nichol},
  {Nicinski}, {Nieto-Santisteban}, {Nitta}, {Okamura}, {Oravetz}, {Ostriker},
  {Owen}, {Padmanabhan}, {Pan}, {Park}, {Pauls}, {Peoples}, {Percival}, {Pier},
  {Pope}, {Pourbaix}, {Price}, {Purger}, {Quinn}, {Raddick}, {Re Fiorentin},
  {Richards}, {Richmond}, {Riess}, {Rix}, {Rockosi}, {Sako}, {Schlegel},
  {Schneider}, {Scholz}, {Schreiber}, {Schwope}, {Seljak}, {Sesar}, {Sheldon},
  {Shimasaku}, {Sibley}, {Simmons}, {Sivarani}, {Allyn Smith}, {Smith},
  {Smol{\v{c}}i{\'c}}, {Snedden}, {Stebbins}, {Steinmetz}, {Stoughton},
  {Strauss}, {SubbaRao}, {Suto}, {Szalay}, {Szapudi}, {Szkody}, {Tanaka},
  {Tegmark}, {Teodoro}, {Thakar}, {Tremonti}, {Tucker}, {Uomoto}, {Vanden
  Berk}, {Vandenberg}, {Vidrih}, {Vogeley}, {Voges}, {Vogt}, {Wadadekar},
  {Watters}, {Weinberg}, {West}, {White}, {Wilhite}, {Wonders}, {Yanny},
  {Yocum}, {York}, {Zehavi}, {Zibetti}, \& {Zucker}}]{2009ApJS..182..543A}
{Abazajian}, K.~N., {Adelman-McCarthy}, J.~K., {Ag{\"u}eros}, M.~A., {et~al.}
  2009, \apjs, 182, 543, \dodoi{10.1088/0067-0049/182/2/543}

\bibitem[{{Abolfathi} {et~al.}(2018){Abolfathi}, {Aguado}, {Aguilar}, {Allende
  Prieto}, {Almeida}, {Ananna}, {Anders}, {Anderson}, {Andrews}, {Anguiano},
  {Arag{\'o}n-Salamanca}, {Argudo-Fern{\'a}ndez}, {Armengaud}, {Ata},
  {Aubourg}, {Avila-Reese}, {Badenes}, {Bailey}, {Balland}, {Barger},
  {Barrera-Ballesteros}, {Bartosz}, {Bastien}, {Bates}, {Baumgarten},
  {Bautista}, {Beaton}, {Beers}, {Belfiore}, {Bender}, {Bernardi}, {Bershady},
  {Beutler}, {Bird}, {Bizyaev}, {Blanc}, {Blanton}, {Blomqvist}, {Bolton},
  {Boquien}, {Borissova}, {Bovy}, {Bradna Diaz}, {Brandt}, {Brinkmann},
  {Brownstein}, {Bundy}, {Burgasser}, {Burtin}, {Busca}, {Ca{\~n}as},
  {Cano-D{\'\i}az}, {Cappellari}, {Carrera}, {Casey}, {Cervantes Sodi}, {Chen},
  {Cherinka}, {Chiappini}, {Choi}, {Chojnowski}, {Chuang}, {Chung}, {Clerc},
  {Cohen}, {Comerford}, {Comparat}, {Correa do Nascimento}, {da Costa},
  {Cousinou}, {Covey}, {Crane}, {Cruz-Gonzalez}, {Cunha}, {da Silva Ilha},
  {Damke}, {Darling}, {Davidson}, {Dawson}, {de Icaza Lizaola}, {de la
  Macorra}, {de la Torre}, {De Lee}, {de Sainte Agathe}, {Deconto Machado},
  {Dell'Agli}, {Delubac}, {Diamond-Stanic}, {Donor}, {Downes}, {Drory}, {du Mas
  des Bourboux}, {Duckworth}, {Dwelly}, {Dyer}, {Ebelke}, {Davis Eigenbrot},
  {Eisenstein}, {Elsworth}, {Emsellem}, {Eracleous}, {Erfanianfar},
  {Escoffier}, {Fan}, {Fern{\'a}ndez Alvar}, {Fernandez-Trincado}, {Fernando
  Cirolini}, {Feuillet}, {Finoguenov}, {Fleming}, {Font-Ribera}, {Freischlad},
  {Frinchaboy}, {Fu}, {G{\'o}mez Maqueo Chew}, {Galbany}, {Garc{\'\i}a
  P{\'e}rez}, {Garcia-Dias}, {Garc{\'\i}a-Hern{\'a}ndez}, {Garma Oehmichen},
  {Gaulme}, {Gelfand}, {Gil-Mar{\'\i}n}, {Gillespie}, {Goddard}, {Gonz{\'a}lez
  Hern{\'a}ndez}, {Gonzalez-Perez}, {Grabowski}, {Green}, {Grier}, {Gueguen},
  {Guo}, {Guy}, {Hagen}, {Hall}, {Harding}, {Hasselquist}, {Hawley}, {Hayes},
  {Hearty}, {Hekker}, {Hernandez}, {Hernandez Toledo}, {Hogg},
  {Holley-Bockelmann}, {Holtzman}, {Hou}, {Hsieh}, {Hunt}, {Hutchinson},
  {Hwang}, {Jimenez Angel}, {Johnson}, {Jones}, {J{\"o}nsson}, {Jullo}, {Khan},
  {Kinemuchi}, {Kirkby}, {Kirkpatrick}, {Kitaura}, {Knapp}, {Kneib},
  {Kollmeier}, {Lacerna}, {Lane}, {Lang}, {Law}, {Le Goff}, {Lee}, {Li}, {Li},
  {Lian}, {Liang}, {Lima}, {Lin}, {Long}, {Lucatello}, {Lundgren}, {Mackereth},
  {MacLeod}, {Mahadevan}, {Maia}, {Majewski}, {Manchado}, {Maraston},
  {Mariappan}, {Marques-Chaves}, {Masseron}, {Masters}, {McDermid}, {McGreer},
  {Melendez}, {Meneses-Goytia}, {Merloni}, {Merrifield}, {Meszaros}, {Meza},
  {Minchev}, {Minniti}, {Mueller}, {Muller-Sanchez}, {Muna}, {Mu{\~n}oz},
  {Myers}, {Nair}, {Nandra}, {Ness}, {Newman}, {Nichol}, {Nidever},
  {Nitschelm}, {Noterdaeme}, {O'Connell}, {Oelkers}, {Oravetz}, {Oravetz},
  {Ort{\'\i}z}, {Osorio}, {Pace}, {Padilla}, {Palanque-Delabrouille},
  {Palicio}, {Pan}, {Pan}, {Parikh}, {P{\^a}ris}, {Park}, {Peirani},
  {Pellejero-Ibanez}, {Penny}, {Percival}, {Perez-Fournon}, {Petitjean},
  {Pieri}, {Pinsonneault}, {Pisani}, {Prada}, {Prakash}, {Queiroz}, {Raddick},
  {Raichoor}, {Barboza Rembold}, {Richstein}, {Riffel}, {Riffel}, {Rix},
  {Robin}, {Rodr{\'\i}guez Torres}, {Rom{\'a}n-Z{\'u}{\~n}iga}, {Ross},
  {Rossi}, {Ruan}, {Ruggeri}, {Ruiz}, {Salvato}, {S{\'a}nchez}, {S{\'a}nchez},
  {Sanchez Almeida}, {S{\'a}nchez-Gallego}, {Santana Rojas}, {Santiago},
  {Schiavon}, {Schimoia}, {Schlafly}, {Schlegel}, {Schneider}, {Schuster},
  {Schwope}, {Seo}, {Serenelli}, {Shen}, {Shen}, {Shetrone}, {Shull}, {Silva
  Aguirre}, {Simon}, {Skrutskie}, {Slosar}, {Smethurst}, {Smith}, {Sobeck},
  {Somers}, {Souter}, {Souto}, {Spindler}, {Stark}, {Stassun}, {Steinmetz},
  {Stello}, {Storchi-Bergmann}, {Streblyanska}, {Stringfellow}, {Su{\'a}rez},
  {Sun}, {Szigeti}, {Taghizadeh-Popp}, {Talbot}, {Tang}, {Tao}, {Tayar},
  {Tembe}, {Teske}, {Thakar}, {Thomas}, {Tissera}, {Tojeiro}, {Tremonti},
  {Troup}, {Urry}, {Valenzuela}, {van den Bosch}, {Vargas-Gonz{\'a}lez},
  {Vargas-Maga{\~n}a}, {Vazquez}, {Villanova}, {Vogt}, {Wake}, {Wang},
  {Weaver}, {Weijmans}, {Weinberg}, {Westfall}, {Whelan}, {Wilcots}, {Wild},
  {Williams}, {Wilson}, {Wood-Vasey}, {Wylezalek}, {Xiao}, {Yan}, {Yang},
  {Ybarra}, {Y{\`e}che}, {Zakamska}, {Zamora}, {Zarrouk}, {Zasowski}, {Zhang},
  {Zhao}, {Zhao}, {Zheng}, {Zheng}, {Zhou}, {Zhu}, {Zinn}, \&
  {Zou}}]{2018ApJS..235...42A}
{Abolfathi}, B., {Aguado}, D.~S., {Aguilar}, G., {et~al.} 2018, \apjs, 235, 42,
  \dodoi{10.3847/1538-4365/aa9e8a}

\bibitem[{{Agarwal} {et~al.}(2019){Agarwal}, {Cullen}, {Khochfar}, {Ceverino},
  \& {Klessen}}]{2019MNRAS.488.3268A}
{Agarwal}, B., {Cullen}, F., {Khochfar}, S., {Ceverino}, D., \& {Klessen},
  R.~S. 2019, \mnras, 488, 3268, \dodoi{10.1093/mnras/stz1347}

\bibitem[{{Agarwal} {et~al.}(2014){Agarwal}, {Dalla Vecchia}, {Johnson},
  {Khochfar}, \& {Paardekooper}}]{2014MNRAS.443..648A}
{Agarwal}, B., {Dalla Vecchia}, C., {Johnson}, J.~L., {Khochfar}, S., \&
  {Paardekooper}, J.-P. 2014, \mnras, 443, 648, \dodoi{10.1093/mnras/stu1112}

\bibitem[{{Agarwal} {et~al.}(2013){Agarwal}, {Davis}, {Khochfar}, {Natarajan},
  \& {Dunlop}}]{2013MNRAS.432.3438A}
{Agarwal}, B., {Davis}, A.~J., {Khochfar}, S., {Natarajan}, P., \& {Dunlop},
  J.~S. 2013, \mnras, 432, 3438, \dodoi{10.1093/mnras/stt696}

\bibitem[{{Agarwal} {et~al.}(2012){Agarwal}, {Khochfar}, {Johnson}, {Neistein},
  {Dalla Vecchia}, \& {Livio}}]{2012MNRAS.425.2854A}
{Agarwal}, B., {Khochfar}, S., {Johnson}, J.~L., {et~al.} 2012, \mnras, 425,
  2854, \dodoi{10.1111/j.1365-2966.2012.21651.x}

\bibitem[{{Alexander} \& {Hickox}(2012)}]{2012NewAR..56...93A}
{Alexander}, D.~M., \& {Hickox}, R.~C. 2012, \nar, 56, 93,
  \dodoi{10.1016/j.newar.2011.11.003}

\bibitem[{Alsing {et~al.}(2019)Alsing, Charnock, Feeney, \&
  Wandelt}]{Alsing_2019}
Alsing, J., Charnock, T., Feeney, S., \& Wandelt, B. 2019, Monthly Notices of
  the Royal Astronomical Society, \dodoi{10.1093/mnras/stz1960}

\bibitem[{{Ananna} {et~al.}(2019){Ananna}, {Treister}, {Urry}, {Ricci},
  {Kirkpatrick}, {LaMassa}, {Buchner}, {Civano}, {Tremmel}, \&
  {Marchesi}}]{2019ApJ...871..240A}
{Ananna}, T.~T., {Treister}, E., {Urry}, C.~M., {et~al.} 2019, \apj, 871, 240,
  \dodoi{10.3847/1538-4357/aafb77}

\bibitem[{{Avni} \& {Bahcall}(1980)}]{1980ApJ...235..694A}
{Avni}, Y., \& {Bahcall}, J.~N. 1980, \apj, 235, 694, \dodoi{10.1086/157673}

\bibitem[{{Ba{\~n}ados} {et~al.}(2018){Ba{\~n}ados}, {Venemans},
  {Mazzucchelli}, {Farina}, {Walter}, {Wang}, {Decarli}, {Stern}, {Fan},
  {Davies}, {Hennawi}, {Simcoe}, {Turner}, {Rix}, {Yang}, {Kelson}, {Rudie}, \&
  {Winters}}]{2018Natur.553..473B}
{Ba{\~n}ados}, E., {Venemans}, B.~P., {Mazzucchelli}, C., {et~al.} 2018, \nat,
  553, 473, \dodoi{10.1038/nature25180}

\bibitem[{{Barrow} {et~al.}(2018){Barrow}, {Wise}, {Aykutalp}, {O'Shea},
  {Norman}, \& {Xu}}]{2018MNRAS.474.2617B}
{Barrow}, K. S.~S., {Wise}, J.~H., {Aykutalp}, A., {et~al.} 2018, \mnras, 474,
  2617, \dodoi{10.1093/mnras/stx2973}

\bibitem[{{Beckmann} {et~al.}(2019){Beckmann}, {Devriendt}, \&
  {Slyz}}]{2019MNRAS.483.3488B}
{Beckmann}, R.~S., {Devriendt}, J., \& {Slyz}, A. 2019, \mnras, 483, 3488,
  \dodoi{10.1093/mnras/sty2890}

\bibitem[{{Beckmann} {et~al.}(2018){Beckmann}, {Slyz}, \&
  {Devriendt}}]{2018MNRAS.478..995B}
{Beckmann}, R.~S., {Slyz}, A., \& {Devriendt}, J. 2018, \mnras, 478, 995,
  \dodoi{10.1093/mnras/sty931}

\bibitem[{{Behroozi} {et~al.}(2013){Behroozi}, {Wechsler}, \&
  {Wu}}]{2013ApJ...762..109B}
{Behroozi}, P.~S., {Wechsler}, R.~H., \& {Wu}, H.-Y. 2013, \apj, 762, 109,
  \dodoi{10.1088/0004-637X/762/2/109}

\bibitem[{{Bentz} \& {Manne-Nicholas}(2018)}]{2018ApJ...864..146B}
{Bentz}, M.~C., \& {Manne-Nicholas}, E. 2018, \apj, 864, 146,
  \dodoi{10.3847/1538-4357/aad808}

\bibitem[{{Bentz} {et~al.}(2009){Bentz}, {Peterson}, {Netzer}, {Pogge}, \&
  {Vestergaard}}]{2009ApJ...697..160B}
{Bentz}, M.~C., {Peterson}, B.~M., {Netzer}, H., {Pogge}, R.~W., \&
  {Vestergaard}, M. 2009, \apj, 697, 160, \dodoi{10.1088/0004-637X/697/1/160}

\bibitem[{{Bernardini} {et~al.}(2020){Bernardini}, {Mayer}, {Reed}, \&
  {Feldmann}}]{2020MNRAS.496.5116B}
{Bernardini}, M., {Mayer}, L., {Reed}, D., \& {Feldmann}, R. 2020, \mnras, 496,
  5116, \dodoi{10.1093/mnras/staa1911}

\bibitem[{{Beskin} \& {Kuznetsova}(2000)}]{2000NCimB.115..795B}
{Beskin}, V.~S., \& {Kuznetsova}, I.~V. 2000, Nuovo Cimento B Serie, 115, 795.
\newblock \doarXiv{astro-ph/0004021}

\bibitem[{{Blandford} \& {McKee}(1982)}]{1982ApJ...255..419B}
{Blandford}, R.~D., \& {McKee}, C.~F. 1982, \apj, 255, 419,
  \dodoi{10.1086/159843}

\bibitem[{{Bower} {et~al.}(2006){Bower}, {Benson}, {Malbon}, {Helly}, {Frenk},
  {Baugh}, {Cole}, \& {Lacey}}]{2006MNRAS.370..645B}
{Bower}, R.~G., {Benson}, A.~J., {Malbon}, R., {et~al.} 2006, \mnras, 370, 645,
  \dodoi{10.1111/j.1365-2966.2006.10519.x}

\bibitem[{Breiman(2001)}]{random_forests}
Breiman, L. 2001, Mach. Learn., 45, 5–32, \dodoi{10.1023/A:1010933404324}

\bibitem[{{Brown} {et~al.}(2019){Brown}, {Duncan}, {Landt}, {Kirk}, {Ricci},
  {Kamraj}, {Salvato}, \& {Ananna}}]{2019MNRAS.489.3351B}
{Brown}, M.~J.~I., {Duncan}, K.~J., {Landt}, H., {et~al.} 2019, \mnras, 489,
  3351, \dodoi{10.1093/mnras/stz2324}

\bibitem[{{Cadiou} {et~al.}(2020){Cadiou}, {Pontzen}, \&
  {Peiris}}]{2020arXiv201202201C}
{Cadiou}, C., {Pontzen}, A., \& {Peiris}, H.~V. 2020, arXiv e-prints,
  arXiv:2012.02201.
\newblock \doarXiv{2012.02201}

\bibitem[{{Chen} {et~al.}(2013){Chen}, {Hickox}, {Alberts}, {Brodwin}, {Jones},
  {Murray}, {Alexander}, {Assef}, {Brown}, {Dey}, {Forman}, {Gorjian},
  {Goulding}, {Le Floc'h}, {Jannuzi}, {Mullaney}, \& {Pope}}]{Chen_2013}
{Chen}, C.-T.~J., {Hickox}, R.~C., {Alberts}, S., {et~al.} 2013, \apj, 773, 3,
  \dodoi{10.1088/0004-637X/773/1/3}

\bibitem[{{Coatman} {et~al.}(2017){Coatman}, {Hewett}, {Banerji}, {Richards},
  {Hennawi}, \& {Prochaska}}]{2017MNRAS.465.2120C}
{Coatman}, L., {Hewett}, P.~C., {Banerji}, M., {et~al.} 2017, \mnras, 465,
  2120, \dodoi{10.1093/mnras/stw2797}

\bibitem[{{Colpi} {et~al.}(2019){Colpi}, {Holley-Bockelmann}, {Bogdanovic},
  {Natarajan}, {Bellovary}, {Sesana}, {Tremmel}, {Schnittman}, {Comerford},
  {Barausse}, {Berti}, {Volonteri}, {Khan}, {McWilliams}, {Burke-Spolaor},
  {Hazboun}, {Conklin}, {Mueller}, \& {Larson}}]{2019arXiv190306867C}
{Colpi}, M., {Holley-Bockelmann}, K., {Bogdanovic}, T., {et~al.} 2019, arXiv
  e-prints, arXiv:1903.06867.
\newblock \doarXiv{1903.06867}

\bibitem[{Cranmer {et~al.}(2019)Cranmer, Galvez, Anderson, Spergel, \&
  Ho}]{cranmer2019modeling}
Cranmer, M.~D., Galvez, R., Anderson, L., Spergel, D.~N., \& Ho, S. 2019,
  Modeling the Gaia Color-Magnitude Diagram with Bayesian Neural Flows to
  Constrain Distance Estimates.
\newblock \doarXiv{1908.08045}

\bibitem[{{Dai} \& {Seljak}(2020)}]{2020arXiv201002926D}
{Dai}, B., \& {Seljak}, U. 2020, arXiv e-prints, arXiv:2010.02926.
\newblock \doarXiv{2010.02926}

\bibitem[{{Davis} {et~al.}(1985){Davis}, {Efstathiou}, {Frenk}, \&
  {White}}]{1985ApJ...292..371D}
{Davis}, M., {Efstathiou}, G., {Frenk}, C.~S., \& {White}, S.~D.~M. 1985, \apj,
  292, 371, \dodoi{10.1086/163168}

\bibitem[{{De Rosa} {et~al.}(2011){De Rosa}, {Decarli}, {Walter}, {Fan},
  {Jiang}, {Kurk}, {Pasquali}, \& {Rix}}]{2011ApJ...739...56D}
{De Rosa}, G., {Decarli}, R., {Walter}, F., {et~al.} 2011, \apj, 739, 56,
  \dodoi{10.1088/0004-637X/739/2/56}

\bibitem[{{De Rosa} {et~al.}(2014){De Rosa}, {Venemans}, {Decarli}, {Gennaro},
  {Simcoe}, {Dietrich}, {Peterson}, {Walter}, {Frank}, {McMahon}, {Hewett},
  {Mortlock}, \& {Simpson}}]{2014ApJ...790..145D}
{De Rosa}, G., {Venemans}, B.~P., {Decarli}, R., {et~al.} 2014, \apj, 790, 145,
  \dodoi{10.1088/0004-637X/790/2/145}

\bibitem[{{Debuhr} {et~al.}(2011){Debuhr}, {Quataert}, \&
  {Ma}}]{2011MNRAS.412.1341D}
{Debuhr}, J., {Quataert}, E., \& {Ma}, C.-P. 2011, \mnras, 412, 1341,
  \dodoi{10.1111/j.1365-2966.2010.17992.x}

\bibitem[{{Di Matteo} {et~al.}(2005){Di Matteo}, {Springel}, \&
  {Hernquist}}]{2005Natur.433..604D}
{Di Matteo}, T., {Springel}, V., \& {Hernquist}, L. 2005, \nat, 433, 604,
  \dodoi{10.1038/nature03335}

\bibitem[{{Ding} {et~al.}(2020){Ding}, {Silverman}, {Treu}, {Schulze},
  {Schramm}, {Birrer}, {Park}, {Jahnke}, {Bennert}, {Kartaltepe}, {Koekemoer},
  {Malkan}, \& {Sanders}}]{2020ApJ...888...37D}
{Ding}, X., {Silverman}, J., {Treu}, T., {et~al.} 2020, \apj, 888, 37,
  \dodoi{10.3847/1538-4357/ab5b90}

\bibitem[{Dinh {et~al.}(2017)Dinh, Sohl-Dickstein, \& Bengio}]{dinh2017density}
Dinh, L., Sohl-Dickstein, J., \& Bengio, S. 2017, Density estimation using Real
  NVP.
\newblock \doarXiv{1605.08803}

\bibitem[{{Djorgovski} {et~al.}(2016){Djorgovski}, {Graham}, {Donalek},
  {Mahabal}, {Drake}, {Turmon}, \& {Fuchs}}]{Djorgovski+2016}
{Djorgovski}, S.~G., {Graham}, M.~J., {Donalek}, C., {et~al.} 2016, arXiv
  e-prints, arXiv:1601.04385.
\newblock \doarXiv{1601.04385}

\bibitem[{{Dubois} {et~al.}(2012){Dubois}, {Devriendt}, {Slyz}, \&
  {Teyssier}}]{2012MNRAS.420.2662D}
{Dubois}, Y., {Devriendt}, J., {Slyz}, A., \& {Teyssier}, R. 2012, \mnras, 420,
  2662, \dodoi{10.1111/j.1365-2966.2011.20236.x}

\bibitem[{{Eftekharzadeh} {et~al.}(2015){Eftekharzadeh}, {Myers}, {White},
  {Weinberg}, {Schneider}, {Shen}, {Font-Ribera}, {Ross}, {Paris}, \&
  {Streblyanska}}]{2015MNRAS.453.2779E}
{Eftekharzadeh}, S., {Myers}, A.~D., {White}, M., {et~al.} 2015, \mnras, 453,
  2779, \dodoi{10.1093/mnras/stv1763}

\bibitem[{{Eilers} {et~al.}(2018){Eilers}, {Hennawi}, \&
  {Davies}}]{2018ApJ...867...30E}
{Eilers}, A.-C., {Hennawi}, J.~F., \& {Davies}, F.~B. 2018, \apj, 867, 30,
  \dodoi{10.3847/1538-4357/aae081}

\bibitem[{{Eilers} {et~al.}(2022){Eilers}, {Simcoe}, {Yue}, {Mackenzie},
  {Matthee}, {Durovcikova}, {Kashino}, {Bordoloi}, \& {Lilly}}]{Eilers+2022}
{Eilers}, A.-C., {Simcoe}, R.~A., {Yue}, M., {et~al.} 2022, arXiv e-prints,
  arXiv:2211.16261, \dodoi{10.48550/arXiv.2211.16261}

\bibitem[{{Event Horizon Telescope Collaboration}
  {et~al.}(2019{\natexlab{a}}){Event Horizon Telescope Collaboration},
  {Akiyama}, {Alberdi}, {Alef}, {Asada}, {Azulay}, {Baczko}, {Ball},
  {Balokovi{\'c}}, {Barrett}, {Bintley}, {Blackburn}, {Boland}, {Bouman},
  {Bower}, {Bremer}, {Brinkerink}, {Brissenden}, {Britzen}, {Broderick},
  {Broguiere}, {Bronzwaer}, {Byun}, {Carlstrom}, {Chael}, {Chan}, {Chatterjee},
  {Chatterjee}, {Chen}, {Chen}, {Cho}, {Christian}, {Conway}, {Cordes}, {Crew},
  {Cui}, {Davelaar}, {De Laurentis}, {Deane}, {Dempsey}, {Desvignes}, {Dexter},
  {Doeleman}, {Eatough}, {Falcke}, {Fish}, {Fomalont}, {Fraga-Encinas},
  {Freeman}, {Friberg}, {Fromm}, {G{\'o}mez}, {Galison}, {Gammie},
  {Garc{\'\i}a}, {Gentaz}, {Georgiev}, {Goddi}, {Gold}, {Gu}, {Gurwell},
  {Hada}, {Hecht}, {Hesper}, {Ho}, {Ho}, {Honma}, {Huang}, {Huang}, {Hughes},
  {Ikeda}, {Inoue}, {Issaoun}, {James}, {Jannuzi}, {Janssen}, {Jeter}, {Jiang},
  {Johnson}, {Jorstad}, {Jung}, {Karami}, {Karuppusamy}, {Kawashima},
  {Keating}, {Kettenis}, {Kim}, {Kim}, {Kim}, {Kino}, {Koay}, {Koch}, {Koyama},
  {Kramer}, {Kramer}, {Krichbaum}, {Kuo}, {Lauer}, {Lee}, {Li}, {Li},
  {Lindqvist}, {Liu}, {Liuzzo}, {Lo}, {Lobanov}, {Loinard}, {Lonsdale}, {Lu},
  {MacDonald}, {Mao}, {Markoff}, {Marrone}, {Marscher}, {Mart{\'\i}-Vidal},
  {Matsushita}, {Matthews}, {Medeiros}, {Menten}, {Mizuno}, {Mizuno}, {Moran},
  {Moriyama}, {Moscibrodzka}, {M{\"u}ller}, {Nagai}, {Nagar}, {Nakamura},
  {Narayan}, {Narayanan}, {Natarajan}, {Neri}, {Ni}, {Noutsos}, {Okino},
  {Olivares}, {Ortiz-Le{\'o}n}, {Oyama}, {{\"O}zel}, {Palumbo}, {Patel}, {Pen},
  {Pesce}, {Pi{\'e}tu}, {Plambeck}, {PopStefanija}, {Porth}, {Prather},
  {Preciado-L{\'o}pez}, {Psaltis}, {Pu}, {Ramakrishnan}, {Rao}, {Rawlings},
  {Raymond}, {Rezzolla}, {Ripperda}, {Roelofs}, {Rogers}, {Ros}, {Rose},
  {Roshanineshat}, {Rottmann}, {Roy}, {Ruszczyk}, {Ryan}, {Rygl},
  {S{\'a}nchez}, {S{\'a}nchez-Arguelles}, {Sasada}, {Savolainen}, {Schloerb},
  {Schuster}, {Shao}, {Shen}, {Small}, {Sohn}, {SooHoo}, {Tazaki}, {Tiede},
  {Tilanus}, {Titus}, {Toma}, {Torne}, {Trent}, {Trippe}, {Tsuda}, {van
  Bemmel}, {van Langevelde}, {van Rossum}, {Wagner}, {Wardle}, {Weintroub},
  {Wex}, {Wharton}, {Wielgus}, {Wong}, {Wu}, {Young}, {Young}, {Younsi},
  {Yuan}, {Yuan}, {Zensus}, {Zhao}, {Zhao}, {Zhu}, {Algaba}, {Allardi},
  {Amestica}, {Anczarski}, {Bach}, {Baganoff}, {Beaudoin}, {Benson},
  {Berthold}, {Blanchard}, {Blundell}, {Bustamente}, {Cappallo},
  {Castillo-Dom{\'\i}nguez}, {Chang}, {Chang}, {Chang}, {Chen}, {Chilson},
  {Chuter}, {C{\'o}rdova Rosado}, {Coulson}, {Crawford}, {Crowley}, {David},
  {Derome}, {Dexter}, {Dornbusch}, {Dudevoir}, {Dzib}, {Eckart}, {Eckert},
  {Erickson}, {Everett}, {Faber}, {Farah}, {Fath}, {Folkers}, {Forbes},
  {Freund}, {G{\'o}mez-Ruiz}, {Gale}, {Gao}, {Geertsema}, {Graham}, {Greer},
  {Grosslein}, {Gueth}, {Haggard}, {Halverson}, {Han}, {Han}, {Hao},
  {Hasegawa}, {Henning}, {Hern{\'a}ndez-G{\'o}mez}, {Herrero-Illana},
  {Heyminck}, {Hirota}, {Hoge}, {Huang}, {Impellizzeri}, {Jiang}, {Kamble},
  {Keisler}, {Kimura}, {Kono}, {Kubo}, {Kuroda}, {Lacasse}, {Laing}, {Leitch},
  {Li}, {Lin}, {Liu}, {Liu}, {Lu}, {Marson}, {Martin-Cocher}, {Massingill},
  {Matulonis}, {McColl}, {McWhirter}, {Messias}, {Meyer-Zhao}, {Michalik},
  {Monta{\~n}a}, {Montgomerie}, {Mora-Klein}, {Muders}, {Nadolski}, {Navarro},
  {Neilsen}, {Nguyen}, {Nishioka}, {Norton}, {Nowak}, {Nystrom}, {Ogawa},
  {Oshiro}, {Oyama}, {Parsons}, {Paine}, {Pe{\~n}alver}, {Phillips}, {Poirier},
  {Pradel}, {Primiani}, {Raffin}, {Rahlin}, {Reiland}, {Risacher}, {Ruiz},
  {S{\'a}ez-Mada{\'\i}n}, {Sassella}, {Schellart}, {Shaw}, {Silva}, {Shiokawa},
  {Smith}, {Snow}, {Souccar}, {Sousa}, {Sridharan}, {Srinivasan}, {Stahm},
  {Stark}, {Story}, {Timmer}, {Vertatschitsch}, {Walther}, {Wei}, {Whitehorn},
  {Whitney}, {Woody}, {Wouterloot}, {Wright}, {Yamaguchi}, {Yu}, {Zeballos},
  {Zhang}, \& {Ziurys}}]{2019ApJ...875L...1E}
{Event Horizon Telescope Collaboration}, {Akiyama}, K., {Alberdi}, A., {et~al.}
  2019{\natexlab{a}}, \apjl, 875, L1, \dodoi{10.3847/2041-8213/ab0ec7}

\bibitem[{{Event Horizon Telescope Collaboration}
  {et~al.}(2019{\natexlab{b}}){Event Horizon Telescope Collaboration},
  {Akiyama}, {Alberdi}, {Alef}, {Asada}, {Azulay}, {Baczko}, {Ball},
  {Balokovi{\'c}}, {Barrett}, {Bintley}, {Blackburn}, {Boland}, {Bouman},
  {Bower}, {Bremer}, {Brinkerink}, {Brissenden}, {Britzen}, {Broderick},
  {Broguiere}, {Bronzwaer}, {Byun}, {Carlstrom}, {Chael}, {Chan}, {Chatterjee},
  {Chatterjee}, {Chen}, {Chen}, {Cho}, {Christian}, {Conway}, {Cordes}, {Crew},
  {Cui}, {Davelaar}, {De Laurentis}, {Deane}, {Dempsey}, {Desvignes}, {Dexter},
  {Doeleman}, {Eatough}, {Falcke}, {Fish}, {Fomalont}, {Fraga-Encinas},
  {Friberg}, {Fromm}, {G{\'o}mez}, {Galison}, {Gammie}, {Garc{\'\i}a},
  {Gentaz}, {Georgiev}, {Goddi}, {Gold}, {Gu}, {Gurwell}, {Hada}, {Hecht},
  {Hesper}, {Ho}, {Ho}, {Honma}, {Huang}, {Huang}, {Hughes}, {Ikeda}, {Inoue},
  {Issaoun}, {James}, {Jannuzi}, {Janssen}, {Jeter}, {Jiang}, {Johnson},
  {Jorstad}, {Jung}, {Karami}, {Karuppusamy}, {Kawashima}, {Keating},
  {Kettenis}, {Kim}, {Kim}, {Kim}, {Kino}, {Koay}, {Koch}, {Koyama}, {Kramer},
  {Kramer}, {Krichbaum}, {Kuo}, {Lauer}, {Lee}, {Li}, {Li}, {Lindqvist}, {Liu},
  {Liuzzo}, {Lo}, {Lobanov}, {Loinard}, {Lonsdale}, {Lu}, {MacDonald}, {Mao},
  {Markoff}, {Marrone}, {Marscher}, {Mart{\'\i}-Vidal}, {Matsushita},
  {Matthews}, {Medeiros}, {Menten}, {Mizuno}, {Mizuno}, {Moran}, {Moriyama},
  {Moscibrodzka}, {M{\"u}ller}, {Nagai}, {Nagar}, {Nakamura}, {Narayan},
  {Narayanan}, {Natarajan}, {Neri}, {Ni}, {Noutsos}, {Okino}, {Olivares},
  {Ortiz-Le{\'o}n}, {Oyama}, {{\"O}zel}, {Palumbo}, {Patel}, {Pen}, {Pesce},
  {Pi{\'e}tu}, {Plambeck}, {PopStefanija}, {Porth}, {Prather},
  {Preciado-L{\'o}pez}, {Psaltis}, {Pu}, {Ramakrishnan}, {Rao}, {Rawlings},
  {Raymond}, {Rezzolla}, {Ripperda}, {Roelofs}, {Rogers}, {Ros}, {Rose},
  {Roshanineshat}, {Rottmann}, {Roy}, {Ruszczyk}, {Ryan}, {Rygl},
  {S{\'a}nchez}, {S{\'a}nchez-Arguelles}, {Sasada}, {Savolainen}, {Schloerb},
  {Schuster}, {Shao}, {Shen}, {Small}, {Sohn}, {SooHoo}, {Tazaki}, {Tiede},
  {Tilanus}, {Titus}, {Toma}, {Torne}, {Trent}, {Trippe}, {Tsuda}, {van
  Bemmel}, {van Langevelde}, {van Rossum}, {Wagner}, {Wardle}, {Weintroub},
  {Wex}, {Wharton}, {Wielgus}, {Wong}, {Wu}, {Young}, {Young}, {Younsi},
  {Yuan}, {Yuan}, {Zensus}, {Zhao}, {Zhao}, {Zhu}, {Cappallo}, {Farah},
  {Folkers}, {Meyer-Zhao}, {Michalik}, {Nadolski}, {Nishioka}, {Pradel},
  {Primiani}, {Souccar}, {Vertatschitsch}, \&
  {Yamaguchi}}]{2019ApJ...875L...3E}
---. 2019{\natexlab{b}}, \apjl, 875, L3, \dodoi{10.3847/2041-8213/ab0c57}

\bibitem[{{Farina} {et~al.}(2022){Farina}, {Schindler}, {Walter},
  {Ba{\~n}ados}, {Davies}, {Decarli}, {Eilers}, {Fan}, {Hennawi},
  {Mazzucchelli}, {Meyer}, {Trakhtenbrot}, {Volonteri}, {Wang}, {Worseck},
  {Yang}, {Gutcke}, {Venemans}, {Bosman}, {Costa}, {De Rosa}, {Drake}, \&
  {Onoue}}]{2022ApJ...941..106F}
{Farina}, E.~P., {Schindler}, J.-T., {Walter}, F., {et~al.} 2022, \apj, 941,
  106, \dodoi{10.3847/1538-4357/ac9626}

\bibitem[{{Ferrarese} \& {Merritt}(2000)}]{2000ApJ...539L...9F}
{Ferrarese}, L., \& {Merritt}, D. 2000, \apjl, 539, L9, \dodoi{10.1086/312838}

\bibitem[{{Gaibler} {et~al.}(2011){Gaibler}, {Khochfar}, \&
  {Krause}}]{2011MNRAS.411..155G}
{Gaibler}, V., {Khochfar}, S., \& {Krause}, M. 2011, \mnras, 411, 155,
  \dodoi{10.1111/j.1365-2966.2010.17674.x}

\bibitem[{{Gammie} {et~al.}(2003){Gammie}, {McKinney}, \&
  {T{\'o}th}}]{2003ApJ...589..444G}
{Gammie}, C.~F., {McKinney}, J.~C., \& {T{\'o}th}, G. 2003, \apj, 589, 444,
  \dodoi{10.1086/374594}

\bibitem[{{Gaspari} {et~al.}(2020){Gaspari}, {Tombesi}, \&
  {Cappi}}]{2020NatAs...4...10G}
{Gaspari}, M., {Tombesi}, F., \& {Cappi}, M. 2020, Nature Astronomy, 4, 10,
  \dodoi{10.1038/s41550-019-0970-1}

\bibitem[{{Gebhardt} {et~al.}(2000){Gebhardt}, {Bender}, {Bower}, {Dressler},
  {Faber}, {Filippenko}, {Green}, {Grillmair}, {Ho}, {Kormendy}, {Lauer},
  {Magorrian}, {Pinkney}, {Richstone}, \& {Tremaine}}]{2000ApJ...539L..13G}
{Gebhardt}, K., {Bender}, R., {Bower}, G., {et~al.} 2000, \apjl, 539, L13,
  \dodoi{10.1086/312840}

\bibitem[{{Genzel}(2014)}]{2014arXiv1410.8717G}
{Genzel}, R. 2014, arXiv e-prints, arXiv:1410.8717.
\newblock \doarXiv{1410.8717}

\bibitem[{{Genzel} {et~al.}(1997){Genzel}, {Eckart}, {Ott}, \&
  {Eisenhauer}}]{1997MNRAS.291..219G}
{Genzel}, R., {Eckart}, A., {Ott}, T., \& {Eisenhauer}, F. 1997, \mnras, 291,
  219, \dodoi{10.1093/mnras/291.1.219}

\bibitem[{{Ghez} {et~al.}(1998){Ghez}, {Klein}, {Morris}, \&
  {Becklin}}]{1998ApJ...509..678G}
{Ghez}, A.~M., {Klein}, B.~L., {Morris}, M., \& {Becklin}, E.~E. 1998, \apj,
  509, 678, \dodoi{10.1086/306528}

\bibitem[{{Ghosh} {et~al.}(2020){Ghosh}, {Urry}, {Wang}, {Schawinski}, {Turp},
  \& {Powell}}]{2020ApJ...895..112G}
{Ghosh}, A., {Urry}, C.~M., {Wang}, Z., {et~al.} 2020, \apj, 895, 112,
  \dodoi{10.3847/1538-4357/ab8a47}

\bibitem[{{Giallongo} {et~al.}(2015){Giallongo}, {Grazian}, {Fiore}, {Fontana},
  {Pentericci}, {Vanzella}, {Dickinson}, {Kocevski}, {Castellano}, {Cristiani},
  {Ferguson}, {Finkelstein}, {Grogin}, {Hathi}, {Koekemoer}, {Newman}, \&
  {Salvato}}]{2015AA...578A..83G}
{Giallongo}, E., {Grazian}, A., {Fiore}, F., {et~al.} 2015, \aap, 578, A83,
  \dodoi{10.1051/0004-6361/201425334}

\bibitem[{{Glikman} {et~al.}(2011){Glikman}, {Djorgovski}, {Stern}, {Dey},
  {Jannuzi}, \& {Lee}}]{2011ApJ...728L..26G}
{Glikman}, E., {Djorgovski}, S.~G., {Stern}, D., {et~al.} 2011, \apjl, 728,
  L26, \dodoi{10.1088/2041-8205/728/2/L26}

\bibitem[{{Grier} {et~al.}(2017){Grier}, {Pancoast}, {Barth}, {Fausnaugh},
  {Brewer}, {Treu}, \& {Peterson}}]{2017ApJ...849..146G}
{Grier}, C.~J., {Pancoast}, A., {Barth}, A.~J., {et~al.} 2017, \apj, 849, 146,
  \dodoi{10.3847/1538-4357/aa901b}

\bibitem[{{G{\"u}ltekin} {et~al.}(2009){G{\"u}ltekin}, {Richstone}, {Gebhardt},
  {Lauer}, {Tremaine}, {Aller}, {Bender}, {Dressler}, {Faber}, {Filippenko},
  {Green}, {Ho}, {Kormendy}, {Magorrian}, {Pinkney}, \&
  {Siopis}}]{2009ApJ...698..198G}
{G{\"u}ltekin}, K., {Richstone}, D.~O., {Gebhardt}, K., {et~al.} 2009, \apj,
  698, 198, \dodoi{10.1088/0004-637X/698/1/198}

\bibitem[{{Habouzit} {et~al.}(2017){Habouzit}, {Volonteri}, \&
  {Dubois}}]{2017MNRAS.468.3935H}
{Habouzit}, M., {Volonteri}, M., \& {Dubois}, Y. 2017, \mnras, 468, 3935,
  \dodoi{10.1093/mnras/stx666}

\bibitem[{{Habouzit} {et~al.}(2022){Habouzit}, {Onoue}, {Ba{\~n}ados},
  {Neeleman}, {Angl{\'e}s-Alc{\'a}zar}, {Walter}, {Pillepich}, {Dav{\'e}},
  {Jahnke}, \& {Dubois}}]{2022MNRAS.511.3751H}
{Habouzit}, M., {Onoue}, M., {Ba{\~n}ados}, E., {et~al.} 2022, \mnras, 511,
  3751, \dodoi{10.1093/mnras/stac225}

\bibitem[{{Haehnelt} {et~al.}(1998){Haehnelt}, {Natarajan}, \&
  {Rees}}]{1998MNRAS.300..817H}
{Haehnelt}, M.~G., {Natarajan}, P., \& {Rees}, M.~J. 1998, \mnras, 300, 817,
  \dodoi{10.1046/j.1365-8711.1998.01951.x}

\bibitem[{{Hahn} \& {Abel}(2011)}]{2011MNRAS.415.2101H}
{Hahn}, O., \& {Abel}, T. 2011, \mnras, 415, 2101,
  \dodoi{10.1111/j.1365-2966.2011.18820.x}

\bibitem[{{H{\"a}ring} \& {Rix}(2004)}]{2004ApJ...604L..89H}
{H{\"a}ring}, N., \& {Rix}, H.-W. 2004, \apjl, 604, L89, \dodoi{10.1086/383567}

\bibitem[{{Harrison} {et~al.}(2018){Harrison}, {Costa}, {Tadhunter},
  {Fl{\"u}tsch}, {Kakkad}, {Perna}, \& {Vietri}}]{2018NatAs...2..198H}
{Harrison}, C.~M., {Costa}, T., {Tadhunter}, C.~N., {et~al.} 2018, Nature
  Astronomy, 2, 198, \dodoi{10.1038/s41550-018-0403-6}

\bibitem[{{Hasinger}(2008)}]{2008A&A...490..905H}
{Hasinger}, G. 2008, \aap, 490, 905, \dodoi{10.1051/0004-6361:200809839}

\bibitem[{{Hawley} {et~al.}(2007){Hawley}, {Beckwith}, \&
  {Krolik}}]{2007Ap&SS.311..117H}
{Hawley}, J.~F., {Beckwith}, K., \& {Krolik}, J.~H. 2007, \apss, 311, 117,
  \dodoi{10.1007/s10509-007-9559-8}

\bibitem[{{Heckman} \& {Best}(2014)}]{2014ARA&A..52..589H}
{Heckman}, T.~M., \& {Best}, P.~N. 2014, \araa, 52, 589,
  \dodoi{10.1146/annurev-astro-081913-035722}

\bibitem[{{Hinshaw} {et~al.}(2013){Hinshaw}, {Larson}, {Komatsu}, {Spergel},
  {Bennett}, {Dunkley}, {Nolta}, {Halpern}, {Hill}, {Odegard}, {Page}, {Smith},
  {Weiland}, {Gold}, {Jarosik}, {Kogut}, {Limon}, {Meyer}, {Tucker}, {Wollack},
  \& {Wright}}]{2013ApJS..208...19H}
{Hinshaw}, G., {Larson}, D., {Komatsu}, E., {et~al.} 2013, \apjs, 208, 19,
  \dodoi{10.1088/0067-0049/208/2/19}

\bibitem[{{Hirschmann} {et~al.}(2010){Hirschmann}, {Khochfar}, {Burkert},
  {Naab}, {Genel}, \& {Somerville}}]{2010MNRAS.407.1016H}
{Hirschmann}, M., {Khochfar}, S., {Burkert}, A., {et~al.} 2010, \mnras, 407,
  1016, \dodoi{10.1111/j.1365-2966.2010.17006.x}

\bibitem[{{Hort{\'u}a} {et~al.}(2020){Hort{\'u}a}, {Volpi}, {Marinelli}, \&
  {Malag{\`o}}}]{2020PhRvD.102j3509H}
{Hort{\'u}a}, H.~J., {Volpi}, R., {Marinelli}, D., \& {Malag{\`o}}, L. 2020,
  \prd, 102, 103509, \dodoi{10.1103/PhysRevD.102.103509}

\bibitem[{Hortúa {et~al.}(2020)Hortúa, Malagò, \& Volpi}]{Hort_a_2020}
Hortúa, H.~J., Malagò, L., \& Volpi, R. 2020, Machine Learning: Science and
  Technology, 1, 035014, \dodoi{10.1088/2632-2153/aba6f1}

\bibitem[{{Jahnke} \& {Macci{\`o}}(2011)}]{2011ApJ...734...92J}
{Jahnke}, K., \& {Macci{\`o}}, A.~V. 2011, \apj, 734, 92,
  \dodoi{10.1088/0004-637X/734/2/92}

\bibitem[{{Jiang} {et~al.}(2007){Jiang}, {Fan}, {Vestergaard}, {Kurk},
  {Walter}, {Kelly}, \& {Strauss}}]{2007AJ....134.1150J}
{Jiang}, L., {Fan}, X., {Vestergaard}, M., {et~al.} 2007, \aj, 134, 1150,
  \dodoi{10.1086/520811}

\bibitem[{{Jiang} {et~al.}(2016){Jiang}, {McGreer}, {Fan}, {Strauss},
  {Ba{\~n}ados}, {Becker}, {Bian}, {Farnsworth}, {Shen}, {Wang}, {Wang},
  {Wang}, {White}, {Wu}, {Wu}, {Yang}, \& {Yang}}]{2016ApJ...833..222J}
{Jiang}, L., {McGreer}, I.~D., {Fan}, X., {et~al.} 2016, \apj, 833, 222,
  \dodoi{10.3847/1538-4357/833/2/222}

\bibitem[{{Johnson} {et~al.}(2011){Johnson}, {Khochfar}, {Greif}, \&
  {Durier}}]{2011MNRAS.410..919J}
{Johnson}, J.~L., {Khochfar}, S., {Greif}, T.~H., \& {Durier}, F. 2011, \mnras,
  410, 919, \dodoi{10.1111/j.1365-2966.2010.17491.x}

\bibitem[{{Juri{\'c}} {et~al.}(2017){Juri{\'c}}, {Kantor}, {Lim}, {Lupton},
  {Dubois-Felsmann}, {Jenness}, {Axelrod}, {Aleksi{\'c}}, {Allsman},
  {AlSayyad}, {Alt}, {Armstrong}, {Basney}, {Becker}, {Becla}, {Biswas},
  {Bosch}, {Boutigny}, {Kind}, {Ciardi}, {Connolly}, {Daniel}, {Daues},
  {Economou}, {Chiang}, {Fausti}, {Fisher-Levine}, {Freemon}, {Gris},
  {Hernandez}, {Hoblitt}, {Ivezi{\'c}}, {Jammes}, {Jevremovi{\'c}}, {Jones},
  {Kalmbach}, {Kasliwal}, {Krughoff}, {Lurie}, {Lust}, {MacArthur}, {Melchior},
  {Moeyens}, {Nidever}, {Owen}, {Parejko}, {Peterson}, {Petravick},
  {Pietrowicz}, {Price}, {Reiss}, {Shaw}, {Sick}, {Slater}, {Strauss},
  {Sullivan}, {Swinbank}, {Van Dyk}, {Vuj{\v{c}}i{\'c}}, {Withers}, \&
  {Yoachim}}]{2017ASPC..512..279J}
{Juri{\'c}}, M., {Kantor}, J., {Lim}, K.~T., {et~al.} 2017, in Astronomical
  Society of the Pacific Conference Series, Vol. 512, Astronomical Data
  Analysis Software and Systems XXV, ed. N.~P.~F. {Lorente}, K.~{Shortridge},
  \& R.~{Wayth}, 279

\bibitem[{{Kamdar} {et~al.}(2016){Kamdar}, {Turk}, \&
  {Brunner}}]{2016MNRAS.457.1162K}
{Kamdar}, H.~M., {Turk}, M.~J., \& {Brunner}, R.~J. 2016, \mnras, 457, 1162,
  \dodoi{10.1093/mnras/stv2981}

\bibitem[{{Kashikawa} {et~al.}(2015){Kashikawa}, {Ishizaki}, {Willott},
  {Onoue}, {Im}, {Furusawa}, {Toshikawa}, {Ishikawa}, {Niino}, {Shimasaku},
  {Ouchi}, \& {Hibon}}]{2015ApJ...798...28K}
{Kashikawa}, N., {Ishizaki}, Y., {Willott}, C.~J., {et~al.} 2015, \apj, 798,
  28, \dodoi{10.1088/0004-637X/798/1/28}

\bibitem[{{Kaspi} {et~al.}(2005){Kaspi}, {Maoz}, {Netzer}, {Peterson},
  {Vestergaard}, \& {Jannuzi}}]{2005ApJ...629...61K}
{Kaspi}, S., {Maoz}, D., {Netzer}, H., {et~al.} 2005, \apj, 629, 61,
  \dodoi{10.1086/431275}

\bibitem[{{Kaspi} {et~al.}(2000){Kaspi}, {Smith}, {Netzer}, {Maoz}, {Jannuzi},
  \& {Giveon}}]{2000ApJ...533..631K}
{Kaspi}, S., {Smith}, P.~S., {Netzer}, H., {et~al.} 2000, \apj, 533, 631,
  \dodoi{10.1086/308704}

\bibitem[{{Katz} {et~al.}(2020){Katz}, {Kelley}, {Dosopoulou}, {Berry},
  {Blecha}, \& {Larson}}]{2020MNRAS.491.2301K}
{Katz}, M.~L., {Kelley}, L.~Z., {Dosopoulou}, F., {et~al.} 2020, \mnras, 491,
  2301, \dodoi{10.1093/mnras/stz3102}

\bibitem[{{Kelley} {et~al.}(2018){Kelley}, {Blecha}, {Hernquist}, {Sesana}, \&
  {Taylor}}]{2018MNRAS.477..964K}
{Kelley}, L.~Z., {Blecha}, L., {Hernquist}, L., {Sesana}, A., \& {Taylor},
  S.~R. 2018, \mnras, 477, 964, \dodoi{10.1093/mnras/sty689}

\bibitem[{{Kelly} \& {Shen}(2013)}]{2013ApJ...764...45K}
{Kelly}, B.~C., \& {Shen}, Y. 2013, \apj, 764, 45,
  \dodoi{10.1088/0004-637X/764/1/45}

\bibitem[{Kelly {et~al.}(2010)Kelly, Vestergaard, Fan, Hopkins, Hernquist, \&
  Siemiginowska}]{Kelly_2010}
Kelly, B.~C., Vestergaard, M., Fan, X., {et~al.} 2010, The Astrophysical
  Journal, 719, 1315, \dodoi{10.1088/0004-637x/719/2/1315}

\bibitem[{Kingma {et~al.}(2017)Kingma, Salimans, Jozefowicz, Chen, Sutskever,
  \& Welling}]{kingma2017improving}
Kingma, D.~P., Salimans, T., Jozefowicz, R., {et~al.} 2017, Improving
  Variational Inference with Inverse Autoregressive Flow.
\newblock \doarXiv{1606.04934}

\bibitem[{{Klypin} \& {Prada}(2019)}]{2019MNRAS.489.1684K}
{Klypin}, A., \& {Prada}, F. 2019, \mnras, 489, 1684,
  \dodoi{10.1093/mnras/stz2194}

\bibitem[{{Kobyzev} {et~al.}(2019){Kobyzev}, {Prince}, \&
  {Brubaker}}]{2019arXiv190809257K}
{Kobyzev}, I., {Prince}, S. J.~D., \& {Brubaker}, M.~A. 2019, arXiv e-prints,
  arXiv:1908.09257.
\newblock \doarXiv{1908.09257}

\bibitem[{{Komissarov}(2005)}]{2005paoa.confE..15K}
{Komissarov}, S. 2005, in KITP Conference: Physics of Astrophysical Outflows
  and Accretion Disks, 15

\bibitem[{{Komissarov}(2001)}]{2001MNRAS.326L..41K}
{Komissarov}, S.~S. 2001, \mnras, 326, L41,
  \dodoi{10.1046/j.1365-8711.2001.04863.x}

\bibitem[{{Kormendy} \& {Ho}(2013)}]{2013ARA&A..51..511K}
{Kormendy}, J., \& {Ho}, L.~C. 2013, \araa, 51, 511,
  \dodoi{10.1146/annurev-astro-082708-101811}

\bibitem[{{Koz{\l}owski}(2017)}]{2017ApJS..228....9K}
{Koz{\l}owski}, S. 2017, \apjs, 228, 9, \dodoi{10.3847/1538-4365/228/1/9}

\bibitem[{{Kulkarni} {et~al.}(2019{\natexlab{a}}){Kulkarni}, {Worseck}, \&
  {Hennawi}}]{Kulkarni+2019}
{Kulkarni}, G., {Worseck}, G., \& {Hennawi}, J.~F. 2019{\natexlab{a}}, VizieR
  Online Data Catalog, J/MNRAS/488/1035

\bibitem[{{Kulkarni} {et~al.}(2019{\natexlab{b}}){Kulkarni}, {Worseck}, \&
  {Hennawi}}]{2019MNRAS.488.1035K}
---. 2019{\natexlab{b}}, \mnras, 488, 1035, \dodoi{10.1093/mnras/stz1493}

\bibitem[{{Kurk} {et~al.}(2007){Kurk}, {Walter}, {Fan}, {Jiang}, {Riechers},
  {Rix}, {Pentericci}, {Strauss}, {Carilli}, \& {Wagner}}]{2007ApJ...669...32K}
{Kurk}, J.~D., {Walter}, F., {Fan}, X., {et~al.} 2007, \apj, 669, 32,
  \dodoi{10.1086/521596}

\bibitem[{{Latif} \& {Khochfar}(2020)}]{2020MNRAS.497.3761L}
{Latif}, M.~A., \& {Khochfar}, S. 2020, \mnras, 497, 3761,
  \dodoi{10.1093/mnras/staa2218}

\bibitem[{{Liska} {et~al.}(2018){Liska}, {Hesp}, {Tchekhovskoy}, {Ingram}, {van
  der Klis}, \& {Markoff}}]{2018MNRAS.474L..81L}
{Liska}, M., {Hesp}, C., {Tchekhovskoy}, A., {et~al.} 2018, \mnras, 474, L81,
  \dodoi{10.1093/mnrasl/slx174}

\bibitem[{{Lodato} \& {Natarajan}(2006)}]{2006MNRAS.371.1813L}
{Lodato}, G., \& {Natarajan}, P. 2006, \mnras, 371, 1813,
  \dodoi{10.1111/j.1365-2966.2006.10801.x}

\bibitem[{{Lodato} \& {Natarajan}(2007)}]{2007MNRAS.377L..64L}
---. 2007, \mnras, 377, L64, \dodoi{10.1111/j.1745-3933.2007.00304.x}

\bibitem[{{LSST Science Collaboration} {et~al.}(2009){LSST Science
  Collaboration}, {Abell}, {Allison}, {Anderson}, {Andrew}, {Angel}, {Armus},
  {Arnett}, {Asztalos}, {Axelrod}, {Bailey}, {Ballantyne}, {Bankert},
  {Barkhouse}, {Barr}, {Barrientos}, {Barth}, {Bartlett}, {Becker}, {Becla},
  {Beers}, {Bernstein}, {Biswas}, {Blanton}, {Bloom}, {Bochanski}, {Boeshaar},
  {Borne}, {Bradac}, {Brandt}, {Bridge}, {Brown}, {Brunner}, {Bullock},
  {Burgasser}, {Burge}, {Burke}, {Cargile}, {Chand rasekharan}, {Chartas},
  {Chesley}, {Chu}, {Cinabro}, {Claire}, {Claver}, {Clowe}, {Connolly}, {Cook},
  {Cooke}, {Cooray}, {Covey}, {Culliton}, {de Jong}, {de Vries}, {Debattista},
  {Delgado}, {Dell'Antonio}, {Dhital}, {Di Stefano}, {Dickinson}, {Dilday},
  {Djorgovski}, {Dobler}, {Donalek}, {Dubois-Felsmann}, {Durech},
  {Eliasdottir}, {Eracleous}, {Eyer}, {Falco}, {Fan}, {Fassnacht}, {Ferguson},
  {Fernandez}, {Fields}, {Finkbeiner}, {Figueroa}, {Fox}, {Francke}, {Frank},
  {Frieman}, {Fromenteau}, {Furqan}, {Galaz}, {Gal-Yam}, {Garnavich},
  {Gawiser}, {Geary}, {Gee}, {Gibson}, {Gilmore}, {Grace}, {Green}, {Gressler},
  {Grillmair}, {Habib}, {Haggerty}, {Hamuy}, {Harris}, {Hawley}, {Heavens},
  {Hebb}, {Henry}, {Hileman}, {Hilton}, {Hoadley}, {Holberg}, {Holman},
  {Howell}, {Infante}, {Ivezic}, {Jacoby}, {Jain}, {R}, {Jedicke}, {Jee},
  {Garrett Jernigan}, {Jha}, {Johnston}, {Jones}, {Juric}, {Kaasalainen},
  {Styliani}, {Kafka}, {Kahn}, {Kaib}, {Kalirai}, {Kantor}, {Kasliwal},
  {Keeton}, {Kessler}, {Knezevic}, {Kowalski}, {Krabbendam}, {Krughoff},
  {Kulkarni}, {Kuhlman}, {Lacy}, {Lepine}, {Liang}, {Lien}, {Lira}, {Long},
  {Lorenz}, {Lotz}, {Lupton}, {Lutz}, {Macri}, {Mahabal}, {Mandelbaum},
  {Marshall}, {May}, {McGehee}, {Meadows}, {Meert}, {Milani}, {Miller},
  {Miller}, {Mills}, {Minniti}, {Monet}, {Mukadam}, {Nakar}, {Neill}, {Newman},
  {Nikolaev}, {Nordby}, {O'Connor}, {Oguri}, {Oliver}, {Olivier}, {Olsen},
  {Olsen}, {Olszewski}, {Oluseyi}, {Padilla}, {Parker}, {Pepper}, {Peterson},
  {Petry}, {Pinto}, {Pizagno}, {Popescu}, {Prsa}, {Radcka}, {Raddick},
  {Rasmussen}, {Rau}, {Rho}, {Rhoads}, {Richards}, {Ridgway}, {Robertson},
  {Roskar}, {Saha}, {Sarajedini}, {Scannapieco}, {Schalk}, {Schindler},
  {Schmidt}, {Schmidt}, {Schneider}, {Schumacher}, {Scranton}, {Sebag},
  {Seppala}, {Shemmer}, {Simon}, {Sivertz}, {Smith}, {Allyn Smith}, {Smith},
  {Spitz}, {Stanford}, {Stassun}, {Strader}, {Strauss}, {Stubbs}, {Sweeney},
  {Szalay}, {Szkody}, {Takada}, {Thorman}, {Trilling}, {Trimble}, {Tyson}, {Van
  Berg}, {Vand en Berk}, {VanderPlas}, {Verde}, {Vrsnak}, {Walkowicz}, {Wand
  elt}, {Wang}, {Wang}, {Warner}, {Wechsler}, {West}, {Wiecha}, {Williams},
  {Willman}, {Wittman}, {Wolff}, {Wood-Vasey}, {Wozniak}, {Young}, {Zentner},
  \& {Zhan}}]{2009arXiv0912.0201L}
{LSST Science Collaboration}, {Abell}, P.~A., {Allison}, J., {et~al.} 2009,
  arXiv e-prints, arXiv:0912.0201.
\newblock \doarXiv{0912.0201}

\bibitem[{{Lucie-Smith} {et~al.}(2020){Lucie-Smith}, {Peiris}, {Pontzen},
  {Nord}, \& {Thiyagalingam}}]{2020arXiv201110577L}
{Lucie-Smith}, L., {Peiris}, H.~V., {Pontzen}, A., {Nord}, B., \&
  {Thiyagalingam}, J. 2020, arXiv e-prints, arXiv:2011.10577.
\newblock \doarXiv{2011.10577}

\bibitem[{{Magorrian} {et~al.}(1998){Magorrian}, {Tremaine}, {Richstone},
  {Bender}, {Bower}, {Dressler}, {Faber}, {Gebhardt}, {Green}, {Grillmair},
  {Kormendy}, \& {Lauer}}]{1998AJ....115.2285M}
{Magorrian}, J., {Tremaine}, S., {Richstone}, D., {et~al.} 1998, \aj, 115,
  2285, \dodoi{10.1086/300353}

\bibitem[{{Mahabal} {et~al.}(2005){Mahabal}, {Djorgovski}, {Graham},
  {Kollipara}, {Granett}, {Krause}, {Williams}, {Bogosavljevic}, {Baltay},
  {Rabinowitz}, {Bauer}, {Andrews}, {Ellman}, {Duffau}, {Jerke}, {Rengstorf},
  {Brunner}, {Musser}, {Mufson}, \& {Gebhard}}]{2005ASPC..347..604M}
{Mahabal}, A.~A., {Djorgovski}, S.~G., {Graham}, M.~J., {et~al.} 2005, in
  Astronomical Society of the Pacific Conference Series, Vol. 347, Astronomical
  Data Analysis Software and Systems XIV, ed. P.~{Shopbell}, M.~{Britton}, \&
  R.~{Ebert}, 604

\bibitem[{{Marziani} {et~al.}(2019){Marziani}, {del Olmo},
  {Mart{\'\i}nez-Carballo}, {Mart{\'\i}nez-Aldama}, {Stirpe}, {Negrete},
  {Dultzin}, {D'Onofrio}, {Bon}, \& {Bon}}]{Marziani+2019}
{Marziani}, P., {del Olmo}, A., {Mart{\'\i}nez-Carballo}, M.~A., {et~al.} 2019,
  \aap, 627, A88, \dodoi{10.1051/0004-6361/201935265}

\bibitem[{{Matsuoka} {et~al.}(2019){Matsuoka}, {Onoue}, {Kashikawa}, {Strauss},
  {Iwasawa}, {Lee}, {Imanishi}, {Nagao}, {Akiyama}, {Asami}, {Bosch},
  {Furusawa}, {Goto}, {Gunn}, {Harikane}, {Ikeda}, {Izumi}, {Kawaguchi},
  {Kato}, {Kikuta}, {Kohno}, {Komiyama}, {Koyama}, {Lupton}, {Minezaki},
  {Miyazaki}, {Murayama}, {Niida}, {Nishizawa}, {Noboriguchi}, {Oguri}, {Ono},
  {Ouchi}, {Price}, {Sameshima}, {Schulze}, {Shirakata}, {Silverman},
  {Sugiyama}, {Tait}, {Takada}, {Takata}, {Tanaka}, {Tang}, {Toba}, {Utsumi},
  {Wang}, \& {Yamashita}}]{2019ApJ...872L...2M}
{Matsuoka}, Y., {Onoue}, M., {Kashikawa}, N., {et~al.} 2019, \apjl, 872, L2,
  \dodoi{10.3847/2041-8213/ab0216}

\bibitem[{{Matsuoka} {et~al.}(2022){Matsuoka}, {Iwasawa}, {Onoue}, {Izumi},
  {Kashikawa}, {Strauss}, {Imanishi}, {Nagao}, {Akiyama}, {Silverman}, {Asami},
  {Bosch}, {Furusawa}, {Goto}, {Gunn}, {Harikane}, {Ikeda}, {Ishimoto},
  {Kawaguchi}, {Kato}, {Kikuta}, {Kohno}, {Komiyama}, {Lee}, {Lupton},
  {Minezaki}, {Miyazaki}, {Murayama}, {Nishizawa}, {Oguri}, {Ono}, {Ouchi},
  {Price}, {Sameshima}, {Sugiyama}, {Tait}, {Takada}, {Takahashi}, {Takata},
  {Tanaka}, {Toba}, {Utsumi}, {Wang}, \& {Yamashita}}]{2022ApJS..259...18M}
{Matsuoka}, Y., {Iwasawa}, K., {Onoue}, M., {et~al.} 2022, \apjs, 259, 18,
  \dodoi{10.3847/1538-4365/ac3d31}

\bibitem[{{Mazzucchelli} {et~al.}(2017){Mazzucchelli}, {Ba{\~n}ados},
  {Venemans}, {Decarli}, {Farina}, {Walter}, {Eilers}, {Rix}, {Simcoe},
  {Stern}, {Fan}, {Schlafly}, {De Rosa}, {Hennawi}, {Chambers}, {Greiner},
  {Burgett}, {Draper}, {Kaiser}, {Kudritzki}, {Magnier}, {Metcalfe}, {Waters},
  \& {Wainscoat}}]{2017ApJ...849...91M}
{Mazzucchelli}, C., {Ba{\~n}ados}, E., {Venemans}, B.~P., {et~al.} 2017, \apj,
  849, 91, \dodoi{10.3847/1538-4357/aa9185}

\bibitem[{{McConnell} \& {Ma}(2013)}]{2013ApJ...764..184M}
{McConnell}, N.~J., \& {Ma}, C.-P. 2013, \apj, 764, 184,
  \dodoi{10.1088/0004-637X/764/2/184}

\bibitem[{{McGibbon} \& {Khochfar}(2021)}]{multi_epoch_ml}
{McGibbon}, R., \& {Khochfar}, S. 2021, arXiv e-prints, arXiv:2112.08424.
\newblock \doarXiv{2112.08424}

\bibitem[{{McGreer} {et~al.}(2013){McGreer}, {Jiang}, {Fan}, {Richards},
  {Strauss}, {Ross}, {White}, {Shen}, {Schneider}, {Myers}, {Brandt}, {DeGraf},
  {Glikman}, {Ge}, \& {Streblyanska}}]{2013ApJ...768..105M}
{McGreer}, I.~D., {Jiang}, L., {Fan}, X., {et~al.} 2013, \apj, 768, 105,
  \dodoi{10.1088/0004-637X/768/2/105}

\bibitem[{{McKinney}(2005)}]{2005ApJ...630L...5M}
{McKinney}, J.~C. 2005, \apjl, 630, L5, \dodoi{10.1086/468184}

\bibitem[{{Meriot} {et~al.}(2022){Meriot}, {Khochfar}, {O{\~n}orbe}, \&
  {Smith}}]{2022MNRAS.512...27M}
{Meriot}, R., {Khochfar}, S., {O{\~n}orbe}, J., \& {Smith}, B. 2022, \mnras,
  512, 27, \dodoi{10.1093/mnras/stac435}

\bibitem[{{Miyoshi} {et~al.}(1995){Miyoshi}, {Moran}, {Herrnstein},
  {Greenhill}, {Nakai}, {Diamond}, \& {Inoue}}]{1995Natur.373..127M}
{Miyoshi}, M., {Moran}, J., {Herrnstein}, J., {et~al.} 1995, \nat, 373, 127,
  \dodoi{10.1038/373127a0}

\bibitem[{{Moews} {et~al.}(2021){Moews}, {Dav{\'e}}, {Mitra}, {Hassan}, \&
  {Cui}}]{2021MNRAS.504.4024M}
{Moews}, B., {Dav{\'e}}, R., {Mitra}, S., {Hassan}, S., \& {Cui}, W. 2021,
  \mnras, 504, 4024, \dodoi{10.1093/mnras/stab1120}

\bibitem[{{Mortlock} {et~al.}(2011){Mortlock}, {Warren}, {Venemans}, {Patel},
  {Hewett}, {McMahon}, {Simpson}, {Theuns}, {Gonz{\'a}les-Solares}, {Adamson},
  {Dye}, {Hambly}, {Hirst}, {Irwin}, {Kuiper}, {Lawrence}, \&
  {R{\"o}ttgering}}]{2011Natur.474..616M}
{Mortlock}, D.~J., {Warren}, S.~J., {Venemans}, B.~P., {et~al.} 2011, \nat,
  474, 616, \dodoi{10.1038/nature10159}

\bibitem[{{Moster} {et~al.}(2020){Moster}, {Naab}, {Lindstr{\"o}m}, \&
  {O'Leary}}]{2020arXiv200512276M}
{Moster}, B.~P., {Naab}, T., {Lindstr{\"o}m}, M., \& {O'Leary}, J.~A. 2020,
  arXiv e-prints, arXiv:2005.12276.
\newblock \doarXiv{2005.12276}

\bibitem[{{Nandra} {et~al.}(2006){Nandra}, {O'Neill}, {George}, {Reeves}, \&
  {Turner}}]{2006AN....327.1039N}
{Nandra}, K., {O'Neill}, P.~M., {George}, I.~M., {Reeves}, J.~N., \& {Turner},
  T.~J. 2006, Astronomische Nachrichten, 327, 1039,
  \dodoi{10.1002/asna.200610641}

\bibitem[{{Natarajan}(2014)}]{2014GReGr..46.1702N}
{Natarajan}, P. 2014, General Relativity and Gravitation, 46, 1702,
  \dodoi{10.1007/s10714-014-1702-6}

\bibitem[{Natarajan {et~al.}(2017)Natarajan, Pacucci, Ferrara, Agarwal,
  Ricarte, Zackrisson, \& Cappelluti}]{Natarajan_2017}
Natarajan, P., Pacucci, F., Ferrara, A., {et~al.} 2017, The Astrophysical
  Journal, 838, 117, \dodoi{10.3847/1538-4357/aa6330}

\bibitem[{{Natarajan} \& {Treister}(2009)}]{PN_UMBH2009}
{Natarajan}, P., \& {Treister}, E. 2009, \mnras, 393, 838,
  \dodoi{10.1111/j.1365-2966.2008.13864.x}

\bibitem[{{Natarajan} {et~al.}(2019){Natarajan}, {Ricarte}, {Baldassare},
  {Bellovary}, {Bender}, {Berti}, {Cappelluti}, {Ferrara}, {Greene}, {Haiman},
  {Holley-Bockelmann}, {Mueller}, {Pacucci}, {Shoemaker}, {Shoemaker},
  {Tremmel}, {Urry}, {Vikhlinin}, \& {Volonteri}}]{2019BAAS...51c..73N}
{Natarajan}, P., {Ricarte}, A., {Baldassare}, V., {et~al.} 2019, \baas, 51, 73.
\newblock \doarXiv{1904.09326}

\bibitem[{{Nelson} {et~al.}(2015){Nelson}, {Pillepich}, {Genel},
  {Vogelsberger}, {Springel}, {Torrey}, {Rodriguez-Gomez}, {Sijacki}, {Snyder},
  {Griffen}, {Marinacci}, {Blecha}, {Sales}, {Xu}, \&
  {Hernquist}}]{2015A&C....13...12N}
{Nelson}, D., {Pillepich}, A., {Genel}, S., {et~al.} 2015, Astronomy and
  Computing, 13, 12, \dodoi{10.1016/j.ascom.2015.09.003}

\bibitem[{{Nelson} {et~al.}(2019){Nelson}, {Springel}, {Pillepich},
  {Rodriguez-Gomez}, {Torrey}, {Genel}, {Vogelsberger}, {Pakmor}, {Marinacci},
  {Weinberger}, {Kelley}, {Lovell}, {Diemer}, \&
  {Hernquist}}]{2019ComAC...6....2N}
{Nelson}, D., {Springel}, V., {Pillepich}, A., {et~al.} 2019, Computational
  Astrophysics and Cosmology, 6, 2, \dodoi{10.1186/s40668-019-0028-x}

\bibitem[{{Netzer}(2015)}]{2015ARA&A..53..365N}
{Netzer}, H. 2015, \araa, 53, 365, \dodoi{10.1146/annurev-astro-082214-122302}

\bibitem[{{Nord} {et~al.}(2019){Nord}, {Connolly}, {Kinney}, {Kubica},
  {Narayan}, {Peek}, {Schafer}, \& {Tollerud}}]{2019BAAS...51g.224N}
{Nord}, B., {Connolly}, A.~J., {Kinney}, J., {et~al.} 2019, in Bulletin of the
  American Astronomical Society, Vol.~51, 224

\bibitem[{{Ntampaka} {et~al.}(2020){Ntampaka}, {Eisenstein}, {Yuan}, \&
  {Garrison}}]{2020ApJ...889..151N}
{Ntampaka}, M., {Eisenstein}, D.~J., {Yuan}, S., \& {Garrison}, L.~H. 2020,
  \apj, 889, 151, \dodoi{10.3847/1538-4357/ab5f5e}

\bibitem[{{Ntampaka} {et~al.}(2019){Ntampaka}, {Avestruz}, {Boada}, {Caldeira},
  {Cisewski-Kehe}, {Di Stefano}, {Dvorkin}, {Evrard}, {Farahi}, {Finkbeiner},
  {Genel}, {Goodman}, {Goulding}, {Ho}, {Kosowsky}, {La Plante}, {Lanusse},
  {Lochner}, {Mandelbaum}, {Nagai}, {Newman}, {Nord}, {Peek}, {Peel}, {Poczos},
  {Rau}, {Siemiginowska}, {Sutherland}, {Trac}, \&
  {Wandelt}}]{2019BAAS...51c..14N}
{Ntampaka}, M., {Avestruz}, C., {Boada}, S., {et~al.} 2019, \baas, 51, 14.
\newblock \doarXiv{1902.10159}

\bibitem[{{Pacucci} {et~al.}(2016){Pacucci}, {Ferrara}, {Grazian}, {Fiore},
  {Giallongo}, \& {Puccetti}}]{2016MNRAS.459.1432P}
{Pacucci}, F., {Ferrara}, A., {Grazian}, A., {et~al.} 2016, \mnras, 459, 1432,
  \dodoi{10.1093/mnras/stw725}

\bibitem[{{Papamakarios} {et~al.}(2019){Papamakarios}, {Nalisnick}, {Jimenez
  Rezende}, {Mohamed}, \& {Lakshminarayanan}}]{2019arXiv191202762P}
{Papamakarios}, G., {Nalisnick}, E., {Jimenez Rezende}, D., {Mohamed}, S., \&
  {Lakshminarayanan}, B. 2019, arXiv e-prints, arXiv:1912.02762.
\newblock \doarXiv{1912.02762}

\bibitem[{Papamakarios {et~al.}(2018)Papamakarios, Pavlakou, \&
  Murray}]{papamakarios2018masked}
Papamakarios, G., Pavlakou, T., \& Murray, I. 2018, Masked Autoregressive Flow
  for Density Estimation.
\newblock \doarXiv{1705.07057}

\bibitem[{{Peng}(2007)}]{2007ApJ...671.1098P}
{Peng}, C.~Y. 2007, \apj, 671, 1098, \dodoi{10.1086/522774}

\bibitem[{{Peterson}(1993)}]{1993PASP..105..247P}
{Peterson}, B.~M. 1993, \pasp, 105, 247, \dodoi{10.1086/133140}

\bibitem[{{Peterson}(2014)}]{2014SSRv..183..253P}
---. 2014, \ssr, 183, 253, \dodoi{10.1007/s11214-013-9987-4}

\bibitem[{{Peterson} {et~al.}(2004){Peterson}, {Ferrarese}, {Gilbert}, {Kaspi},
  {Malkan}, {Maoz}, {Merritt}, {Netzer}, {Onken}, {Pogge}, {Vestergaard}, \&
  {Wandel}}]{2004ApJ...613..682P}
{Peterson}, B.~M., {Ferrarese}, L., {Gilbert}, K.~M., {et~al.} 2004, \apj, 613,
  682, \dodoi{10.1086/423269}

\bibitem[{{Powell} {et~al.}(2020){Powell}, {Urry}, {Cappelluti}, {Johnson},
  {LaMassa}, {Ananna}, \& {Kollmann}}]{2020ApJ...891...41P}
{Powell}, M.~C., {Urry}, C.~M., {Cappelluti}, N., {et~al.} 2020, \apj, 891, 41,
  \dodoi{10.3847/1538-4357/ab6e65}

\bibitem[{Powell {et~al.}(2018)Powell, Cappelluti, Urry, Koss, Finoguenov,
  Ricci, Trakhtenbrot, Allevato, Ajello, Oh, Schawinski, \&
  Secrest}]{Powell_2018}
Powell, M.~C., Cappelluti, N., Urry, C.~M., {et~al.} 2018, The Astrophysical
  Journal, 858, 110, \dodoi{10.3847/1538-4357/aabd7f}

\bibitem[{{Reynolds}(2020)}]{2020arXiv201108948R}
{Reynolds}, C.~S. 2020, arXiv e-prints, arXiv:2011.08948.
\newblock \doarXiv{2011.08948}

\bibitem[{Rezende \& Mohamed(2016)}]{rezende2016variational}
Rezende, D.~J., \& Mohamed, S. 2016, Variational Inference with Normalizing
  Flows.
\newblock \doarXiv{1505.05770}

\bibitem[{{Ricarte} \& {Natarajan}(2018{\natexlab{a}})}]{2018MNRAS.474.1995R}
{Ricarte}, A., \& {Natarajan}, P. 2018{\natexlab{a}}, \mnras, 474, 1995,
  \dodoi{10.1093/mnras/stx2851}

\bibitem[{{Ricarte} \& {Natarajan}(2018{\natexlab{b}})}]{2018MNRAS.481.3278R}
---. 2018{\natexlab{b}}, \mnras, 481, 3278, \dodoi{10.1093/mnras/sty2448}

\bibitem[{{Ricarte} {et~al.}(2019){Ricarte}, {Tremmel}, {Natarajan}, \&
  {Quinn}}]{2019MNRAS.489..802R}
{Ricarte}, A., {Tremmel}, M., {Natarajan}, P., \& {Quinn}, T. 2019, \mnras,
  489, 802, \dodoi{10.1093/mnras/stz2161}

\bibitem[{{Richards} {et~al.}(2002){Richards}, {Fan}, {Newberg}, {Strauss},
  {Vanden Berk}, {Schneider}, {Yanny}, {Boucher}, {Burles}, {Frieman}, {Gunn},
  {Hall}, {Ivezi{\'c}}, {Kent}, {Loveday}, {Lupton}, {Rockosi}, {Schlegel},
  {Stoughton}, {SubbaRao}, \& {York}}]{Richards+2002}
{Richards}, G.~T., {Fan}, X., {Newberg}, H.~J., {et~al.} 2002, \aj, 123, 2945,
  \dodoi{10.1086/340187}

\bibitem[{{Robertson} {et~al.}(2006){Robertson}, {Bullock}, {Cox}, {Di Matteo},
  {Hernquist}, {Springel}, \& {Yoshida}}]{2006ApJ...645..986R}
{Robertson}, B., {Bullock}, J.~S., {Cox}, T.~J., {et~al.} 2006, \apj, 645, 986,
  \dodoi{10.1086/504412}

\bibitem[{{Ross} {et~al.}(2013){Ross}, {McGreer}, {White}, {Richards}, {Myers},
  {Palanque-Delabrouille}, {Strauss}, {Anderson}, {Shen}, {Brandt},
  {Y{\`e}che}, {Swanson}, {Aubourg}, {Bailey}, {Bizyaev}, {Bovy}, {Brewington},
  {Brinkmann}, {DeGraf}, {Di Matteo}, {Ebelke}, {Fan}, {Ge}, {Malanushenko},
  {Malanushenko}, {Mandelbaum}, {Maraston}, {Muna}, {Oravetz}, {Pan},
  {P{\^a}ris}, {Petitjean}, {Schawinski}, {Schlegel}, {Schneider}, {Silverman},
  {Simmons}, {Snedden}, {Streblyanska}, {Suzuki}, {Weinberg}, \&
  {York}}]{2013ApJ...773...14R}
{Ross}, N.~P., {McGreer}, I.~D., {White}, M., {et~al.} 2013, \apj, 773, 14,
  \dodoi{10.1088/0004-637X/773/1/14}

\bibitem[{{Schawinski} {et~al.}(2017){Schawinski}, {Zhang}, {Zhang}, {Fowler},
  \& {Santhanam}}]{2017MNRAS.467L.110S}
{Schawinski}, K., {Zhang}, C., {Zhang}, H., {Fowler}, L., \& {Santhanam}, G.~K.
  2017, \mnras, 467, L110, \dodoi{10.1093/mnrasl/slx008}

\bibitem[{{Schawinski} {et~al.}(2006){Schawinski}, {Khochfar}, {Kaviraj}, {Yi},
  {Boselli}, {Barlow}, {Conrow}, {Forster}, {Friedman}, {Martin}, {Morrissey},
  {Neff}, {Schiminovich}, {Seibert}, {Small}, {Wyder}, {Bianchi}, {Donas},
  {Heckman}, {Lee}, {Madore}, {Milliard}, {Rich}, \&
  {Szalay}}]{2006Natur.442..888S}
{Schawinski}, K., {Khochfar}, S., {Kaviraj}, S., {et~al.} 2006, \nat, 442, 888,
  \dodoi{10.1038/nature04934}

\bibitem[{{Schaye} {et~al.}(2015){Schaye}, {Crain}, {Bower}, {Furlong},
  {Schaller}, {Theuns}, {Dalla Vecchia}, {Frenk}, {McCarthy}, {Helly},
  {Jenkins}, {Rosas-Guevara}, {White}, {Baes}, {Booth}, {Camps}, {Navarro},
  {Qu}, {Rahmati}, {Sawala}, {Thomas}, \& {Trayford}}]{2015MNRAS.446..521S}
{Schaye}, J., {Crain}, R.~A., {Bower}, R.~G., {et~al.} 2015, \mnras, 446, 521,
  \dodoi{10.1093/mnras/stu2058}

\bibitem[{{Schneider} {et~al.}(2010){Schneider}, {Richards}, {Hall}, {Strauss},
  {Anderson}, {Boroson}, {Ross}, {Shen}, {Brandt}, {Fan}, {Inada}, {Jester},
  {Knapp}, {Krawczyk}, {Thakar}, {Vanden Berk}, {Voges}, {Yanny}, {York},
  {Bahcall}, {Bizyaev}, {Blanton}, {Brewington}, {Brinkmann}, {Eisenstein},
  {Frieman}, {Fukugita}, {Gray}, {Gunn}, {Hibon}, {Ivezi{\'c}}, {Kent}, {Kron},
  {Lee}, {Lupton}, {Malanushenko}, {Malanushenko}, {Oravetz}, {Pan}, {Pier},
  {Price}, {Saxe}, {Schlegel}, {Simmons}, {Snedden}, {SubbaRao}, {Szalay}, \&
  {Weinberg}}]{2010AJ....139.2360S}
{Schneider}, D.~P., {Richards}, G.~T., {Hall}, P.~B., {et~al.} 2010, \aj, 139,
  2360, \dodoi{10.1088/0004-6256/139/6/2360}

\bibitem[{{Sexton} {et~al.}(2019){Sexton}, {Canalizo}, {Hiner}, {Komossa},
  {Woo}, {Treister}, \& {Hiner Dimassimo}}]{2019ApJ...878..101S}
{Sexton}, R.~O., {Canalizo}, G., {Hiner}, K.~D., {et~al.} 2019, \apj, 878, 101,
  \dodoi{10.3847/1538-4357/ab21d5}

\bibitem[{{Shao} {et~al.}(2017){Shao}, {Wang}, {Jones}, {Carilli}, {Walter},
  {Fan}, {Riechers}, {Bertoldi}, {Wagg}, {Strauss}, {Omont}, {Cox}, {Jiang},
  {Narayanan}, \& {Menten}}]{2017ApJ...845..138S}
{Shao}, Y., {Wang}, R., {Jones}, G.~C., {et~al.} 2017, \apj, 845, 138,
  \dodoi{10.3847/1538-4357/aa826c}

\bibitem[{{Shen}(2013)}]{2013BASI...41...61S}
{Shen}, Y. 2013, Bulletin of the Astronomical Society of India, 41, 61.
\newblock \doarXiv{1302.2643}

\bibitem[{{Shen} \& {Liu}(2012)}]{2012ApJ...753..125S}
{Shen}, Y., \& {Liu}, X. 2012, \apj, 753, 125,
  \dodoi{10.1088/0004-637X/753/2/125}

\bibitem[{Shen \& Liu(2012)}]{Shen_2012}
Shen, Y., \& Liu, X. 2012, The Astrophysical Journal, 753, 125,
  \dodoi{10.1088/0004-637x/753/2/125}

\bibitem[{{Shen} {et~al.}(2011){Shen}, {Richards}, {Strauss}, {Hall},
  {Schneider}, {Snedden}, {Bizyaev}, {Brewington}, {Malanushenko},
  {Malanushenko}, {Oravetz}, {Pan}, \& {Simmons}}]{2011ApJS..194...45S}
{Shen}, Y., {Richards}, G.~T., {Strauss}, M.~A., {et~al.} 2011, \apjs, 194, 45,
  \dodoi{10.1088/0067-0049/194/2/45}

\bibitem[{{Shen} {et~al.}(2019{\natexlab{a}}){Shen}, {Hall}, {Horne}, {Zhu},
  {McGreer}, {Simm}, {Trump}, {Kinemuchi}, {Brandt}, {Green}, {Grier}, {Guo},
  {Ho}, {Homayouni}, {Jiang}, {I-Hsiu Li}, {Morganson}, {Petitjean},
  {Richards}, {Schneider}, {Starkey}, {Wang}, {Chambers}, {Kaiser},
  {Kudritzki}, {Magnier}, \& {Waters}}]{2019ApJS..241...34S}
{Shen}, Y., {Hall}, P.~B., {Horne}, K., {et~al.} 2019{\natexlab{a}}, \apjs,
  241, 34, \dodoi{10.3847/1538-4365/ab074f}

\bibitem[{{Shen} {et~al.}(2019{\natexlab{b}}){Shen}, {Wu}, {Jiang},
  {Ba{\~n}ados}, {Fan}, {Ho}, {Riechers}, {Strauss}, {Venemans}, {Vestergaard},
  {Walter}, {Wang}, {Willott}, {Wu}, \& {Yang}}]{2019ApJ...873...35S}
{Shen}, Y., {Wu}, J., {Jiang}, L., {et~al.} 2019{\natexlab{b}}, \apj, 873, 35,
  \dodoi{10.3847/1538-4357/ab03d9}

\bibitem[{{Sijacki} {et~al.}(2007){Sijacki}, {Springel}, {Di Matteo}, \&
  {Hernquist}}]{2007MNRAS.380..877S}
{Sijacki}, D., {Springel}, V., {Di Matteo}, T., \& {Hernquist}, L. 2007,
  \mnras, 380, 877, \dodoi{10.1111/j.1365-2966.2007.12153.x}

\bibitem[{{Sijacki} {et~al.}(2015){Sijacki}, {Vogelsberger}, {Genel},
  {Springel}, {Torrey}, {Snyder}, {Nelson}, \&
  {Hernquist}}]{2015MNRAS.452..575S}
{Sijacki}, D., {Vogelsberger}, M., {Genel}, S., {et~al.} 2015, \mnras, 452,
  575, \dodoi{10.1093/mnras/stv1340}

\bibitem[{{Small} \& {Blandford}(1992)}]{1992MNRAS.259..725S}
{Small}, T.~A., \& {Blandford}, R.~D. 1992, \mnras, 259, 725,
  \dodoi{10.1093/mnras/259.4.725}

\bibitem[{{Springel} {et~al.}(2005{\natexlab{a}}){Springel}, {Di Matteo}, \&
  {Hernquist}}]{2005MNRAS.361..776S}
{Springel}, V., {Di Matteo}, T., \& {Hernquist}, L. 2005{\natexlab{a}}, \mnras,
  361, 776, \dodoi{10.1111/j.1365-2966.2005.09238.x}

\bibitem[{{Springel} {et~al.}(2005{\natexlab{b}}){Springel}, {Di Matteo}, \&
  {Hernquist}}]{springel+2005}
---. 2005{\natexlab{b}}, \mnras, 361, 776,
  \dodoi{10.1111/j.1365-2966.2005.09238.x}

\bibitem[{{Springel} {et~al.}(2006){Springel}, {Frenk}, \&
  {White}}]{2006Natur.440.1137S}
{Springel}, V., {Frenk}, C.~S., \& {White}, S. D.~M. 2006, \nat, 440, 1137,
  \dodoi{10.1038/nature04805}

\bibitem[{{Springel} {et~al.}(2001){Springel}, {White}, {Tormen}, \&
  {Kauffmann}}]{2001MNRAS.328..726S}
{Springel}, V., {White}, S. D.~M., {Tormen}, G., \& {Kauffmann}, G. 2001,
  \mnras, 328, 726, \dodoi{10.1046/j.1365-8711.2001.04912.x}

\bibitem[{Steidel {et~al.}(1999)Steidel, Adelberger, Giavalisco, Dickinson, \&
  Pettini}]{Steidel_1999}
Steidel, C.~C., Adelberger, K.~L., Giavalisco, M., Dickinson, M., \& Pettini,
  M. 1999, The Astrophysical Journal, 519, 1, \dodoi{10.1086/307363}

\bibitem[{{Storchi-Bergmann} \&
  {Schnorr-M{\"u}ller}(2019)}]{2019NatAs...3...48S}
{Storchi-Bergmann}, T., \& {Schnorr-M{\"u}ller}, A. 2019, Nature Astronomy, 3,
  48, \dodoi{10.1038/s41550-018-0611-0}

\bibitem[{{Tachibana} {et~al.}(2020){Tachibana}, {Graham}, {Kawai},
  {Djorgovski}, {Drake}, {Mahabal}, \& {Stern}}]{2020ApJ...903...54T}
{Tachibana}, Y., {Graham}, M.~J., {Kawai}, N., {et~al.} 2020, \apj, 903, 54,
  \dodoi{10.3847/1538-4357/abb9a9}

\bibitem[{{Taylor} {et~al.}(2019){Taylor}, {Burke-Spolaor}, {Baker}, {Charisi},
  {Islo}, {Kelley}, {Madison}, {Simon}, {Vigeland}, \& {Nanograv
  Collaboration}}]{2019BAAS...51c.336T}
{Taylor}, S., {Burke-Spolaor}, S., {Baker}, P.~T., {et~al.} 2019, \baas, 51,
  336.
\newblock \doarXiv{1903.08183}

\bibitem[{{Tchekhovskoy} {et~al.}(2010){Tchekhovskoy}, {Narayan}, \&
  {McKinney}}]{2010ApJ...711...50T}
{Tchekhovskoy}, A., {Narayan}, R., \& {McKinney}, J.~C. 2010, \apj, 711, 50,
  \dodoi{10.1088/0004-637X/711/1/50}

\bibitem[{{Tchekhovskoy} {et~al.}(2011){Tchekhovskoy}, {Narayan}, \&
  {McKinney}}]{2011MNRAS.418L..79T}
---. 2011, \mnras, 418, L79, \dodoi{10.1111/j.1745-3933.2011.01147.x}

\bibitem[{{Trakhtenbrot} \& {Netzer}(2012)}]{2012MNRAS.427.3081T}
{Trakhtenbrot}, B., \& {Netzer}, H. 2012, \mnras, 427, 3081,
  \dodoi{10.1111/j.1365-2966.2012.22056.x}

\bibitem[{{Trakhtenbrot} {et~al.}(2011){Trakhtenbrot}, {Netzer}, {Lira}, \&
  {Shemmer}}]{2011ApJ...730....7T}
{Trakhtenbrot}, B., {Netzer}, H., {Lira}, P., \& {Shemmer}, O. 2011, \apj, 730,
  7, \dodoi{10.1088/0004-637X/730/1/7}

\bibitem[{{Treister} {et~al.}(2011){Treister}, {Schawinski}, {Volonteri},
  {Natarajan}, \& {Gawiser}}]{2011Natur.474..356T}
{Treister}, E., {Schawinski}, K., {Volonteri}, M., {Natarajan}, P., \&
  {Gawiser}, E. 2011, \nat, 474, 356, \dodoi{10.1038/nature10103}

\bibitem[{{Tremaine} {et~al.}(1994){Tremaine}, {Richstone}, {Byun}, {Dressler},
  {Faber}, {Grillmair}, {Kormendy}, \& {Lauer}}]{1994AJ....107..634T}
{Tremaine}, S., {Richstone}, D.~O., {Byun}, Y.-I., {et~al.} 1994, \aj, 107,
  634, \dodoi{10.1086/116883}

\bibitem[{{Tremaine} {et~al.}(2002){Tremaine}, {Gebhardt}, {Bender}, {Bower},
  {Dressler}, {Faber}, {Filippenko}, {Green}, {Grillmair}, {Ho}, {Kormendy},
  {Lauer}, {Magorrian}, {Pinkney}, \& {Richstone}}]{2002ApJ...574..740T}
{Tremaine}, S., {Gebhardt}, K., {Bender}, R., {et~al.} 2002, \apj, 574, 740,
  \dodoi{10.1086/341002}

\bibitem[{{Urry} \& {Padovani}(1995)}]{1995PASP..107..803U}
{Urry}, C.~M., \& {Padovani}, P. 1995, \pasp, 107, 803, \dodoi{10.1086/133630}

\bibitem[{{Venemans} {et~al.}(2015){Venemans}, {Ba{\~n}ados}, {Decarli},
  {Farina}, {Walter}, {Chambers}, {Fan}, {Rix}, {Schlafly}, {McMahon},
  {Simcoe}, {Stern}, {Burgett}, {Draper}, {Flewelling}, {Hodapp}, {Kaiser},
  {Magnier}, {Metcalfe}, {Morgan}, {Price}, {Tonry}, {Waters}, {AlSayyad},
  {Banerji}, {Chen}, {Gonz{\'a}lez-Solares}, {Greiner}, {Mazzucchelli},
  {McGreer}, {Miller}, {Reed}, \& {Sullivan}}]{2015ApJ...801L..11V}
{Venemans}, B.~P., {Ba{\~n}ados}, E., {Decarli}, R., {et~al.} 2015, \apjl, 801,
  L11, \dodoi{10.1088/2041-8205/801/1/L11}

\bibitem[{{Vestergaard}(2019)}]{2019NatAs...3...11V}
{Vestergaard}, M. 2019, Nature Astronomy, 3, 11,
  \dodoi{10.1038/s41550-018-0670-2}

\bibitem[{{Vestergaard} \& {Osmer}(2009)}]{2009ApJ...699..800V}
{Vestergaard}, M., \& {Osmer}, P.~S. 2009, \apj, 699, 800,
  \dodoi{10.1088/0004-637X/699/1/800}

\bibitem[{{Vestergaard} \& {Peterson}(2006)}]{2006ApJ...641..689V}
{Vestergaard}, M., \& {Peterson}, B.~M. 2006, \apj, 641, 689,
  \dodoi{10.1086/500572}

\bibitem[{{Villaescusa-Navarro} {et~al.}(2022){Villaescusa-Navarro}, {Ding},
  {Genel}, {Tonnesen}, {La Torre}, {Spergel}, {Teyssier}, {Li}, {Heneka},
  {Lemos}, {Angl{\'e}s-Alc{\'a}zar}, {Nagai}, \&
  {Vogelsberger}}]{2022arXiv220102202V}
{Villaescusa-Navarro}, F., {Ding}, J., {Genel}, S., {et~al.} 2022, arXiv
  e-prints, arXiv:2201.02202.
\newblock \doarXiv{2201.02202}

\bibitem[{{Vogelsberger} {et~al.}(2014){Vogelsberger}, {Genel}, {Springel},
  {Torrey}, {Sijacki}, {Xu}, {Snyder}, {Nelson}, \&
  {Hernquist}}]{2014MNRAS.444.1518V}
{Vogelsberger}, M., {Genel}, S., {Springel}, V., {et~al.} 2014, \mnras, 444,
  1518, \dodoi{10.1093/mnras/stu1536}

\bibitem[{{Volonteri}(2012)}]{2012Sci...337..544V}
{Volonteri}, M. 2012, Science, 337, 544, \dodoi{10.1126/science.1220843}

\bibitem[{{Warren} {et~al.}(1991){Warren}, {Hewett}, {Irwin}, \&
  {Osmer}}]{Warren+1991a}
{Warren}, S.~J., {Hewett}, P.~C., {Irwin}, M.~J., \& {Osmer}, P.~S. 1991,
  \apjs, 76, 1, \dodoi{10.1086/191563}

\bibitem[{{Wechsler} \& {Tinker}(2018)}]{Wechsler+2018}
{Wechsler}, R.~H., \& {Tinker}, J.~L. 2018, \araa, 56, 435,
  \dodoi{10.1146/annurev-astro-081817-051756}

\bibitem[{{Weinberger} {et~al.}(2017){Weinberger}, {Springel}, {Hernquist},
  {Pillepich}, {Marinacci}, {Pakmor}, {Nelson}, {Genel}, {Vogelsberger},
  {Naiman}, \& {Torrey}}]{2017MNRAS.465.3291W}
{Weinberger}, R., {Springel}, V., {Hernquist}, L., {et~al.} 2017, \mnras, 465,
  3291, \dodoi{10.1093/mnras/stw2944}

\bibitem[{{Weinberger} {et~al.}(2018){Weinberger}, {Springel}, {Pakmor},
  {Nelson}, {Genel}, {Pillepich}, {Vogelsberger}, {Marinacci}, {Naiman},
  {Torrey}, \& {Hernquist}}]{2018MNRAS.479.4056W}
{Weinberger}, R., {Springel}, V., {Pakmor}, R., {et~al.} 2018, \mnras, 479,
  4056, \dodoi{10.1093/mnras/sty1733}

\bibitem[{{Willott} {et~al.}(2010{\natexlab{a}}){Willott}, {Albert},
  {Arzoumanian}, {Bergeron}, {Crampton}, {Delorme}, {Hutchings}, {Omont},
  {Reyl{\'e}}, \& {Schade}}]{2010AJ....140..546W}
{Willott}, C.~J., {Albert}, L., {Arzoumanian}, D., {et~al.} 2010{\natexlab{a}},
  \aj, 140, 546, \dodoi{10.1088/0004-6256/140/2/546}

\bibitem[{{Willott} {et~al.}(2010{\natexlab{b}}){Willott}, {Delorme},
  {Reyl{\'e}}, {Albert}, {Bergeron}, {Crampton}, {Delfosse}, {Forveille},
  {Hutchings}, {McLure}, {Omont}, \& {Schade}}]{2010AJ....139..906W}
{Willott}, C.~J., {Delorme}, P., {Reyl{\'e}}, C., {et~al.} 2010{\natexlab{b}},
  \aj, 139, 906, \dodoi{10.1088/0004-6256/139/3/906}

\bibitem[{{Wise} {et~al.}(2019){Wise}, {Regan}, {O'Shea}, {Norman}, {Downes},
  \& {Xu}}]{2019Natur.566...85W}
{Wise}, J.~H., {Regan}, J.~A., {O'Shea}, B.~W., {et~al.} 2019, \nat, 566, 85,
  \dodoi{10.1038/s41586-019-0873-4}

\bibitem[{{Woo} {et~al.}(2015){Woo}, {Yoon}, {Park}, {Park}, \&
  {Kim}}]{2015ApJ...801...38W}
{Woo}, J.-H., {Yoon}, Y., {Park}, S., {Park}, D., \& {Kim}, S.~C. 2015, \apj,
  801, 38, \dodoi{10.1088/0004-637X/801/1/38}

\bibitem[{{Woods} {et~al.}(2019){Woods}, {Agarwal}, {Bromm}, {Bunker}, {Chen},
  {Chon}, {Ferrara}, {Glover}, {Haemmerl{\'e}}, {Haiman}, {Hartwig}, {Heger},
  {Hirano}, {Hosokawa}, {Inayoshi}, {Klessen}, {Kobayashi}, {Koliopanos},
  {Latif}, {Li}, {Mayer}, {Mezcua}, {Natarajan}, {Pacucci}, {Rees}, {Regan},
  {Sakurai}, {Salvadori}, {Schneider}, {Surace}, {Tanaka}, {Whalen}, \&
  {Yoshida}}]{2019PASA...36...27W}
{Woods}, T.~E., {Agarwal}, B., {Bromm}, V., {et~al.} 2019, \pasa, 36, e027,
  \dodoi{10.1017/pasa.2019.14}

\bibitem[{{Yang} {et~al.}(2016){Yang}, {Wang}, {Wu}, {Fan}, {McGreer}, {Bian},
  {Yi}, {Yang}, {Ai}, {Dong}, {Zuo}, {Green}, {Jiang}, {Wang}, {Wang}, \&
  {Yue}}]{2016ApJ...829...33Y}
{Yang}, J., {Wang}, F., {Wu}, X.-B., {et~al.} 2016, \apj, 829, 33,
  \dodoi{10.3847/0004-637X/829/1/33}

\bibitem[{{Yao-Yu Lin} {et~al.}(2020){Yao-Yu Lin}, {Pandya}, {Pratap}, {Liu},
  \& {Carrasco Kind}}]{2020arXiv201115095Y}
{Yao-Yu Lin}, J., {Pandya}, S., {Pratap}, D., {Liu}, X., \& {Carrasco Kind}, M.
  2020, arXiv e-prints, arXiv:2011.15095.
\newblock \doarXiv{2011.15095}

\bibitem[{{Znajek}(1977)}]{1977MNRAS.179..457Z}
{Znajek}, R.~L. 1977, \mnras, 179, 457, \dodoi{10.1093/mnras/179.3.457}

\bibitem[{{Zuo} {et~al.}(2015){Zuo}, {Wu}, {Fan}, {Green}, {Wang}, \&
  {Bian}}]{2015ApJ...799..189Z}
{Zuo}, W., {Wu}, X.-B., {Fan}, X., {et~al.} 2015, \apj, 799, 189,
  \dodoi{10.1088/0004-637X/799/2/189}

\end{thebibliography}
